\documentclass[twocolumn,
showpacs,
showkeys,
preprintnumbers,
nofootinbib,
superscriptaddress,
amsmath,
amssymb,
floatfix,
secnumarabic,
aps,
pra,
a4paper,
notitlepage,
final,
]{revtex4}%

\usepackage[colorlinks=true,urlcolor=blue]{hyperref}
\usepackage{graphicx}
\usepackage{epsfig}

\newcommand{\beq}{\begin{equation}}
\newcommand{\eeq}{\end{equation}}
\newcommand{\beqar}{\begin{eqnarray}}
\newcommand{\eeqar}{\end{eqnarray}}
\newcommand{\bal}{\begin{aligned}}
\newcommand{\eal}{\end{aligned}}

\def\dalam{\hbox
{\vrule\vbox{\hrule\hbox to 1ex{ \hfill}\kern 1 ex\hrule}\vrule}}

\def\1/2{\hbox{$ {1 \over 2}$ }}
\def\tr{\hbox{Tr}}

\def\h{\hbar}
\def\i/h{{i \over \h}}

\def\sh{\sinh}

\def\ctg{\hbox{ctg}}

\def\inf{\infty}

\def\a{\alpha} 
\def\b{\beta} 
 
\def\g{\gamma}  
\def\d{\delta} 
\def\l{\lambda} \def\L{\Lambda}
\def\e{\epsilon} \def\E{\hbox{$\cal E $}}

\def\s{\sigma}
\def\r{\rho}

\def\p{\psi}

\def\tt{\theta}

\def\<{\langle}
\def\>{\rangle}

\def\({\left(}
\def\[{\left[}
\def\){\right)}
\def\]{\right]}

\usepackage{subfigure}

\usepackage[notcite,notref,color]{showkeys}  

\usepackage{tikz}
\usetikzlibrary{decorations.pathreplacing}
\usetikzlibrary{decorations.pathmorphing}    
\usepackage{multirow}
\usepackage{dcolumn}		
\newcolumntype{.}{D{.}{.}{-1}}
\newcolumntype{i}[1]{D{.}{.}{#1}}

\newcommand{\myfrac}[2]{{\ifmmode{}^{#1}\!/_{\!#2}\else${}^{#1}\!/_{\!#2}$\fi}}

\begin{document}

\sloppy

\title{Casimir force variability in one-dimensional QED systems}

\author{Yu.~Voronina}
\email{voroninayu@physics.msu.ru} \affiliation{Department of Physics and
Institute of Theoretical Problems of MicroWorld, Moscow State
University, 119991, Leninsky Gory, Moscow, Russia}

\author{I.~Komissarov}
\email{ i.komissarov@columbia.edu} \affiliation{Department of Physics, Columbia University, New York, NY 10027, USA}

\author{K.~Sveshnikov}
\email{costa@bog.msu.ru} \affiliation{Department of Physics and
Institute of Theoretical Problems of MicroWorld, Moscow State
University, 119991, Leninsky Gory, Moscow, Russia}

\date{\today}


\begin{abstract}
The Casimir force between two short-range   charge sources,  embedded in a background of one dimensional massive Dirac fermions, is explored  by means of the original $\ln\text{[Wronskian]}$ contour integration techniques. For identical sources with the same (positive) charge we find that in the non-perturbative region  the Casimir interaction between them can reach  sufficiently large negative values and simultaneously reveal the features of a long-range force  in spite of nonzero fermion mass, that could significantly influence the properties of such  quasi-one-dimensional QED systems.  For large distances $s$ between   sources   we recover that  their mutual interaction is governed first of all by the structure of the discrete spectrum  of a single source, in dependence on which it can be tuned to give an attractive, a repulsive, or an (almost) compensated   Casimir force with various rates of the exponential fall-down, quite different from the standard $\exp (-2 m s)$ law. By means of the same $\ln\text{[Wronskian]}$  techniques the case of two $\d$-sources is also considered in a self-consistent manner with similar results for variability of the Casimir force. A quite different behavior of the Casimir force is found for the antisymmetric source-anti-source system. In particular, in this case there is no possibility for a long-range interaction between sources. The asymptotics of the Casimir force follows the standard $\exp (-2 m s)$ law. Moreover, for small separations between sources the Casimir force for symmetric and antisymmetric cases turns out to be of opposite sign.
\end{abstract}


\maketitle

\section{Introduction}\label{sec:intro}

There is now a lot of interest to essentially non-perturbative vacuum polarization (Casimir)  effects in  quasi-one-dimensional QED systems caused by charged impurities. Actually, one-dimensional QED systems with impurities appear nowadays
in many  situations, which fill  the range from  relativistic H-like atoms in a strong homogenous magnetic field ~\cite{atom, davydov2017, sveshnikov2017, voronina2017}
 up to charged impurities in low-dimensional nanostructures like semiconductor quantum wires,  carbon nanotubes, in conducting
polymers, etc.~\cite{giamarchi2004},  fermionic atoms in ultracold
gases~\cite{moritz2003, recati2005a, kolomeisky2008} and defects in one-dimensional fermionic quantum  liquids~\cite{recati2005b, fuchs2007, romeo2016}. Impurities have a profound effect on the physical properties of these low-dimensional systems. In certain exceptionally
clean systems, impurities can be created and controlled up to the Casimir forces between them mediated by  fermions. The general literature on the Casimir effect is
vast and the reader may consult Ref.~\cite{14} for some experimental
results and Refs.~\cite{15}-\cite{19} for reviews and background
work. The Casimir interaction mediated by fermions has been intensively studied from different points of view and in different geometries during the last two decades in Refs.~\cite{20}. The main result is that for Dirac fermions we have a Casimir force whose strength and sign can
be tuned by the impurity separation and their internal structure. This provides a
physical situation where the Casimir interaction could be continuously tunable from attractive through almost completely compensated to the repulsive one by variation of an internal control parameter, realizing the known
bounds for the one dimensional Casimir interaction as
two limiting cases. In the light of
proofs showing the absence of repulsive Casimir interactions
for the photonic field in vacuum, this is a quite remarkable situation.
Moreover, in Ref.~\cite{tanaka2013} it was shown that the electronic Casimir force between two impurities on a one-dimensional semiconductor quantum wire can be of a very long range, despite nonzero effective mass of the mediator.

Of special interest in the fermionic Casimir effect is the situation, when for some reasons the impurities should be modeled as $\d$-like sources, since the Dirac equation (DE) is inconsistent with direct inserting of external $\d$-potentials.  This problem   was explored in Ref.~\cite{Jaffe2004} in terms of the energy density and interaction between two ``Dirac spikes'' as a function of a single ``spike'' parameters and the distance between them. In this model each ``spike'' is represented by a square barrier, which enters the fermion dynamics as an additional mass term, and the  $\d$-limit is considered via transfer-matrix, which in this limit allows for a self-consistent treatment. In Refs.~\cite{nanotubes} the Casimir interaction between two square potential barriers  (``scatterers''), mediated by the massless fermions, has been considered. The Casimir force between the scatterers was found for both the case of finite width and strength of the barriers and in the  $\d$-limit. The result of both works is that for identical $\d$-scatterers, separated by a large distance $d$, the interaction  force between them reveals the conventional attractive asymptotics $\sim 1/d^2$. At the same time,   for a more general case of inequivalent scatterers the magnitude and sign of the force depend on their relative spinor polarizations  ~\cite{nanotubes}.

In this paper within the framework of general quasi-one-dimensional QED system we consider the Casimir interaction of two short-range   Coulomb sources, either extended or $\d$-like, which enter the fermion dynamics as localized electrostatic potentials. In the case of the scalar coupling, considered in Refs.~\cite{Jaffe2004,nanotubes}, the scatterers affect equally the positive- and negative-frequency fermionic modes. In the  case of vector coupling the behavior of electronic and positronic components  is principally different and leads to a number of new effects, the most significant of which is the discrete levels diving into the lower continuum and related non-perturbative effects of vacuum reconstruction, when the positively charged sources  attain the overcritical region. The main question of interest is how these non-perturbative  effects, including the effects of super-criticality, manifest themselves in Casimir forces between such sources. For identical positively charged sources, by means of the original  $\ln\text{[Wronskian]}$ contour integration techniques,  we find that   the interaction energy between sources can exceed sufficiently large negative values and simultaneously reveal the features of a long-range force  in spite of nonzero fermion mass, which could significantly influence the properties of such  quasi-one-dimensional QED systems.  Moreover,  the Casimir force shows up a highly nontrivial behavior with increasing distance between sources, which includes separate vertical jumps at finite distances, caused by the effects of discrete levels creation-annihilation  at the lower threshold, as well as different  exponent rates and signs in the asymptotics.  The  case of two $\d$-like sources is also considered in detail. To  the contrary, the antisymmetric source-anti-source system reveals quite different features. In particular, in this case there is no possibility for the long-range interaction between sources. The asymptotics of the Casimir force follows the standard $\exp (-2 m s)$ law. Moreover,  in the symmetric case the Casimir force between sources for small separations is attractive, while in the antisymmetric one it turns into sufficiently strong repulsion. Remarkably enough, the classic electrostatic force for  such Coulomb sources should be of opposite sign. There is no evident explanation for this effect. However, the set of parameters used is quite wide to consider this effect as a general one. These results may be relevant for indirect interactions between charged defects and adsorbed species in  quasi-one-dimensional QED systems mentioned above.

The single short-range positively charged Coulomb source is described as a potential square well of width $2 a$ and depth $V_0$
\begin{equation}\label{v}
V(x) =-V_0\, \tt (a-|x|) \ .
\end{equation}
Actually the potential (\ref{v}) could be interpreted  as created by the charged impurity considered as a spherical shell of radius  $R_0$ and charge $Z$, strongly screened for $|x|>R_0$. For this case
\begin{equation}
V_0=Z \a/R_0 \ .
\end{equation}

In this work we consider the system of two such sources, separated by the distance $s$. The  component of the vacuum polarization energy $\E_{vac}$, responsible for their interaction, is defined as
\begin{equation}
\E_{vac}^{int}(s)=\E_{vac,2}(s)-\E_{vac,2}(s \to \inf) \ ,
\label{casint}\end{equation}
where  $\E_{vac,2}(s)$ is the total vacuum polarization energy for the system containing two potentials like (\ref{v}), located at the distance $s$ from each other, while  $\E_{vac,2}(s \to \inf)=2\, \E_{vac,1}$ with the latter being the vacuum energy of a single source.

It would be worth to note that since we consider here the sources with  several parameters (for a single well these are the depth $V_0$ and the half-width $a$), the subcritical and overcritical regions for a concrete level are defined by a set of pairs $(V_0,~a)$, rather than by a single quantity $Z_{cr}$, as it occurs  whenever a concrete relation between the size and charge of the Coulomb source is implied. In the case of a single source (\ref{v}) in the diagram  $(V_0,~a)$ the subcritical and overcritical regions are separated by a curve (see Fig. \ref{dcr}). Therefore under the notion of the  ``critical charge''  $Z_{cr,i}$ for the $i$-th discrete level we'll imply the whole set of the source parameters, which separate the sub- and overcritical regions from each other, rather than one definite quantity.

As in other works on vacuum polarization in  strong  Coulomb fields ~\cite{wk1956}-\cite{davydov2018}, the radiative corrections from virtual photons are neglected. Henceforth, if it is not stipulated separately, the units  $\hbar=m_e=c=1$ are used. Thence the coupling constant $\a=e^2$ is also dimensionless, and the numerical calculations, illustrating the general picture, are performed for  $\a=1/137.036$. However, it would be worthwhile to note that for the effective electron-hole vacuum in the quasi-one-dimensional systems like nanotubes and wires, as in graphene, the actual value of the finite structure constant and hence, the magnitude of the Casimir effects could be quite different from the one in the pure  QED.

\subsection*{2. Evaluation of the Casimir energy via $\ln$[Wronskian] contour integration}

The starting point for the essentially non-perturbative evaluation of the vacuum energy in QED systems is the Schwinger vacuum average  ~\cite{wk1956}-\cite{21},~\cite{mohr1998}
\begin{equation}\begin{gathered}
\E_{vac}=  \dfrac{1}{2}\left(\sum\limits_{\e_n<\e_F}\e_n-\sum\limits_{\e_n \geqslant \e_F}\e_n\right)_V \ - \\
 - \ \dfrac{1}{2}\left(\sum\limits_{\e_n<\e_F}\e_n-\sum\limits_{\e_n \geqslant \e_F}\e_n\right)_0  \ ,
\label{eterms}
\end{gathered}\end{equation}
\normalsize
with $\e_n$ being the eigenvalues of the corresponding DE
\begin{equation}
\[\alpha p + \beta + V(x)\] \psi(x)= \epsilon\,\psi(x) \ ,
\label{deq}
\end{equation}
while for the positively charged sources $\e_F$ should be chosen at the lower threshold, i.e. $\e_F=-1$. The label $V$ indicates the presence of the external potential, while the label $0$ corresponds to the free case. Throughout the paper by solving DE the representation $\a=\s_2\, ,  \b=\s_3$ is used.

For the subsequent analysis it is convenient to separate in  (\ref{eterms}) the contributions from the discrete spectrum and both continua and apply to the latter the well-known tool, which replaces it by the scattering phase $\delta(\epsilon)$ integration (see e.g., Refs.~\cite{raja1982, sv1991, Jaffe2004} and refs. therein). After a number of almost evident steps  one obtains~\cite{sveshnikov2017}
\begin{equation}
\E_{vac}=\dfrac{1}{2 \pi}\int\limits_1^{+\infty} \delta_{tot}(\epsilon)~d\epsilon +\frac{1}{2}\sum\limits_{-1 \leqslant \e_n < 1} (1-\e_n) \ ,
\label{evac}
\end{equation}
where $\delta_{tot}(\epsilon)$ is the total phase shift for the given  $|\e|$, including the contributions from scattering states in both continua, while in the discrete spectrum the (effective) electron rest mass is subtracted from each level in order to retain in $\E_{vac}$ the interaction effects only.

Such approach to calculation of $ \E_{vac} $ turns out to be quite effective, since the total  phase shift $\d_{tot}(\e)$ behaves in both (IR and UV) limits much better, than each of the elastic phases separately, and is automatically an even function of the external potential. More concretely, in the Coulomb potentials with non-vanishing source size  $\d_{tot}(\e)$ is finite for $\epsilon \rightarrow 1$ and decreases  $\sim 1/\epsilon^3$ for $\e \rightarrow \infty$, that provides the convergence of the phase integral in  (\ref{evac}) ~\cite{davydov2017}-\cite{voronina2017},~\cite{Jaffe2004, davydov2018, voronina2018}. The sum over bound energies $1-\e_n$ of discrete levels  in the case of short-range sources like (\ref{v}) is finite from the very beginning, since such potentials allow for a finite number of discrete levels. As a result, the expression (\ref{evac}) turns out to be finite without any additional renormalization.

However, the convergence of $\E_{vac}$ in the form (\ref{evac}) is completely caused by the specifics of 1+1 D and in no way means no need for a renormalization.  Renormalization via  fermionic loop is required on account of the analysis of the vacuum charge density $\rho_{vac}(x)$, from which there follows that without such UV-renormalization the integral induced charge will not acquire the value that follows from quite obvious physical arguments~\cite{davydov2017}-\cite{voronina2017},~\cite{gyul1975}-\cite{mohr1998}. For the system under consideration such analysis is performed in Refs.~\cite{annphys, tmf} for both cases including  the singlet and doublet of sources like (\ref{v}).

Another obvious requirement  is that for $V_0 \rightarrow 0$ the value of $\E_{vac}$ should coincide with $\E_{vac}^{(1)}$, obtained in the first order of the QED perturbation theory (PT).   The latter is found quite similar to the  unscreened case considered in Refs.~\cite{davydov2017}-\cite{voronina2017} and for a single source like (\ref{v}) equals to
\begin{equation}
\E_{vac,1}^{(1)}=\frac{2V_0^2}{\pi^2}\, \int\limits_{0}^{+\infty}dq~\frac{\sin^2{qa}}{q^2}
\left(1-2{ \mathrm{arcsinh} (q/2) \over q \sqrt{1+(q/2)^2}}\right) \ ,
\label{evacperturb}
\end{equation}
while for the configuration containing a doublet of such identical sources, separated by the distance $d$, it is given by the following expression
\begin{equation}\begin{gathered}
\E_{vac,2}^{(1)}=\frac{2V_0^2}{\pi^2}\int\limits_{0}^{+\infty}dq~\frac{\[\sin{(q(a+d))]}-
\sin{(qd)}\]^2}{q^2} \times \\ \times \left(1-2{ \mathrm{arcsinh} (q/2) \over q \sqrt{1+(q/2)^2}}\right) \ . \label{evacperturb2}
\end{gathered}\end{equation}
It is easy to verify that the non-renormalized vacuum energy (\ref{evac}) doesn't satisfy this requirement, since the introduced below renormalization coefficient (\ref{ren}), which provides also the correspondence between $\E_{vac}^R$  and $\E_{vac}^{(1)}$ for $V_0 \rightarrow 0$, in general case  turns out to be non-zero with the only exception for certain parameter sets.

For these reasons, in complete analogy with the renormalization of the charge density, considered in Refs.~\cite{davydov2017}-\cite{voronina2017},~\cite{gyul1975}-\cite{voronina2018},   we should pass from $\E_{vac}$ to the renormalized vacuum energy $\E_{vac}^R$. In the practically useful form $\E_{vac}^R$ should be represented as follows~\cite{davydov2017}-\cite{voronina2017},~\cite{davydov2018, voronina2018}
\begin{equation}
\E_{vac}^R=\E_{vac}+\lambda V_0^2 \ ,
\label{evren}
\end{equation}
where the renormalization coefficient
\begin{equation}
\lambda=\lim\limits_{V_0\rightarrow 0} \dfrac{\E_{vac}^{(1)}(V_0)-\E_{vac}(V_0)}{V_0^2} \ .
\label{ren}
\end{equation}
depends solely on the shape of the external potential and in the general 1+1 D case is a sign-alternating function with zeros \cite{davydov2017}-\cite{voronina2017}, \cite{annphys}.

\begin{figure}[b]
	\begin{center}
		\includegraphics[width=0.5\textwidth]{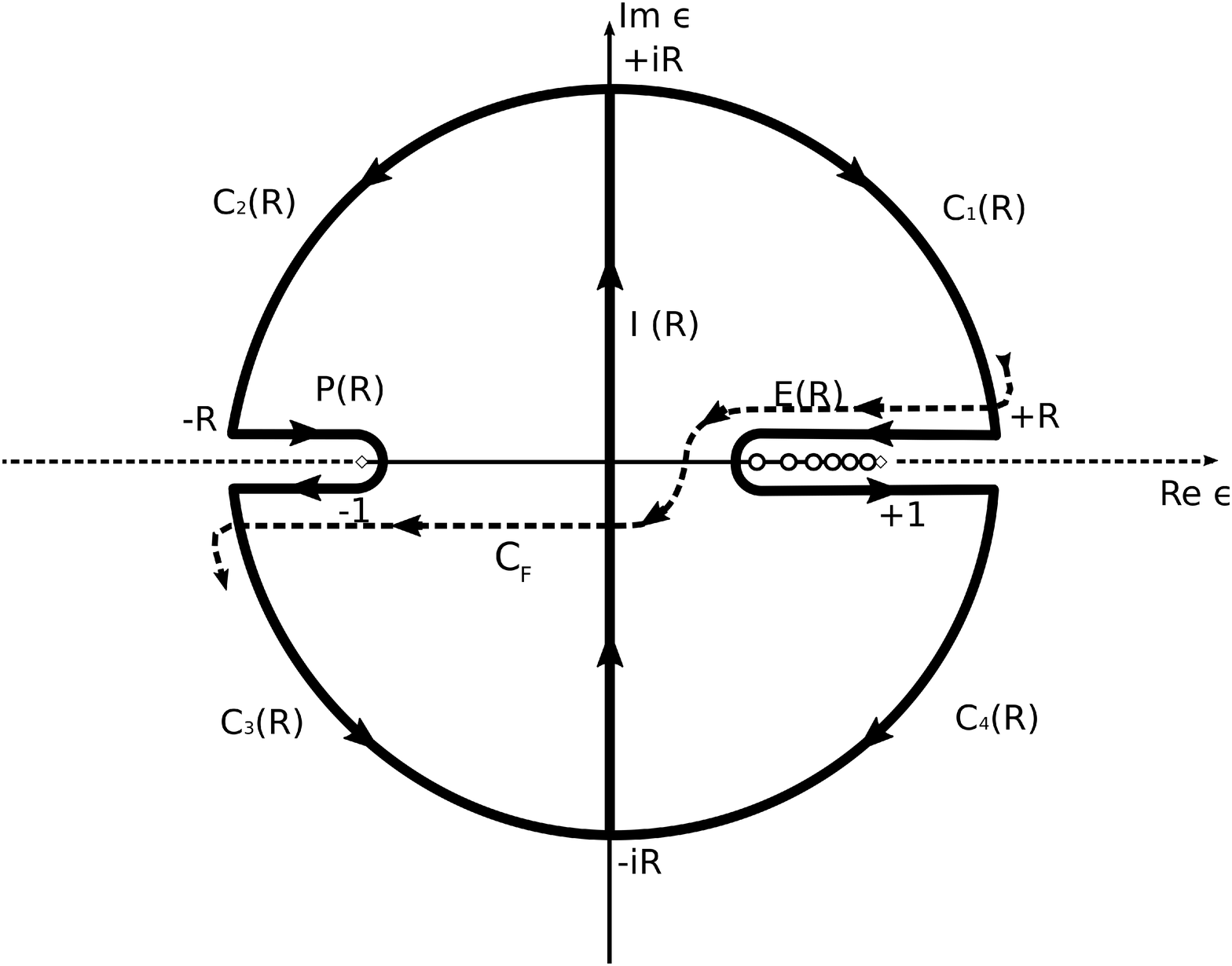}
	\end{center}
	\caption{The WK-contours in the complex energy plane, used for the representation of the vacuum charge density and vacuum energy via contour integrals. }
	\label{contour}
\end{figure}

The evaluation of $\E_{vac}^R$  via the sum of the phase integral and discrete levels is considered in detail in Refs.~\cite{davydov2017}-\cite{voronina2017},~\cite{davydov2018, voronina2018} for the unscreened or partially screened extended Coulomb source, and in the present case will differ only by certain  technical details. However, for our purposes of a detailed study of Casimir interaction between the localized Coulomb-like external sources  an alternative approach for evaluation of the non-renormalized $\E_{vac}$ turns out to be more efficient. This approach is quite similar to the calculation of the vacuum density  $\rho_{vac}(x)$ via integration of specially constructed function along the Wichmann-Kroll (WK)  contours \cite{wk1956, gyul1975, mohr1998}, which are shown in  Fig.\ref{contour}.
Namely, it is easy to see that the function
\begin{equation}
F(\epsilon,V_0)=\dfrac{\epsilon \left(\mathrm{d} J(\epsilon)/ \mathrm{d} \epsilon \right)}{J(\epsilon)} \ ,
\label{F}
\end{equation}
where $J(\e)$ is the Wronskian for the spectral problem (\ref{deq}), reveals all the pole properties, which are required for the representation  of the expression (\ref{eterms}) via integrals along the WK contours, since actually $J(\e)$  is nothing else, but the Jost function of the spectral problem (\ref{deq}) \cite{sveshnikov2017}. The real zeros of  $J(\e)$ lie in the interval $-1 \leqslant \e_n < 1$ and coincide with discrete energy levels, while the complex ones reside on the second sheet of the Riemann energy surface with negative imaginary part of the wavenumber $k=\sqrt{\e^2-1}$ and for  $\mathrm{Re} \ k >0 $ correspond to the elastic resonances. Moreover, both $J(\e)$ and  $\tr G$ as functions of the wavenumber $k$ reveal the same reflection symmetry $f^{\ast}(k)=f(-k^{\ast})$ of the Jost function.

To represent $\E_{vac}$ via integration along the WK contours, it suffices to pass to the difference  \begin{equation}
\mathcal{H}(\e,V_0)=F(\e,V_0)-F(\e,0) \ ,
\label{H}
\end{equation}
normalized on the free case. As a result, the non-renormalized induced vacuum energy can be represented as
\begin{equation}
\E_{vac}=-\frac{1}{4 \pi i}\lim\limits_{R\rightarrow\infty} \left( \int\limits_{P(R)}d\epsilon ~ \mathcal{H}(\epsilon,V_0) +  \int\limits_{E(R)}d\epsilon~ \mathcal{H}(\epsilon,V_0)\right) \ .
\label{41}\end{equation}
In the next step one finds by means of the analysis of the asymptotics  of the function $\mathcal{H}(\epsilon,V_0)$ on  the large circle in Fig.\ref{contour} that the initial integration along the contours $P(R)$ and $E(R)$ for the singlet or doublet of external potentials like  (\ref{v}) can be reduced to integration along the imaginary axis \cite{annphys}. Upon taking into account the (possible) existence of negative discrete levels and proceeding further in complete analogy with the  corresponding treatment of the vacuum density, performed in Refs. \cite{davydov2017}-\cite{voronina2017},\cite{gyul1975}, one finds the final expression for $\E_{vac}$ in the following form
\begin{equation}
\E_{vac}=\dfrac{1}{2 \pi} \int\limits_{-\infty}^{+\infty} dy~ \mathcal{H}(i y,V_0) - \sum\limits_{-1 \leqslant \e_n <0} \e_n \ .
\label{econt}
\end{equation}
For the single source (\ref{v}) the integrand in (\ref{econt}) takes the form
\begin{widetext}\begin{equation}
\begin{aligned}
\mathcal{H}(i y,V_0)=
\dfrac{iV_0\, y\,(V_0[V_0+2iy]\g(iy)+2aj^2(iy,V_0)\g^2(iy)\sin[2aj(iy,V_0)])}{j^2(iy,V_0)\g^3(iy)(j(iy,V_0)\g(iy)\cos[2aj(iy,V_0)]+(1-iV_0\,y+y^2)\sin[2aj(iy,V_0)])} - \\
- \dfrac{2ia\,V_0\,y\,j(iy,V_0)\,\g^3(iy)\,\cos[2aj(iy,V_0)]}{j^2(iy,V_0)\g^3(iy)\,(j(iy,V_0)\g(iy)\cos[2aj(iy,V_0)]+(1-iV_0\,y+y^2)\sin[2aj(iy,V_0)])} \  ,
\end{aligned}
\end{equation}\end{widetext}
where
\begin{equation}\label{j}
 j(\epsilon,V_0)=\sqrt{(V_0+\epsilon)^2-1} \ , \quad \gamma(\epsilon)=\sqrt{1-\epsilon^2} \ .
\end{equation}
For the doublet configuration the corresponding expression will be presented below.
The direct numerical calculation shows that both approaches to  the vacuum energy (\ref{evac}) and (\ref{econt}) lead with a high precision to the same results.

Besides  $\E_{vac}$, in 1+1 D the renormalization term $\l V_0^2$ in the expression (\ref{evren}) turns out to be quite important, especially in the non-perturbative region. For a single source (\ref{v}) the dependence  of the renormalization term on the source parameters is determined first of all by the coefficient $\lambda(a)$, which can be represented as
\begin{equation}
\l(a)=\l_1(a)-\l_2(a) \ ,
\label{renpart}
\end{equation}
where $\l_1$ emerges from the PT contribution $\E_{vac}^{(1)}$ to the vacuum energy
\begin{multline}
\l_1(a)=\frac{a}{\pi}-I_1(a) \ , \\ I_1(a)=\frac{4}{\pi^2}\int\limits_0^\infty dq ~ \frac{\sin^2(qa)}{q^2}\, \left(1-2{ \mathrm{arcsinh} (q/2) \over q \sqrt{1+(q/2)^2}}\right) \ ,
\label{l1}
\end{multline}
while $\l_2$ corresponds to the first  (quadratic in $V_0$) term in $\E_{vac}$, which is found from the Born series  (see Refs.~\cite{davydov2017}-\cite{voronina2017},~\cite{gyul1975},~\cite{davydov2018, voronina2018})
\begin{equation}
\lambda_2(a)=\frac{a}{\pi}-\frac{1}{16}+I_2(a) \ , \quad I_2(a)=\frac{1}{4\pi} \int\limits_{0}^{\infty} dy~\frac{e^{-4a\sqrt{1+y^2}}}{(1+y^2)^2} \ .
\label{l2}
\end{equation}
By means of the relation $\lambda_1(a)+\lambda_2(a)=a/\pi$,  whose derivation  requires some additional considerations and so is presented separately \cite{tmf}, one obtains
\begin{equation}
\lambda(a)=\dfrac{a}{\pi}-2\lambda_2(a)=\dfrac{1}{8}-\dfrac{a}{\pi}-2I_2(a) \ .
\label{lambdaf1}
\end{equation}
The asymptotics of $\l(a)$ for $a \ll 1$ and $a \gg 1$, which are important for the further analysis of the Casimir interaction between separate sources, are considered in detail in Ref.~\cite{annphys}. So here we present only the required results. Namely, the asymptotics  of $\l(a)$ for  $a \ll 1$ reads
\begin{equation}\label{limlambda}
\lambda(a\rightarrow 0)=\dfrac{a}{\pi}-2 a^2+O(a^3) \ ,
\end{equation}
while for large  $a$  neglecting the exponentially small corrections it is given by
\begin{equation}
\lambda(a \rightarrow \infty)=\dfrac{1}{8}-\dfrac{a}{\pi} \ .
\end{equation}
As  a result, for small $a$ the coefficient $\lambda (a)$ grows linearly, while for large $a$ it decreases  with the same modulus slope $1/\pi$. Hence, the renormalization coefficient $\lambda (a)$ should vanish at certain $a=a_{cr}$. In the  case of the single well (\ref{v}) it  has a unique zero when $a_{cr}\simeq 0.297 $. More details concerning the behavior of $\lambda (a)$ are given in Ref.~\cite{annphys}.

\subsection*{3. Casimir energy  of two identical positively charged short-range Coulomb sources}

Now -- having dealt with the general formulation for calculation of $\E_{vac}$ this way -- let us  turn to the configuration of two such identical positively charged short-range Coulomb sources, described by the potential
\begin{equation}
V_2(x) =-V_0 \, \tt\(|x|- d\)\,\tt\(d+a -|x|\)  \ .
\label{v2}
\end{equation}
Let us note that now the separate sources have the total width $a$, that provides the restoration of the initial potential well (\ref{v}) with the width  $2a$ for $d \to 0$.

Further procedure of calculation and renormalization of the vacuum  energy repeats completely the case of  the single source and so doesn't need any special discussion besides the structure of the renormalization term in $\E_{vac,2}^R$. As in the case of one potential well, the calculation of $\E_{vac,2}^R$ requires the renormalization in the second order with respect to the external potential
\begin{equation}
\E_{vac,2}^R=\E_{vac,2}+\Lambda(a,d) V_0^2 \ ,
\label{evren2}
\end{equation}
where
\begin{equation}
\Lambda(a,d)=\Lambda_1(a,d)-\Lambda_2(a,d) \ .
\end{equation}
The components of the renormalization coefficient $\Lambda_i(a,d)$,  $i=1\, , 2\,$,  are expressed in terms of the corresponding coefficients $\lambda_i(a)$ for the single source as follows
\begin{equation}
\Lambda_i(a,d)=\lambda_i(a+d)+\lambda_i(d)+2\lambda_i(a/2)-2\lambda_i(d+a/2) \ .
\label{lambig12}
\end{equation}
From (\ref{lambig12}) by means of the relation $\lambda_1(a)+\lambda_2(a)=a/\pi$ one finds that $\Lambda_i(a,d)$ are subject of the same relation
\begin{equation}
\Lambda_1(a,d)+\Lambda_2(a,d)=a/\pi \ .
\end{equation}
As a result, the renormalization coefficient for the two-well problem (\ref{v2}) can be represented as
\begin{multline}
\Lambda(a,d)=a/\pi-2\Lambda_2(a,d)
= \\ = a/\pi-2\lambda_2(a+d)-2\lambda_2(d)-4\lambda_2(a/2)+4\lambda_2(d+a/2) \ .
\label{lambifin}
\end{multline}

\begin{figure}[h]
	\begin{center}
	\includegraphics[width=1.0\linewidth]{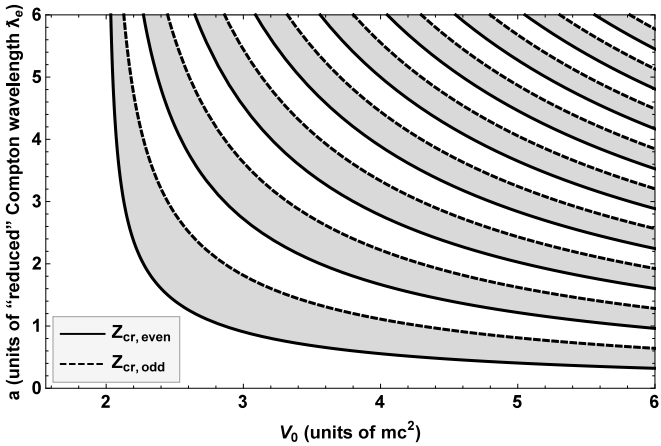}
	\end{center}
	\caption{The shaded regions correspond to those sets of $(V_0\, , a)$, when by varying  $d$ it is possible to provide the diving of the lowest even level into the continuum. The solid lines correspond to even, while the dashed ones --- to odd ``critical charges''  for a single source (\ref{v}).}
	\label{dcr}
\end{figure}

Now let us list the results for found  this way $\E_{vac,2}^R$, which are necessary for the further analysis of the Casimir interaction between separate sources. The most significant here is the dependence  on the parameter $d\, , \ 0 \leq d \leq \inf$, which defines the separation of the sources in such a way that  the distance $s$ between the centers of the wells is given by
\begin{equation}
s=a+2\, d \ .
\end{equation}
At first let us explore the features of the discrete spectrum of DE with the potential (\ref{v2}). For $d \to \inf $ the wells become independent, while the spectrum of DE -- twice degenerate. More concretely, with growing  $d$ the even levels increase, while the odd ones, in contrast, decrease, and so for $d \rightarrow \infty$ the neighboring even and odd levels seek each other. To analyze the role of   $d$ in the overcritical region the equations for the critical parameters of the source (\ref{v2}) (i.e., the set $[V_0$, $a$ , $d]$, for which the discrete levels  approach the threshold of  the lower continuum) should be considered. For  odd levels the  ``critical charges'' are determined from the equation
\begin{equation}
\sin[a\sqrt{(V_0-1)^2-1}]=0 \ ,
\label{codd}\end{equation}
which coincides with the similar equation for a single potential well up to replacement  $a \to 2 a$. Since (\ref{codd}) doesn't depend on $d$, any change of $d$ for fixed  $(V_0\, , a)$ doesn't yield any diving of odd levels into the lower continuum. At the same time, for even levels the equation for their ``critical charges'' takes the form
\begin{multline}
\sqrt{(V_0-1)^2-1} \cos[a\sqrt{(V_0-1)^2-1}] \ + \\ +  \ 2d\,V_0 \sin[a\sqrt{(V_0-1)^2-1}]=0 \ .
\label{creven}
\end{multline}
So for even levels the ``critical charges''  depend on the distance between the sources. The parameter $d$ can be easily found from (\ref{creven}), and so the dependence $d(V_0$, $a)$ together with condition $d > 0$ defines the critical values of $d$ for even levels in the potential (\ref{v2}). The regions  $(V_0\, , a)$, where the even levels diving into the lower continuum takes place by certain $d>0$, are shown as shaded ones in Fig.\ref{dcr}.  The non-shaded regions in Fig.\ref{dcr} correspond to those sets of  $(V_0\, , a)$, when the eq.(\ref{creven}) doesn't possess any solutions with  $d >0 $. The solid and dashed curves in Fig.\ref{dcr} determine the sets $(V_0\, , a)$, when $d_{cr}=0$, and so correspond to the critical charges for a single well.

\begin{figure}[ht!]
\subfigure[]{
		\includegraphics[width=\columnwidth]{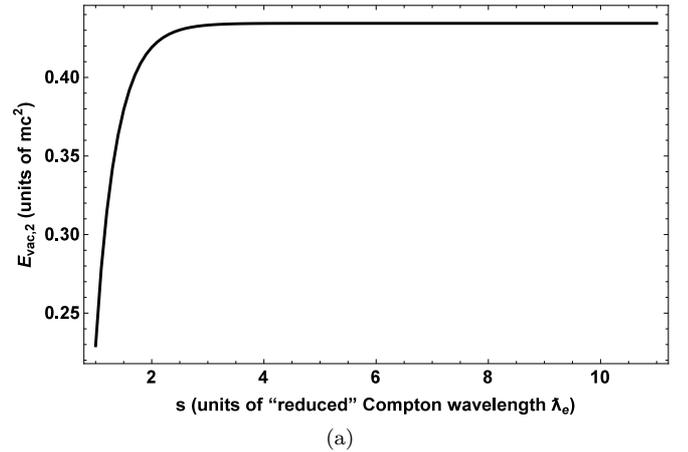}
}
\hfill
\subfigure[]{
		\includegraphics[width=\columnwidth]{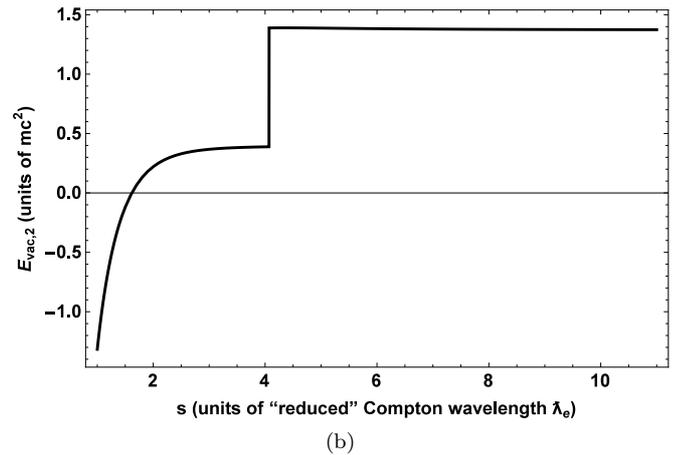}
}
\caption{ (a):  $\E_{vac,2}^{R}(s)$ for (a): $a=1$, $V_0=2$; (b): for $a=1$, $V_0=4.08$.}
	\label{vac2}	
\end{figure}

This way there appear two essentially different regimes for behavior of  $\E_{vac,2}^R$ as a function of the source separation. The first one corresponds to the situation, when in the whole interval  $0< d \leqslant \infty$ no discrete level attains the lower continuum,  nor does any one return back from the continuum (the parameters $(V_0\, , a)$ correspond to  $d<0$ in the eq. (\ref{creven})). In this case the integral vacuum charge $Q_{vac,2}$ keeps its value, while the vacuum energy $\E_{vac,2}^R$, the jumps in which are entirely due to creation-annihilation of discrete levels from the lower continuum,  is a continuous function of  $d$ and $s$. This regime for  $\E_{vac,2}^R(s)$ is shown in Fig.\ref{vac2}a, calculated for $V_0=2$, $a=1$, which correspond to the lowest unpainted region in Fig.\ref{dcr}.
Numerical integration confirms that in this case the integral induced charge $Q_{vac}$ doesn't depend on  $s$ and vanishes, since the parameters $(V_0\, , a)$ lie in the subcritical region and so varying $s$ doesn't lead to appearance of new levels at the lower threshold, while the dependence of the renormalized vacuum energy  $\E_{vac,2}^R(s)$, as it follows from the Fig.\ref{vac2}a, is given by a continuous curve.

The second regime for    $\E_{vac,2}^R(s)$  is realized for $(V_0\, , a)$, which lie in the shaded regions in Fig.\ref{dcr}. For such values of $(V_0\, , a)$  with growing  $s$ one (or several) discrete levels emerge from  the lower continuum  by passing through the corresponding $s_{cr}=a+ 2 d_{cr}$. During this process the vacuum charge $Q_{vac}$ each time grows by $+|e|$, while  $\E_{vac,2}^R(s)$ acquires a specific jump upwards by $+1$.
The direct calculation of  $\E_{vac,2}^R(s)$ (see Fig.\ref{vac2}b) for the parameters $V_0=4.08$ and $a=1$, which reside in the first from below shaded region in  Fig.\ref{dcr}, confirms these effects completely. In particular, the numerical check shows that due to emergence at $s_{cr}\simeq 4.0709$ ($d_{cr} \simeq 1.5354$) of one even level from  the lower continuum the integral vacuum charge grows  from $Q_{vac}=-|e|$ for $s<s_{cr}$ up to $Q_{vac}=0$ for $s>s_{cr}$. Simultaneously for $s=s_{cr}$ there takes place a jump in the vacuum energy $\E_{vac,2}^R(s)$ by $+1$, as it follows from  Fig.\ref{vac2}b. More details concerning the behavior of the charge density for these two regimes are considered in Refs.~\cite{annphys, tmf}.

\subsection*{4. Casimir forces between two identical positively charged short-range Coulomb sources}

The Casimir interaction energy $\E_{vac}^{int}(d)$ for the system of two identical short-range Coulomb sources (\ref{v2}) is determined through the relation (\ref {casint}) with subsequent renormalization. In what follows we'll use the parametrization of the source separation in terms of $d$ instead of $s$ as the most pertinent one.  Indeed here the efficiency of the method (\ref{41}), based on the integration of the logarithmic derivative of the Wronskian along the WK contours (Fig.\ref{contour})  compared to evaluation of  $\E_{vac}^R$  via the sum of the phase integral and discrete levels (\ref{evac}),(\ref{evren}), shows up most clearly, since it provides for $\E_{vac}^{int}(d)$ more analytic details, at least for $d \gg 1$.

Upon subtraction of $2\E_{vac}^R(V_0,\,a/2)$ from the expression (\ref{evren2}) the general structure of $\E_{vac}^{int}(d)$ takes the form
\beq\label{Evacint}
\E_{vac}^{int}(d)=I_{int}(d)-S_{int}(d)+\Lambda_{int}(d)\,V_0^2 \ ,
\eeq
with $I_{int}(d)$ being the contribution from the integral term, $S_{int}(d)$ -- from the negative discrete levels, while  $\Lambda_{int}(d)\,V_0^2$ -- from the renormalization term, respectively.

It would be pertinent to start with the renormalization term, the asymptotics of which for large $d$ can be explored most simply and in the general form. By means of (\ref{lambdaf1}) and (\ref{lambifin}) the renormalization coefficient  $\Lambda_{int}(d)=\Lambda(a,d)-2\lambda(a/2)$ can be represented as
\begin{equation}
\begin{aligned}
&\Lambda_{int}(d)=a/\pi-2\lambda_2(a+d)-2\lambda_2(d)-4\lambda_2(a/2) \ + \\ &
+ \ 4\lambda_2(d+a/2)-2(a/(2\pi)-2\lambda_2(a/2))= \\
 &=4\lambda_2(d+a/2)-2\lambda_2(a+d)-2\lambda_2(d) \ .
\label{renint}
\end{aligned}
\end{equation}
\normalsize
In the next step, inserting the definition of $\lambda_2$ (\ref{l2}) into (\ref{renint}), one finds
\small
\begin{equation}\label{lambda-int}
\begin{gathered}
\Lambda_{int}(d)=4I_2(d+a/2)-2I_2(a+d)-2I_2(d)=\\ =-\dfrac{1}{2\pi}\int\limits_0^\infty \dfrac{(1-e^{-2a\sqrt{1+y^2}})^2 e^{-4d\sqrt{1+y^2}}}{(1+y^2)^2}\,dy \leqslant 0 \ .
\end{gathered}
\end{equation}
\normalsize
So the contribution to the interaction energy $\E_{vac}^{int}(d)$ from the renormalization term turns out to be strictly negative and exponentially decreasing for  $d \gg 1$.
The exact form of the asymptotics of $\Lambda_{int}(d \gg 1)$ can be found from  the expression (\ref{renint}) via triple integration of the MacDonald function asymptotics in the way, quite similar to the evaluation of the asymptotics of $\lambda(a\rightarrow \infty)$, considered in Ref.~\cite{annphys}, and takes the form
\begin{multline}
\Lambda_{int}(d \gg 1)=-\dfrac{e^{-4 d}}{\sqrt{2 \pi d}}\,e^{-2 a}\,\Big(\sinh^2a \ +  \\ + \ \dfrac{\sinh a\,(8 a e^{-a}-13 \sinh a)}{32 d} + O\left(\dfrac{1}{d^2}\Big)\right) \ .
\label{aslam}
\end{multline}

Now let us consider the behavior of the integral term in (\ref{Evacint}) for $d \gg 1$, at first without subtracting the contribution from infinitely separated wells. Upon integration by parts it can be written as follows
\beq\label{IntWronskReg}
I(d)=-{1\over \pi}\int\limits_0^{\infty} dy~ \mathrm{Re}\left[\ln \(J_{red}(d,i y)\)\right] \ ,
\eeq
where the ``reduced'' Wronskian
\beq\label{Jred}
J_{red}(d,\e)=J(d,\e)/J_0(\e)
\eeq
contains in the nominator the Wronskian $J(d,\e)$ for the double-well potential (\ref{v2})
\beq
J(d,\e)={2\, e^{-2 a \sqrt{1-\e^2}}\over \sqrt{1-\e^2}}\left[ f_1^2(\e)-e^{-4 d \sqrt{1-\e^2}} f_2^2(\e) \right] \ ,
\eeq
in which
\begin{equation}
\begin{gathered}
f_1(\e)=\sqrt{1-\e^2}\, \cos(a\sqrt{(V_0+\e)^2-1})- \\ - (\e^2-1+ \e\,V_0)\, \sin(a\sqrt{(V_0+\e)^2-1})/ \sqrt{(V_0+\e)^2-1} \ , \\
f_2(\e)=V_0 \sin(a\sqrt{(V_0+\e)^2-1})/ \sqrt{(V_0+\e)^2-1} \ ,
\end{gathered}
\end{equation}
while in the denominator the Wronskian $J_0(\e)=2\, \sqrt{1-\e^2}$, corresponding to the free case $V_0=0$.

The behavior of the integral (\ref{IntWronskReg}) for large $d$ is found via the following expansion of the integrand
\beq\label{74}\begin{gathered}
\ln \[J_{red}(d,i y)\]=\ln\left( f^2_1(i y) {e^{-2 a \sqrt{1+y^2}}\over 1+y^2}\right) \ - \\ - \ e^{-4 d \sqrt{1+y^2}}\,\({f_2(i y) \over f_1(i y)}\)^2+O\(e^{-8 d \sqrt{1+y^2}}\) \ .
\end{gathered}\eeq
Upon substituting the expansion (\ref{74}) into the integral (\ref{IntWronskReg}) one obtains two first leading terms in the asymptotics of $I(d)$ for $d \gg 1$
\beq\label{75}\begin{gathered}
I(d) \simeq -{1\over \pi}\int\limits_0^{\infty} dy~ \mathrm{Re}\left[\ln \(f^2_1(i y) {e^{-2 a \sqrt{1+y^2}}\over 1+y^2}\)\right] \ + \\ + \ {1\over \pi}\int\limits_0^{\infty} dy~ \mathrm{Re}\left[ e^{-4 d \sqrt{1+y^2}}\,\({f_2(i y) \over f_1(i y)}\)^2\right] \ .
\end{gathered}\eeq
Since the first term in (\ref{75})  doesn't depend on $d$,  the leading term in the asymptotics of the integral term in $\E_{vac}^{int}(d)$ for  $d \gg 1$ takes the form
\begin{equation}\label{76}
\begin{gathered}
I_{int}(d)=I(d)-I(d \to \infty)= \\ = -{1\over \pi}\int\limits_0^{\infty} dy~ \mathrm{Re}\left[ \ln \((1+y^2){e^{2 a \sqrt{1+y^2}}\, \over f^2_1(i y) }\,J_{red}(d,i y)\) \right] \\ \simeq {1\over \pi}\int\limits_0^{\infty} dy~ \mathrm{Re}\left[ e^{-4 d \sqrt{1+y^2}}\,\({f_2(i y) \over f_1(i y)}\)^2\right] \ .
\end{gathered}
\end{equation}

For large $d$ the integrand in (\ref{76})  decreases rapidly with growing $y$, hence, the main contribution to the integral is provided by small $y$. Therefore it turns out to be efficient to rewrite the expression (\ref{76})  in the form
\beq\label{77}\begin{gathered}
I_{int}(d)\simeq \\ {e^{-4 d}\over \pi}\int\limits_0^{\infty} dy~\mathrm{Re}\left[ e^{-4 d (\sqrt{1+y^2}-1-y^2/2)} \({f_2(i y) \over f_1(i y)}\)^2\right]e^{-2 d y^2} \ ,
\end{gathered}\eeq
and thereafter to expand the square brackets in the integrand in the power series in  $y$. All the integrals, emerging this way,  can be calculated analytically. The final expansion of $I_{int}(d)$ for  $d \gg 1$ reads
\beq\label{78}\begin{gathered}
I_{int}(d)= V^2_0\, {e^{-4 d}\over \sqrt{2 \pi d}} \ \times \\ \times \  \Bigg( {A^2 \over 2} +  {1\over 8 d} \left( {3 A^2 \over 8} + B \right)  + O\({1\over d^2} \) \Bigg) \ ,
\end{gathered}\eeq
where
\begin{equation}\label{79}
\begin{gathered}
z_0=\sqrt{V^2_0 - 1} \ , \quad A={1\over 1+ z_0 \ctg (a z_0)} \ ,    \\
B=A^3\Bigg[ -3 V^2_0 \left( 1 - {\ctg(a z_0)\over z_0} + {a \over \sin^2(a z_0)} \right)^2 A \ - \\ - \ 2-{(1+z^4_0)\over z^3_0}\,\ctg(a z_0) \ + \\
+ \ {a\over z^2_0 \sin^2(a z_0)} \left( 1- 2 V_0^2(1- a z_0 \ctg(a z_0))\right) \Bigg] \ .
\end{gathered}
\end{equation}
It should be specially noted that the formulae (\ref{79}) work equally well both for $V_0 > 1$ and  $V_0 <1$. For $V_0=1$  upon taking in (\ref{79}) the limit $z_0 \to 0$ the expressions for $A$ and $B$ are replaced by
\begin{equation}\label{80}
\begin{gathered}
A={a\over 1+ a} \ , \quad
B=-{a^2\over 45(1+ a)^4} \times \\ \times (45+135 a +255 a^2+210 a^3 + 68 a^4+ 8 a^5) \ .
\end{gathered}
\end{equation}

So the asymptotics of the integral term in $\E_{vac}^{int}(d)$ for  $d \gg 1$ turns out to be  $\sim e^{-4 d}/\sqrt{d}$, which is quite similar to the behavior of the renormalization term (\ref{aslam}). It should be mentioned that  the expansion (\ref{78}) can be used also for finite $d$ in the case, when the each next term in the expansion (\ref{74}) is much less than the previous one. At the same time, there might occur  an alternative situation, similar to the case $a=1$, $V_0=8$, considered below, when the coefficients  $A$ and $B$ turn out to be quite large. The reason is that the zero denominator in $A$  is nothing else, but the condition for existence of the level with $\e_0=0$ in the single well. For $a=1$, $V_0=8$ the lowest level is $\e_0 \simeq 0.02085$, and so by sufficiently small variation of the well parameters this level can be made strictly zero. It follows whence that in the general case the expansion given above doesn't hold for the case, when there exists  in the well the level close to $\e_0=0$, since in this case the expansion coefficients $A$ and $B$ become large.

In the latter case it should be taken into account by expanding the square bracket in (\ref{77}) in the power series in $y$ that the expansion of the function $f_1(i y)$ starts now from the linear in $y$ term, since the first term of the series  $\cos(a z_0)+\sin(a z_0)/z_0$ vanishes. As a result, for the case  $\e_0=0$ one obtains
\beq\label{e0=0}
I_{int}(d)=- {V^2_0-1 \over (1+a) V^2_0 }\, e^{-2 d}+ O\(e^{-4 d}\) \ ,
\eeq
whence it follows that for this special case the rate of decrease of the integral term in (\ref{Evacint}) for  $d \gg 1$ becomes sufficiently less. It should be mentioned in addition that the multiplier before the leading exponent in (\ref{e0=0}) is strictly negative, since the zero level might appear in the single well only for $V_0>1$.

Now let us consider the (possible) contribution to (\ref{Evacint}) from negative discrete levels for $d \gg 1$. In the general case, the discrete levels are determined by  the corresponding zeros of the Wronskian $J(d,\e)$ and satisfy the equation
\beq\label{DiscrWronsk}
f_1^2(\e)-e^{-4 d \sqrt{1-\e^2}} f_2^2(\e) =0 \ .
\eeq
For $d \to \infty$ the eq.(\ref{DiscrWronsk}) transforms into $f_1(\e)=0$, which  is obviously the equation for degenerate by parity levels in the system with two infinitely separated wells, or, equivalently, for the levels of the single well.
Let us consider one of the levels $\e_0$  in the single well, for which $f_1(\e_0)=0$. In the limit $d \to \inf$ the value  $\e_0$ serves as the zero approximation for corresponding even and odd levels in the double-well potential (\ref{v2}). To find the splitting of $\e_0$ into the even and odd components for finite $d \gg 1$, let us seek the solution of (\ref{DiscrWronsk}) in the form $\e=\e_0+\delta\e$, where $\delta\e$  is a small correction to $\e_0$. Inserting this expansion into (\ref{DiscrWronsk}) and decomposing  the l.h.s. in $\delta\e$ including the third order  with account of $f_1(\e_0)=0$, one obtains a cubic equation
\beq\label{levels}
- A_1 e^{-4 d \sqrt{1-\e_0^2}} + B_1 e^{-4 d \sqrt{1-\e_0^2}} \delta\e + C_1 \delta\e^2 + D_1 \delta\e^3= 0 \ ,
\eeq
where
\beq\label{levels1}\begin{gathered}
A_1=f^2_2(\e_0) \ , \\ B_1=-{2 f_2(\e_0) \over \sqrt{1-\e_0^2}}\[2 d \e_0 f_2(\e_0) + \sqrt{1-\e_0^2} f'_2(\e_0)\] \ , \\
C_1=\[f'_1(\e_0)\]^2 \ , \quad
D_1=f'_1(\e_0)f''_1(\e_0) \ .
\end{gathered}\eeq
Solving further the eq. (\ref{levels}) by means of successive iterations, one finds the following splitting of the unperturbed level $\e_0$
\beq\label{corr}\begin{gathered}
\delta\e = \pm |K_1(a)| e^{-2 d \sqrt{1-\e_0^2}} + K_2(a,d) e^{-4 d \sqrt{1-\e_0^2}} +  \\ + O\(e^{-6 d \sqrt{1-\e_0^2}}\) \ ,
\end{gathered}\eeq
where
\beq\label{corr1}
q_0=V_0+\e_0 \ , \quad K_1(a)={(1-\e^2_0)\,(1-q^2_0)\over V_0(\e_0+q_0+a q_0\,\sqrt{1-\e_0^2})} \ ,
\eeq
\begin{widetext}\begin{equation}
\label{corr2}
K_2(a,d)={(1-\e_0^2)^{3/2}(q^2_0-1)^2 \over 2\, V^2_0 (\e_0+q_0+a q_0\sqrt{1-\e_0^2})^2} \times
\eeq
$$
\times \Big[4 d \e_0 \ +
{2 a^2 q^2_0\, (1-\e_0^2)\,(\e_0 q_0-1)+(2-\e^2_0-q^2_0)\,(\e_0 q_0+1) + a \sqrt{1-\e_0^2}\,(2\e_0 q_0 \, (q^2_0-1)+(\e_0^2-1)\,(2 q^2_0-1))\over \sqrt{1-\e_0^2}\,(q^2_0-1)\,(\e_0+q_0+a q_0\,\sqrt{1-\e_0^2})}\Big] \ ,
$$
\end{widetext}
whereby the upper sign in (\ref{corr}) corresponds to the odd level, while the lower -- to the even one. Here is worth to note that for discrete levels in the single well like (\ref{v}) there always holds the relation $q_0 >1$ (for details see Ref.~\cite{greiner2000}). So both $K_{1,2}$ are always well-defined, since their denominators are strictly positive.

In  the case of $\e_0<0$ for sufficiently large  $d$ both levels $\e_{odd}$ and $\e_{even}$  become also negative, therefore  their total contribution to $\E_{vac}^{int}(d)$ equals to
\beq\label{odd+even}\begin{gathered}
\e_{odd}+\e_{even}= \\ = 2\, \e_0 + 2\, K_2(a,d)\, e^{-4 d \sqrt{1-\e_0^2}} + O\(e^{-6 d \sqrt{1-\e_0^2}}\) \ .
\end{gathered}\eeq
So in this case  the contribution to  $\E_{vac}^{int}(d)$ for large  $d$, caused by negative discrete level $\e_0< 0$ in the single well, takes the form
\beq\label{levels2}\begin{gathered}
S_{int}(d)=\e_{odd}+\e_{even}-2 \e_0 = \\ = 2\, K_2(a,d)\, e^{-4 d \sqrt{1-\e_0^2}}+ O\(e^{-6 d \sqrt{1-\e_0^2}}\) \ .
\end{gathered}\eeq
At the same time, the zero level $\e_0=0$ splits for finite $d$ into a pair, where  only the even one is negative, which gives  the following term in $\E_{vac}^{int}$
\beq\label{levels3}
S_{int}(d)=\e_{even} = -{V^2_0-1 \over(1+a)\, V^2_0} e^{-2 d} + O\(e^{-4 d}\) \ .
\eeq
It should be mentioned that the analysis  performed above for the discrete levels contribution to the interaction energy has the correct status only subject to condition   $d \sqrt{1-\e_0^2} \gg 1$. The latter means that whenever the single well parameters are such that the level $\e_0$ lies arbitrary close to the lower threshold, the expressions (\ref{levels2})-(\ref{levels3}) could be valid only for such separations, which provide the fulfillment of this condition.

So the resulting behavior of $\E_{vac}^{int}(d)$ for $d \gg 1$ to a high degree turns out to be  subject of the  single well configuration. If there are only positive levels in the single well, the asymptotics of $\E_{vac}^{int}(d)$ should be  $O\(e^{-4 d}/\sqrt{d}\)$ due to the integral and renormalization terms. The strictly zero level $\e_0=0$ yields the contributions to  $I_{int}(d)$ and $S_{int}(d)$ with twice less exponent rates   (\ref{e0=0}) and (\ref{levels3}), but in $\E_{vac}^{int}(d)$ these terms exactly cancel each other, hence, there remains the same exponential law of decrease  $\sim e^{-4 d}$.

In presence of negative levels in the spectrum of the single well the leading term in the asymptotics  of $\E_{vac}^{int}(d)$ becomes different, namely, the main contribution to the asymptotics  of $\E_{vac}^{int}(d)$ will be given by the lowest  $\e_0$
\beq\label{e0<0}
\E_{vac}^{int}(d) =- 2\, K_2(a,d)\, e^{-4 d \sqrt{1-\e_0^2}}+ O\(e^{-6 d \sqrt{1-\e_0^2}}\) \ .
\eeq
It should be mentioned that  $\e_0$ can be arbitrarily close to $\e_F=-1$, hence $\sqrt{1-\e_0^2}$ -- arbitrarily small (but nonzero). In this case the exponential fall-down of $\E_{vac}^{int}(d)$ takes place only at extremely large $d$ subject to condition  $d \sqrt{1-\e_0^2} \gg 1$ and so the Casimir interaction between such wells acquires the features of a long-range force. It is noteworthy that this effect arises  due to  the lowest discrete level, rather than due to replacement of the exponential asymptotics by a power-like, what could happen only for a massless mediator similar to considered  in Refs.~\cite{Jaffe2004,nanotubes}.

\begin{figure*}[ht!]
\subfigure[]{
		\includegraphics[width=\columnwidth]{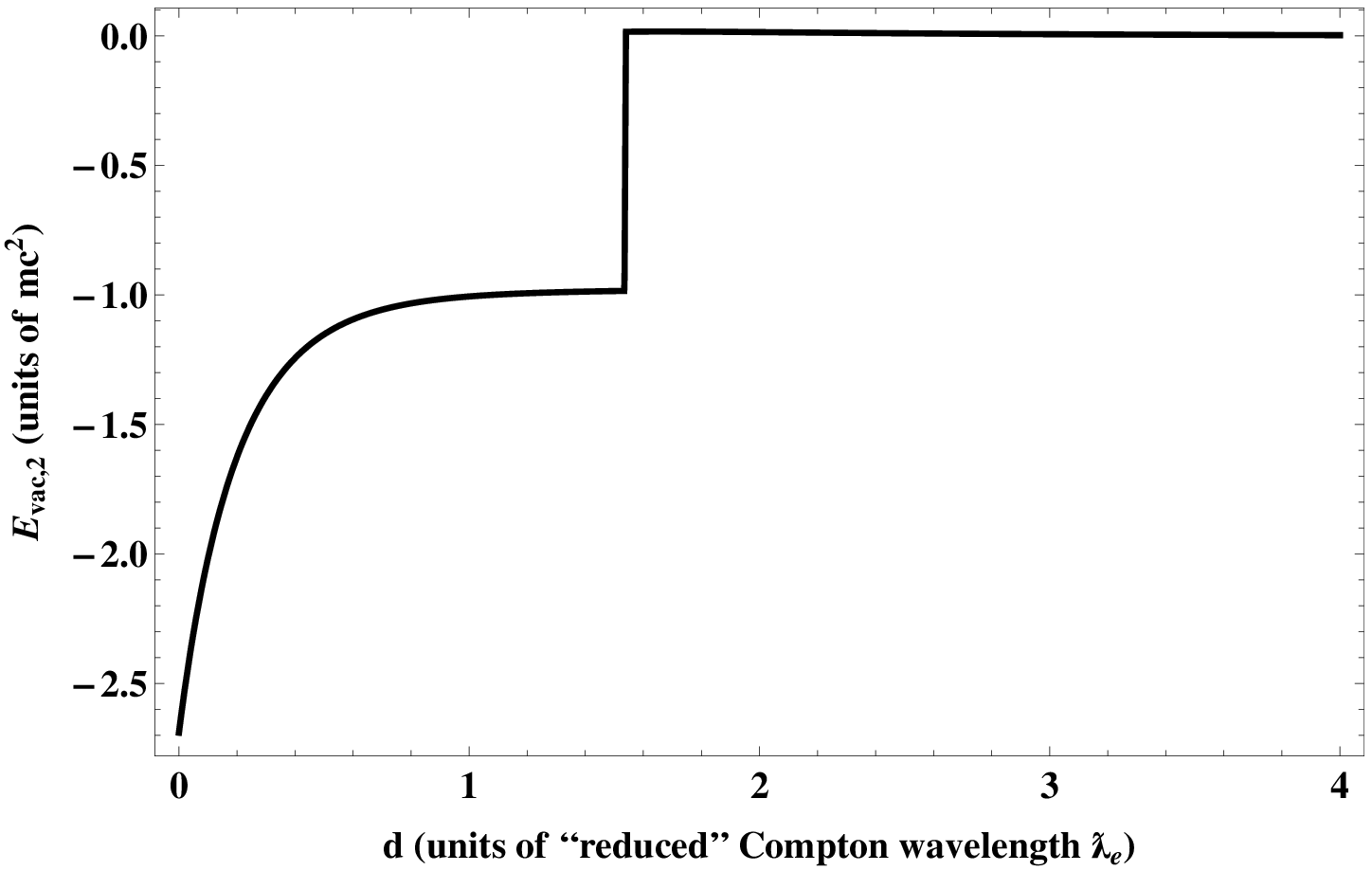}
}
\hfill
\subfigure[]{
		\includegraphics[width=\columnwidth]{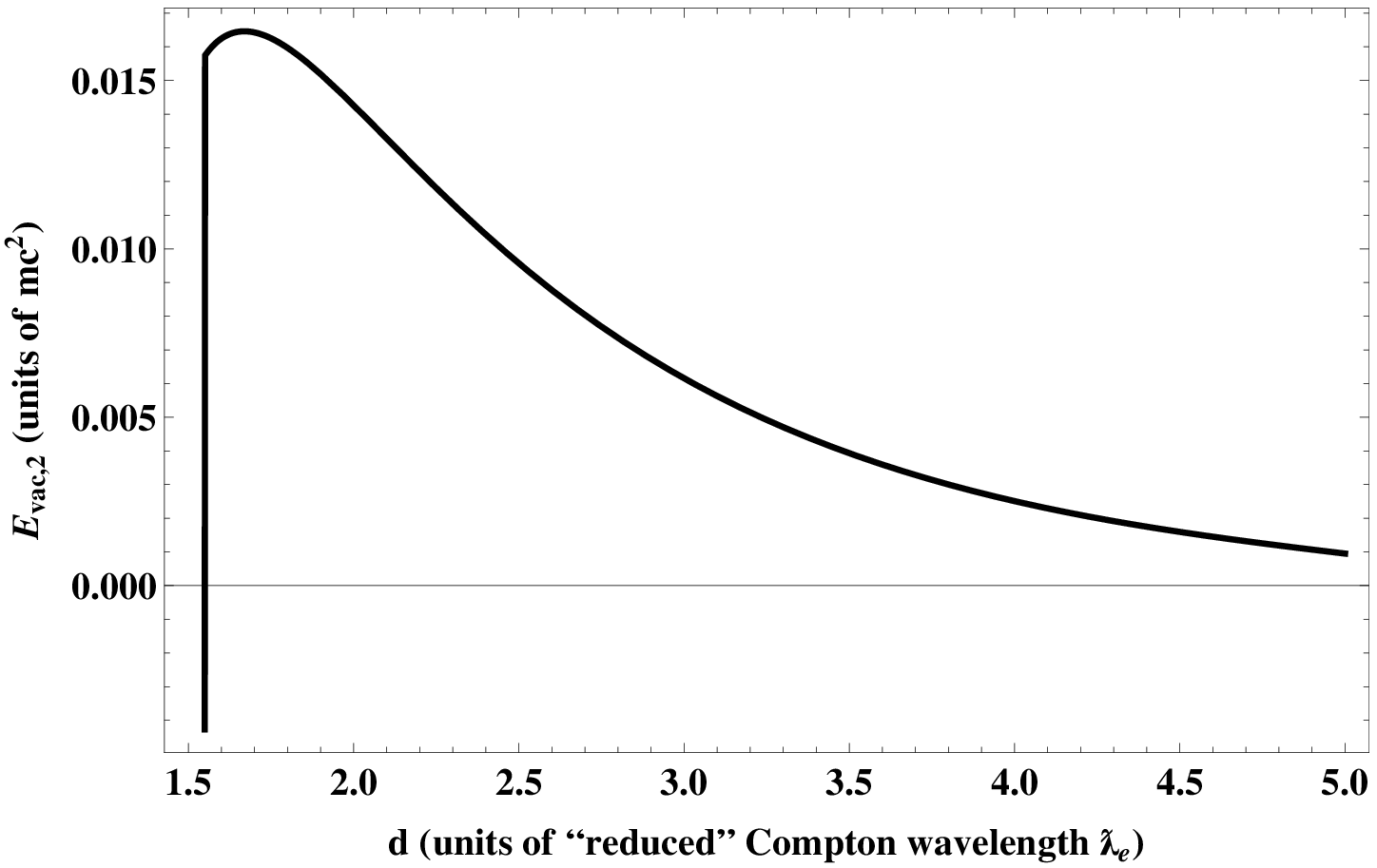}
}
\vfill
\subfigure[]{
		\includegraphics[width=\columnwidth]{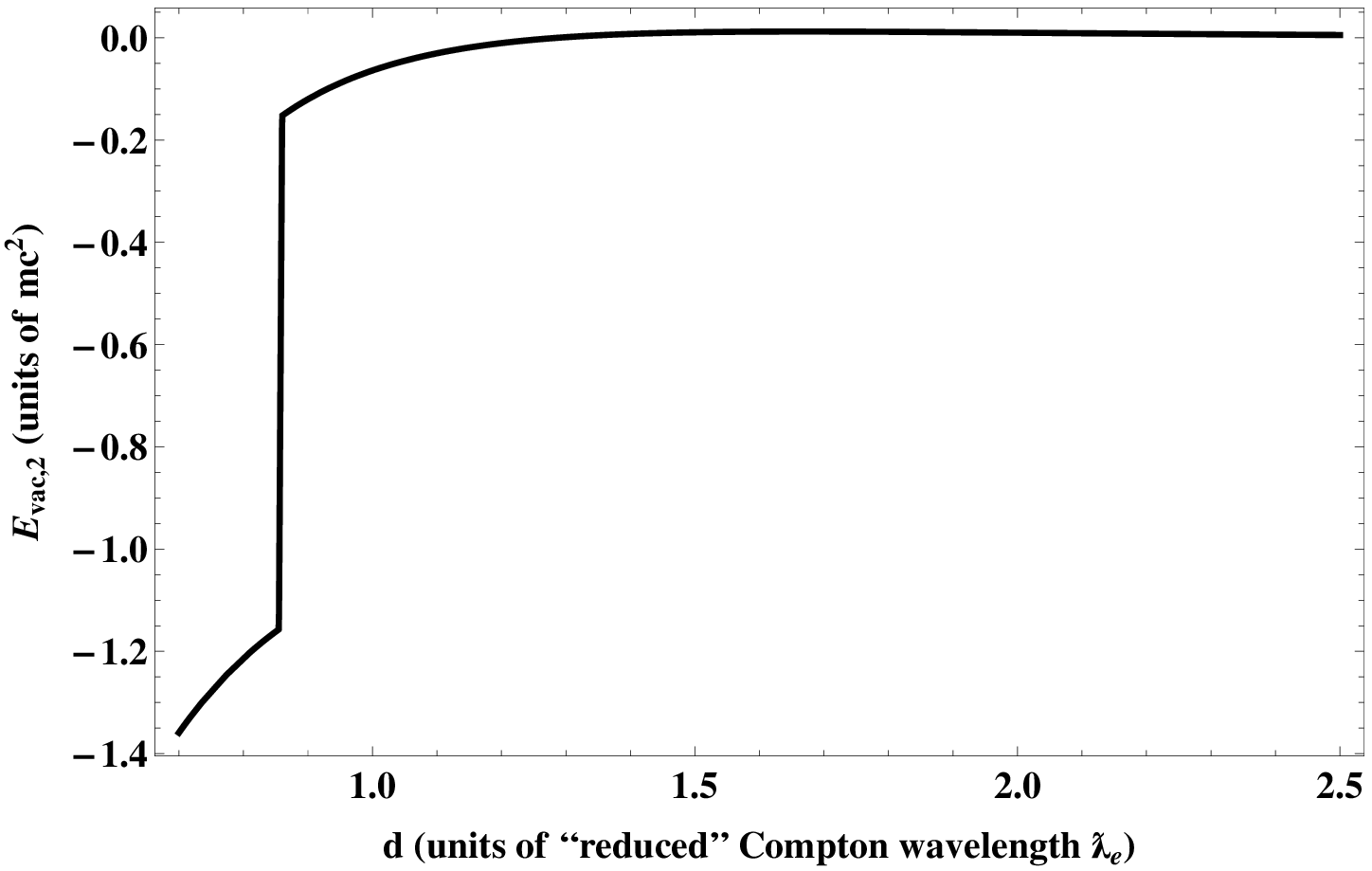}
}
\hfill
\subfigure[]{
		\includegraphics[width=\columnwidth]{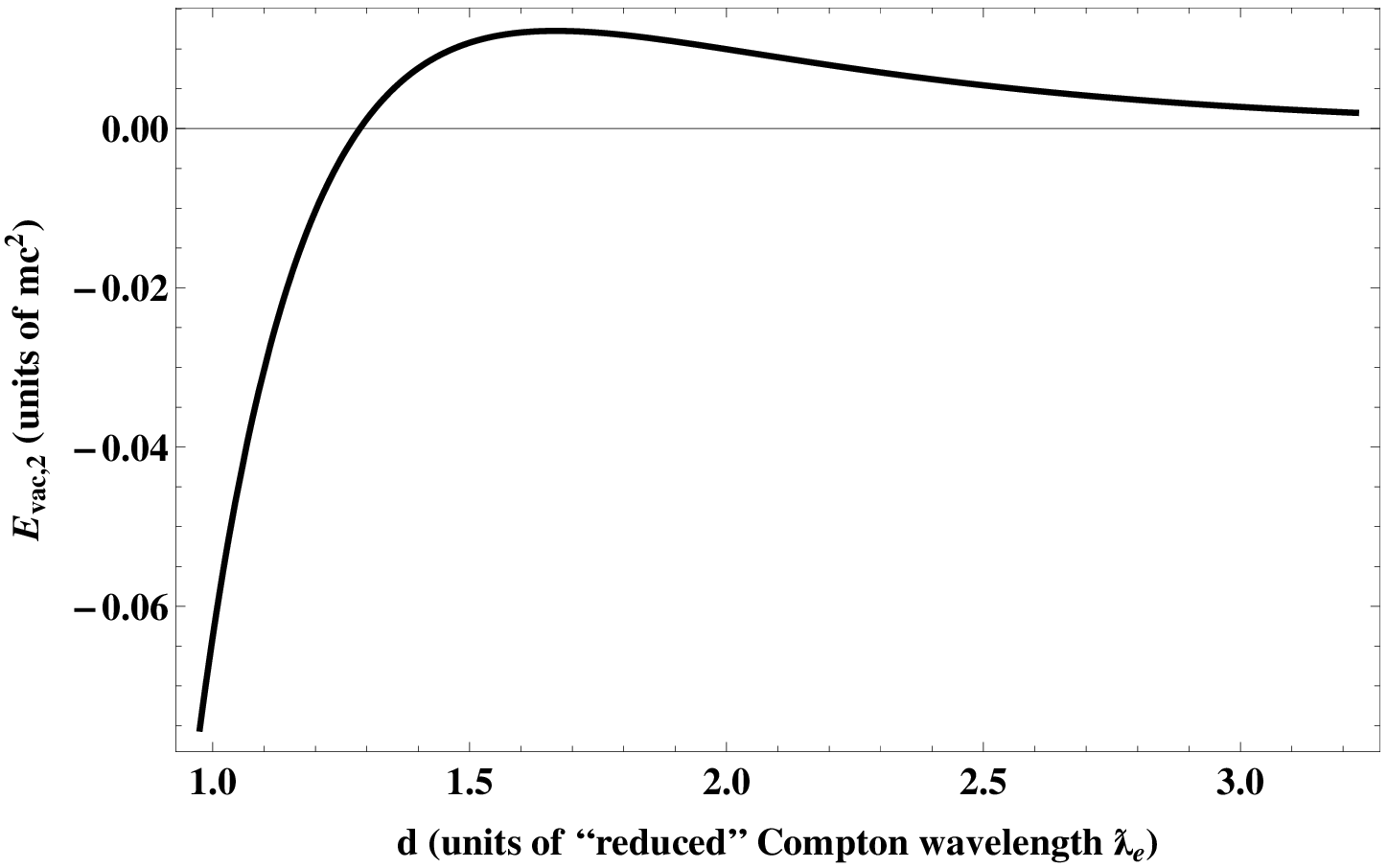}
}
\caption{The dependence of the Casimir interaction energy between two wells on the distance $d$ between them for $a=1$ and (a,b): $V_0=4.08$; (c,d):  $V_0=10$. }
	\label{Eint1}	
\end{figure*}

The same effect is found in the work~\cite{tanaka2013}, where it was shown that the electronic Casimir force between two impurities on a one-dimensional semiconductor quantum wire can be of a very long range, despite nonzero effective mass of the mediator. It should be emphasized that in this work  the electronic Casimir-Polder effect is interpreted in terms of the radiation
reaction field, where one of the two sources creates a virtual
cloud of the field around itself, and the interaction of this field
with the other atom induces the  Casimir-Polder force. So in contrast to our approach based on the QED vacuum polarization  there is no need to
utilize the idea of vacuum fluctuations of the field as a cause
of the electronic Casimir-Polder effect. Although these two interpretations look
qualitatively different, Milonni et al. revealed that they are two
sides of the same coin about the Casimir effect~\cite{milonni1994}-\cite{milonni}. Moreover, in the present case the analogy between these two approaches can be illustrated by means of the similarity in the answers for the origin of the long-distance behavior of  Casimir force. In our case it is the negative discrete level in the single well, which lies close to the lower threshold, while in  Ref.~\cite{tanaka2013} it is the single-impurity ground-state energy, which could be very small
as one of the striking features of the Van Hove singularity, which causes the appearance of the bound state just below the band edge
regardless of the bare impurity energy~\cite{tanaka2006}. And in both cases we deal with the effect,  which cannot be
described by means of the perturbative methods.

\begin{figure*}[ht!]
\subfigure[]{
		\includegraphics[width=\columnwidth]{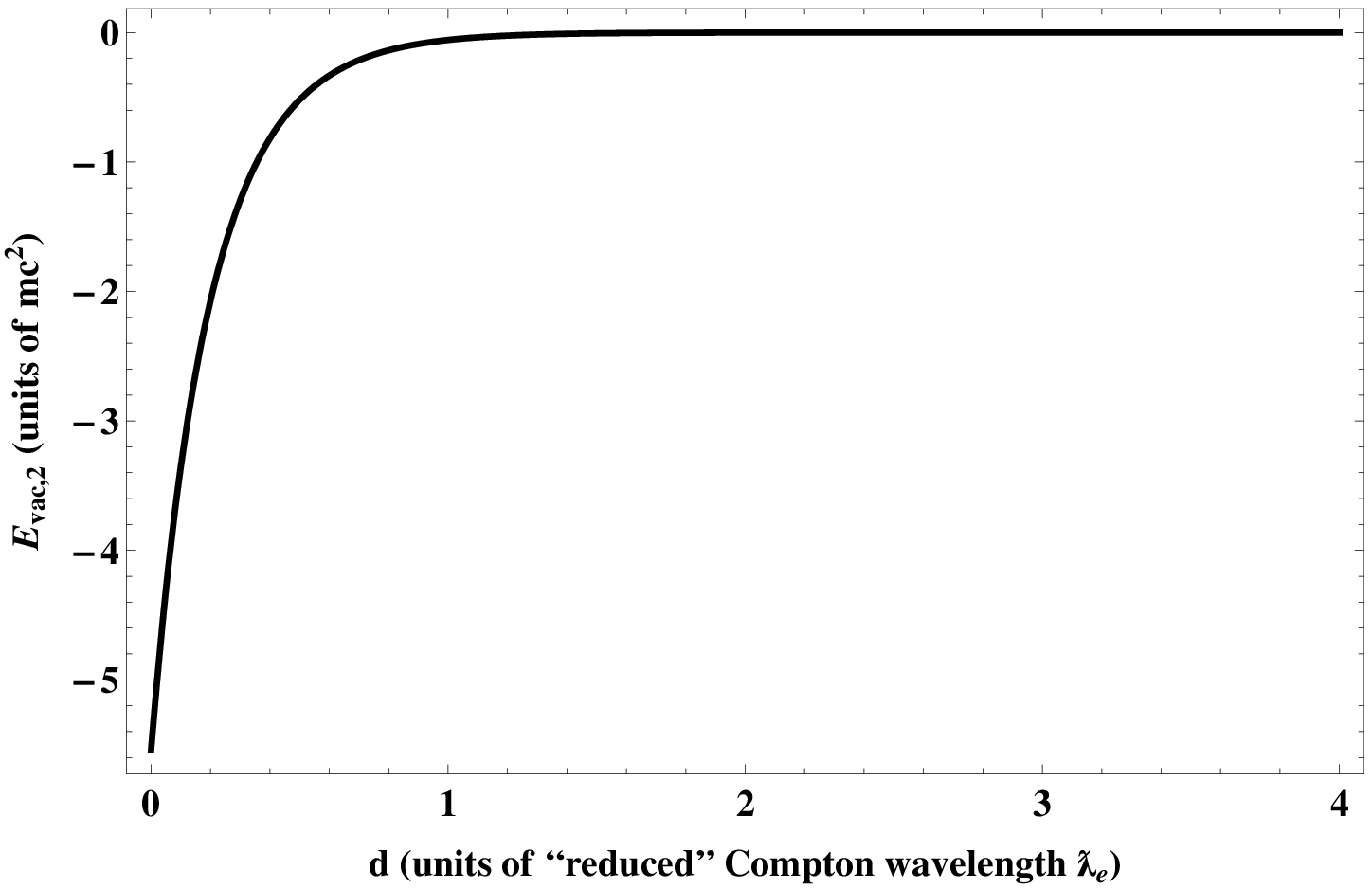}
}
\hfill
\subfigure[]{
		\includegraphics[width=\columnwidth]{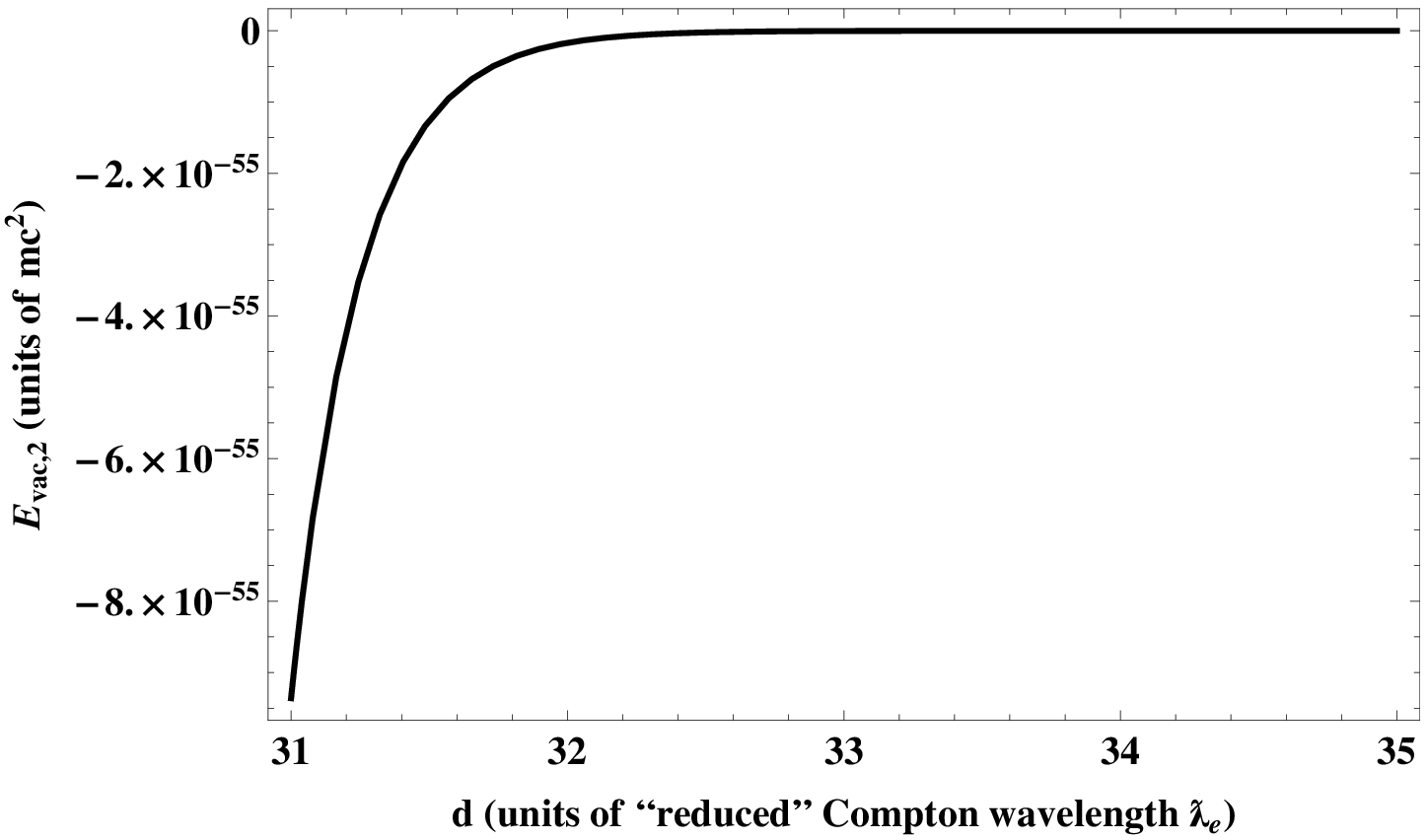}
}
\vfill
\subfigure[]{
		\includegraphics[width=\columnwidth]{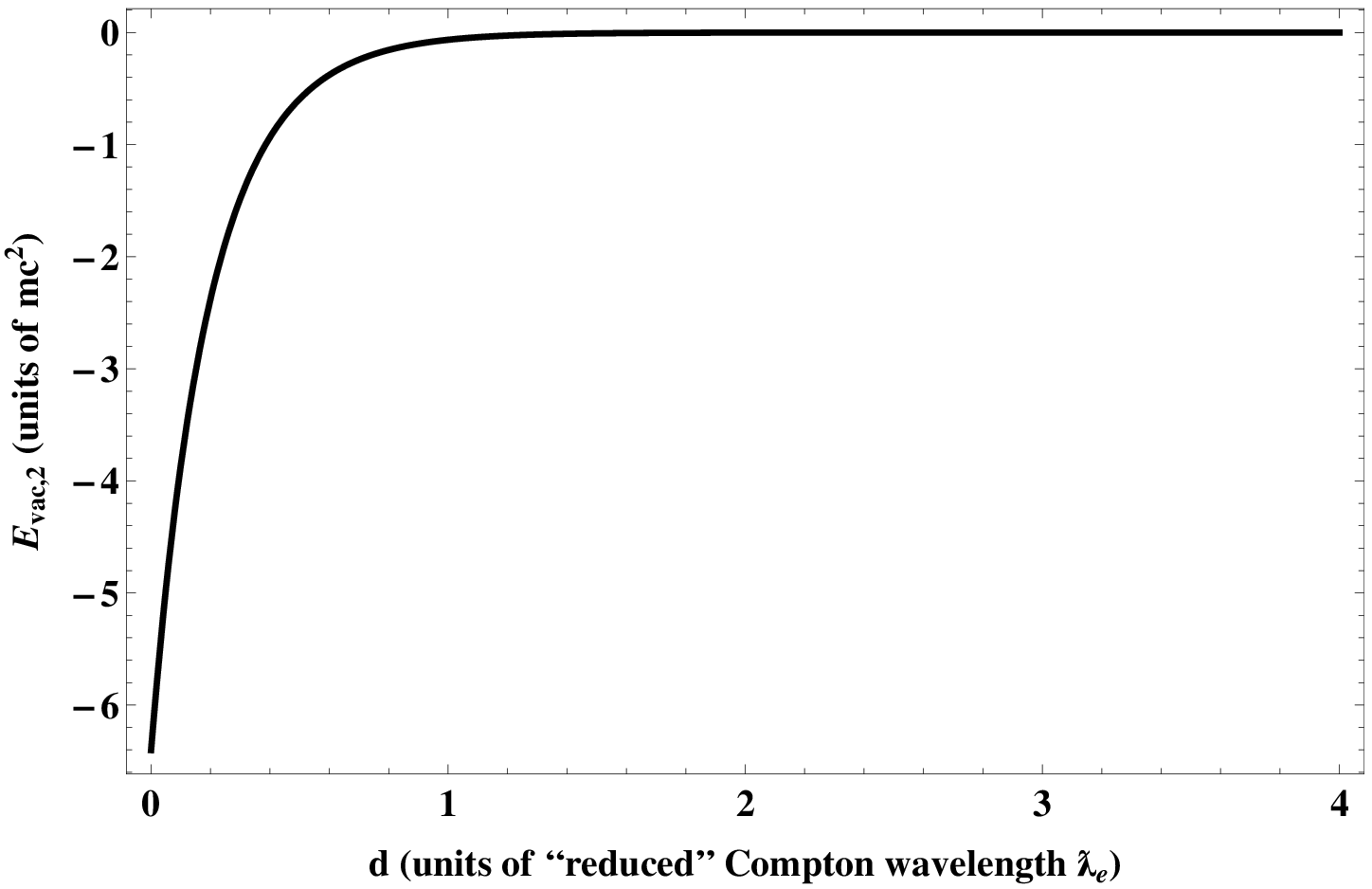}
}
\hfill
\subfigure[]{
		\includegraphics[width=\columnwidth]{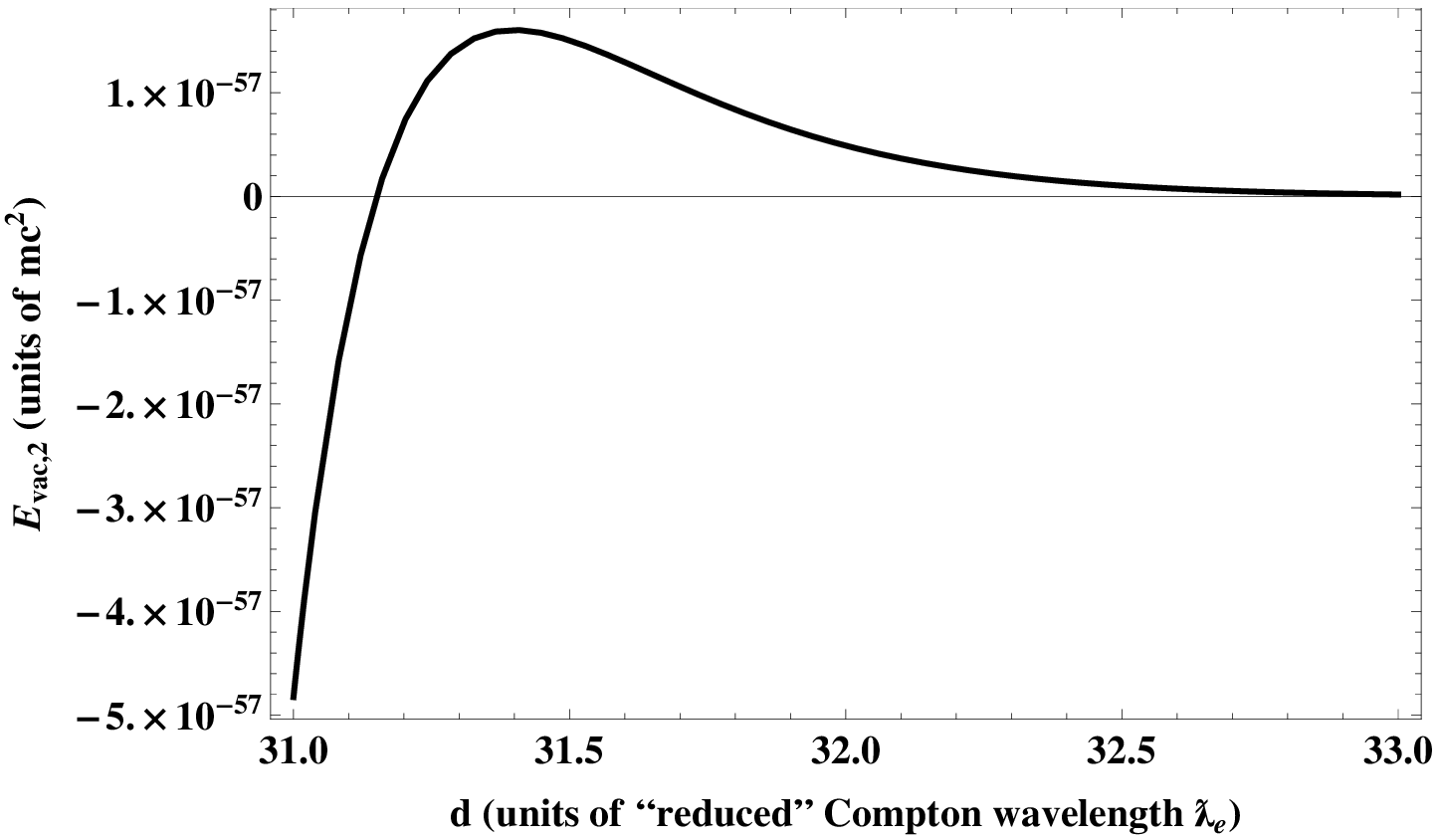}
}
\caption{ The dependence of the Casimir interaction energy between two wells on the distance $d$ between them for $a=1$ and (a,b): $V_0=7.4$; (c,d): $V_0=8$. }
	\label{Eint2}	
\end{figure*}

The concrete type of interaction between the wells can be quite different subject to the single well parameters  $V_0$ and $a$, both in the asymptotics and for finite distances between the wells. In Figs.\ref{Eint1}-\ref{Eint2}  $\E_{vac}^{int}(d)$ is presented for $a=1$ and $V_0=4.08,\ 7.4,\ 8,\ 10$. It follows that for $d \gg 1$ and $V_0=4.08,\ 8, 10$ the interaction energy is positive (reflecting wells), whereas for $V_0=7.4$ the energy at large distances becomes negative  (attracting wells).

Such behavior can be easily understood by means of the analysis presented above. Actually, for  $V_0=4.08,\ 10$ (Fig.\ref{Eint1}) in the corresponding single well the lowest discrete level is negative ($-0.9648$ and $-0.90811$, respectively). As a result, for growing $d$ in $\E_{vac}^{int}(d)$ there takes place  firstly the jump by $+1$, provided by emergence of the discrete level  from the lower continuum by passing through the corresponding $d_{cr}$ (quite similar to the picture shown in Fig.\ref{vac2}b), while  for $d \gg 1$ the behavior of $\E_{vac}^{int}(d)$ is defined primarily by the contribution from the discrete spectrum, which in this case has the form
\beq\label{e0=-1}\begin{gathered}
\E_{vac}^{int}(d) \simeq  -2 K_2(a,d) e^{-4 d \sqrt{1-\e_0^2}} \to \\ \to -4 d \e_0\, {(1-\e_0^2)^{3/2}\,(q_0^2-1)^2 \over V^2_0\,(\e_0+q_0+a q_0\,\sqrt{1-\e^2_0})^2} e^{-4 d \sqrt{1-\e_0^2}}\, >0 \ ,
\end{gathered}\end{equation}
since in the coefficient $K_2(a,d)$ under the condition $d \sqrt{1-\e_0^2} \gg 1$ the main term in the square bracket in (\ref{corr2}) will be  $4 d \e_0$.
So in presence of a negative level $\e_0<0$ in the ``initial'' single well  the interaction energy  becomes positive for sufficiently large distances between wells.

For $V_0=7.4,\ 8$ (Fig.\ref{Eint2}) the negative levels in the single well are absent, therefore the behavior of $\E_{vac}^{int}(d)$ for $d \gg 1$ is defined by the following expression
\beq\label{e0>0}\begin{gathered}
\E_{vac}^{int}(d \gg 1) = I_{int}(d)+V_0^2\Lambda_{int}(d) \simeq \\ V^2_0\,{e^{-4 d} \over \sqrt {2\pi d}}  \left[ {1\over 2}\left({1\over 1+ z_0 \ctg(a z_0)}\right)^2-e^{-2 a}\sinh^2(a)\right] \ .
 \end{gathered}\eeq
The sign of $\E_{vac}^{int}(d \gg 1)$ depends on the sign of the square bracket in (\ref{e0>0}). For $V_0=7.4$ the square bracket in (\ref{e0>0}) is negative, and hence, for $d \gg 1$  the wells attract each other (Fig.\ref{Eint2}b). For $V_0=8$ it is positive, since  for these values of $(V_0\, , a)$ the expression  $1+ z_0 \ctg(a z_0)$ is close to zero, as it was already mentioned above, and so the asymptotics of the Casimir force is repulsive, but at the same time takes place for sufficiently larger $d$ (see Fig.\ref{Eint2}d).

\subsection*{5. Casimir forces between two $\d$-wells}

Now let us explore separately the Casimir interaction between two $\d$-wells, for which the width and depth are related via  $a=C/V_0$ with $V_0\to \infty\, , a \to 0$ and $C>0$ being some constant, proportional to the  charge of the source. It is well-known that the direct inserting of $\d$-potentials into DE leads to contradictions, since  DE is first order (see e.g., Ref.~\cite{Jaffe2004}). More concretely, the terms involving a $\d$-function are only well defined if $\p$ is continuous at the points, where the $\d$-peaks are located. However, the first equation in (\ref{deq}) implies a jump in the lower component of the Dirac WF $\p_2$ for continuous upper one $\p_1$, while  the second
requires a jump in $\p_1$ for continuous $\p_2$. Thus the equations are not consistent. In Ref.~\cite{Jaffe2004} this problem was solved in terms of the transfer-matrix, which in the $\d$-limit remains well-defined. Here we present another approach for dealing with $\d$-potentials, based on the  $\ln\text{[Wronskian]}$ contour integration, described in the previous Sections.

First we consider the case of a  single $\d$-well, where in order to keep the correspondence with the case of finite wells, considered above,  it is implied that this $\d$-well is twice ``wider''. Direct evaluation of the corresponding limits for separate components in (\ref{econt}) yields the following contributions to the renormalized vacuum energy of a single $\d$-well. The integral term in (\ref{econt}) gives
\beq\label{delta-int}\begin{gathered}
I\to -{1\over \pi}\int\limits_0^{\infty} dy~ \hbox{Re}\left[\ln \left( \cos(2 C)-{i y \over \sqrt{ 1 + y^2 }}\sin(2 C)\right)\right] = \\ = {1-|\cos(2 C)| \over 2} \ .
\end{gathered}\eeq
The equation for the discrete spectrum takes the form
\beq
\label{eq:discr1delta}
\cos(2 C)-{\e \over \sqrt{ 1 - \e^2 }}\sin(2 C)=0 \ ,
\eeq
which possesses a single root
\beq\label{eq:root}
\e_0=\hbox{sign}(\sin(4 C))\,|\cos(2 C)| \ .
\eeq
 Depending on the sign of $\sin(4 C)$ this root can be either positive or negative, and hence, doesn't contribute or contribute to the vacuum energy of the single $\d$-well. So in the general case the non-renormalized vacuum energy of a single  $\d$-well can be represented as follows
\beq
\label{eq:nonrezdelta}
\E_{vac}=I-S=I-\theta(-\e_0)\e_0={1-\hbox{sign}(\sin(4 C))\,|\cos(2 C)| \over 2}  \ .
\eeq
Proceeding further,  on account of  the asymptotics for the  renormalization coefficients  $\l_1(a)$ and $\l_2(a)$ for infinitely small width of the well, which can be easily derived from formulae (\ref{renpart})-(\ref{limlambda}), one finds
\beq
\label{eq:limits}
V^2_0\lambda_1(a)\to {V_0 C \over \pi}-C^2\to \infty ,  \quad V^2_0\lambda_2(a)\to C^2 \ .
\eeq
So in contrast to all the others terms, the PT contribution to the renormalization term doesn't possess any finite $\d$-limit, and hence,  $\E^R_{vac}$ for the  single $\d$-well is divergent:
\beq
\label{eq:endelta}\begin{gathered}
\E^R_{vac}=I-S+\l\, V^2_0 \to \\ \to {1-\hbox{sign}(\sin(4 C))\,|\cos(2 C)| \over 2}-2\, C^2+{V_0 C \over \pi}\to\infty \ .
\end{gathered}\eeq

Actually, this result should be expected from general considerations, since for  discontinuous potentials the Fouriet-transform  $\text{\~{A}}_0(q)$ of the external potential $A_0^{ext}(x)$ decreases in the momentum space too slow and so the one-loop perturbative energy  diverges. The same in essence effect appears also in more spatial dimensions by screening of the Coulomb asymptotics through the simple vertical cutoff, and it is necessary to introduce additional smoothing in order to maintain the convergence of the perturbative contribution to the energy \cite{voronina2018}. It should be clear that in the considered case of a $\d$-well such smoothing would also lead to the finite answer.

However, the Casimir interaction energy between two  $\d$-wells turns out to be  well-defined quantity without any additional smoothing, since the divergent parts doesn't depend on the distance between wells. Namely, the integral component in  (\ref{Evacint}) will give in this case the following contribution to $\E_{vac}^{int}(d)$
\begin{equation}
\begin{gathered}\label{delta-int1}
I_{int}(d)=I(d)-I(d \to \infty)
\to \\ \to -{1\over 2 \pi}\,\int\limits_0^{\infty} dy~ \ln \left[ 1+(\sin C)^4\, \hbox{e}^{-4 d \sqrt{1+y^2}} \times \right. \\ \left. \times {\hbox{e}^{-4 d \sqrt{1+y^2}}  - \ 2 ((1+y^2)(\ctg C)^2 -y^2)\over ((\cos C)^2 + y^2)^2} \right] \ .
\end{gathered}
\end{equation}
Here it should be mentioned that   in this case each $\d$-well should be  twice ``narrower'' compared to the  single $\d$-well, considered in (\ref{delta-int})-(\ref{eq:endelta}), what implies $C \to C/2$ in all the subsequent expressions, defining  separate components in (\ref{econt}) for the two $\d$-wells configuration.

In particular, the eq. for the discrete spectrum (\ref{DiscrWronsk}) splits now into two equations for two levels $\e_\pm$
\beq\label{95}
\ctg\,C\, \sqrt{1-\e^2_\pm}\, -\e_\pm=\mp \hbox{e}^{-2 d \sqrt{1-\e^2_\pm}} \ ,
\eeq
whence  there follows the next contribution to $\E_{vac}^{int}(d)$ from the negative part of the discrete spectrum
$$ S_{int}\to \theta(-\e_+)\e_+ + \theta(-\e_-)\e_- -2 \theta(-\e_0)\e_0 \ , $$
with $\e_0$ being now the single level of a separated $\d$-well, which differs from (\ref{eq:root}) by $C \to C/2$, namely
\beq\label{eq:root1}
\e_0=\hbox{sign}(\sin(2 C))\,|\cos(C)| \ .
\eeq
Proceeding further, from (\ref{lambda-int}) one finds the following limit for the renormalization coefficient in $\E_{vac}^{int}(d)$
\beq\label{eq:lambda1delta}
\Lambda_{int}(d) V^2_0 \to -{2\, C^2\over \pi}\int\limits_0^{\infty} dy~ {\hbox{e}^{-4 d \sqrt{1+y^2}} \over 1+y^2} \ .
\eeq
As a result, within the $\ln\text{[Wronskian]}$ contour integration  the renormalized Casimir interaction energy between two $\d$-wells turns out to be a well-defined quantity.

Compared to the case  of finite wells, the Casimir interaction between two $\d$-sources turns out to be no less rich in the variability of the Casimir force both at finite distances and in asymptotic behavior. Namely, for $d \gg 1$ the components of  $\E^{int}_{vac}(d)$ behave as follows. The integral part
(\ref{delta-int1}) turns out to be
\beq
\label{eq:deltaIintBigd}\begin{gathered}
I_{int}\simeq e^{-4 d}\, {\hbox{tg}^2 C \over 2\sqrt{2 \pi d}} \times \\ \times  \left(1 + {1\over 4}\left({19\over 8}-{3\over \cos^2 C}\right){1\over d} + O\({1\over d^2}\)\right) \ ,
\end{gathered}\eeq
the renormalization term (\ref{eq:lambda1delta}) equals to
\beq
\label{eq:deltaLambdaBigd}
 \Lambda_{int}(d) V^2_0 \simeq -e^{-4 d}\, {C^2 \over \sqrt{2\pi d}}\left(1 - {5\over 32}{1\over d} + O\({1\over d^2}\)\right) \ ,
\eeq
while the asymptotics of  discrete levels is given by
\beq
\label{eq:deltaDiscrBigd}\begin{gathered}
\e_\pm \simeq \e_0 \pm e^{-2 d\sqrt{1-\e^2_0}}\,(1-\e^2_0)-  e^{-4 d\sqrt{1-\e^2_0}}\,\e_0\, (1-\e^2_0) \ \times \\ \times \ \left(1-4 d \sqrt{1-\e^2_0}\right)/2 + O\left(e^{-6 d\sqrt{1-\e^2_0}}\right) \ ,
\end{gathered}\eeq
approaching the level in the single  $\d$-well (\ref{eq:root1}) from above and from below, respectively.

\begin{figure*}[ht!]
\subfigure[]{
		\includegraphics[width=\columnwidth]{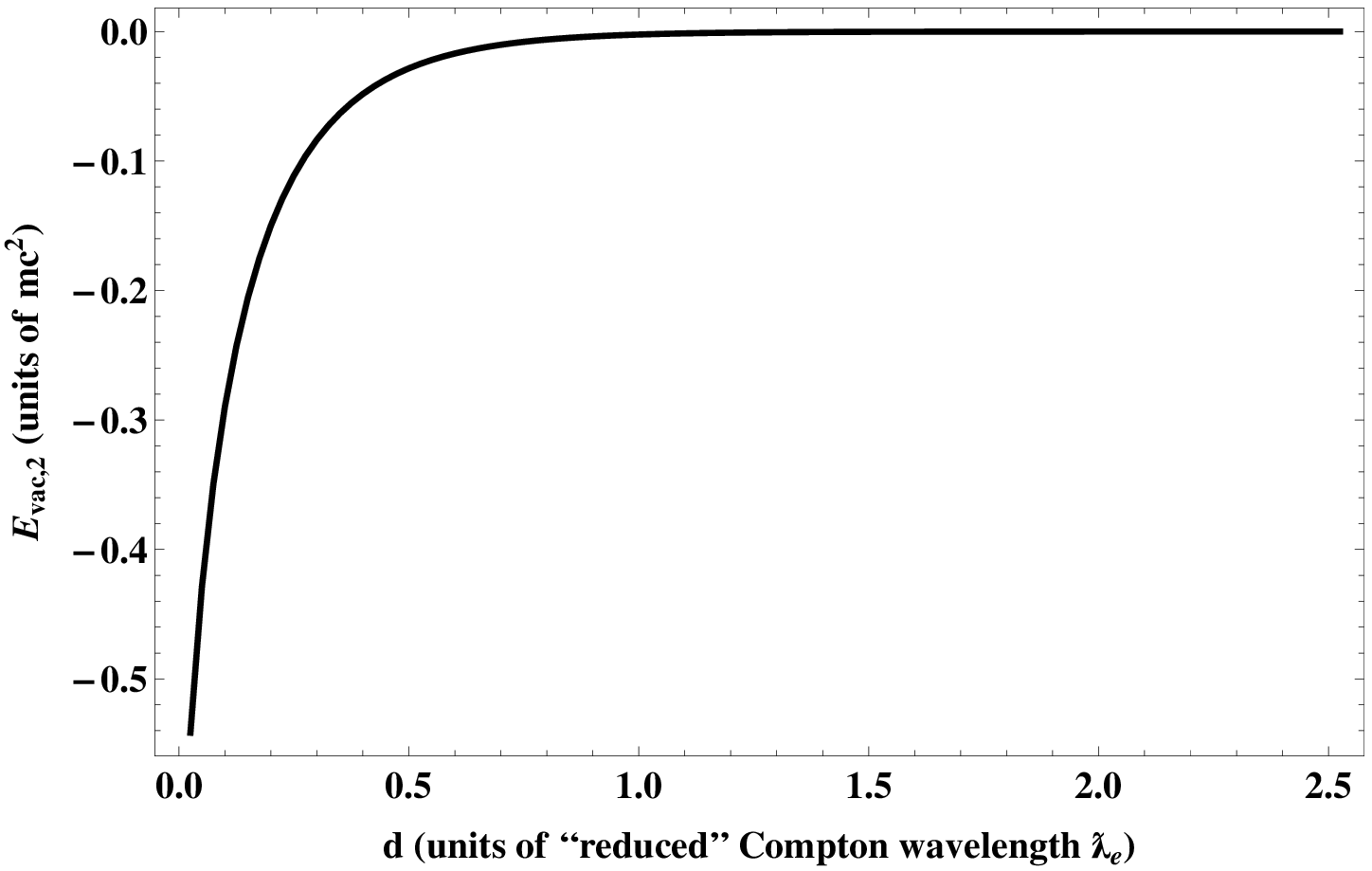}
}
\hfill
\subfigure[]{
		\includegraphics[width=\columnwidth]{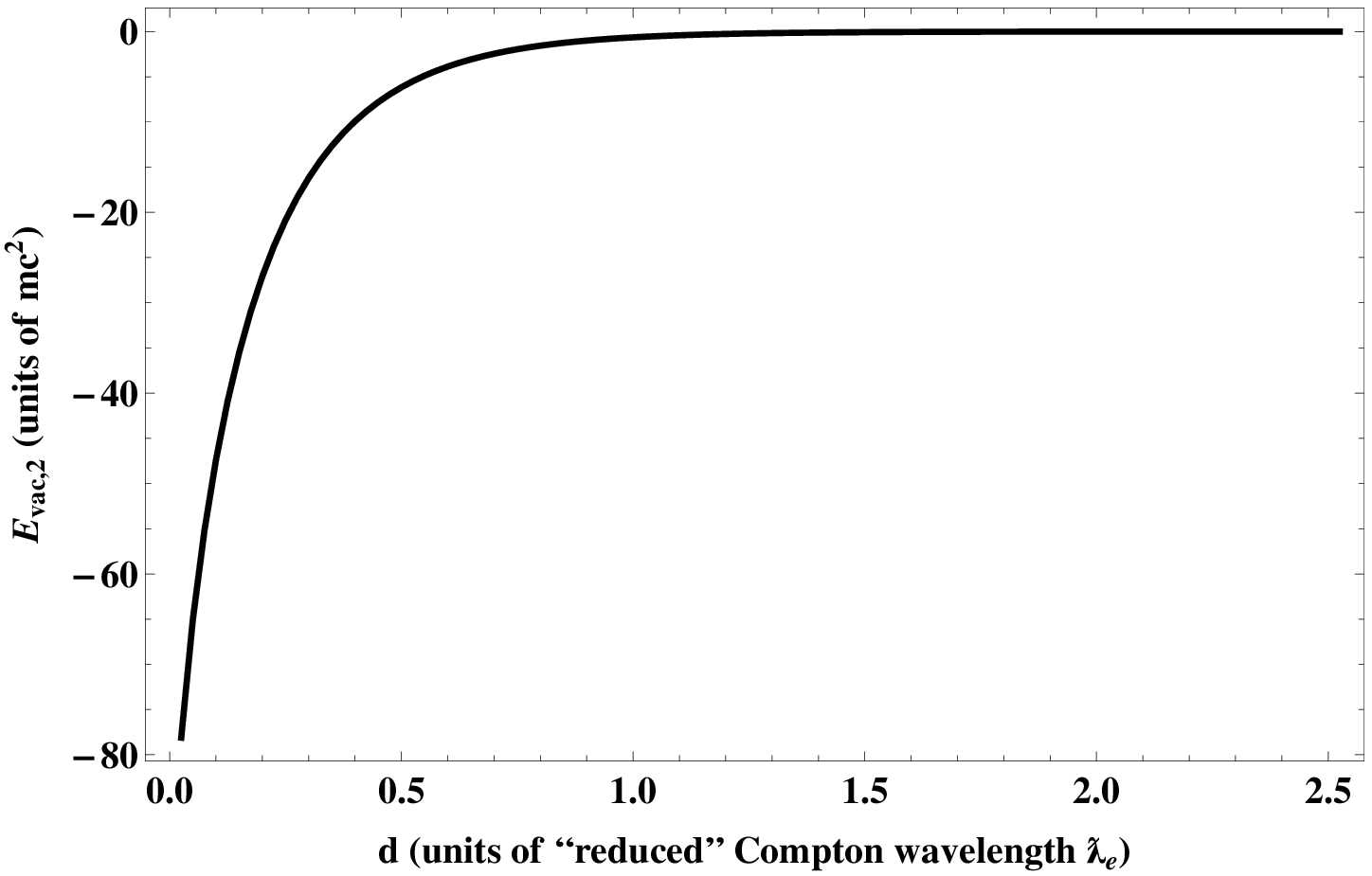}
}
\vfill
\subfigure[]{
		\includegraphics[width=\columnwidth]{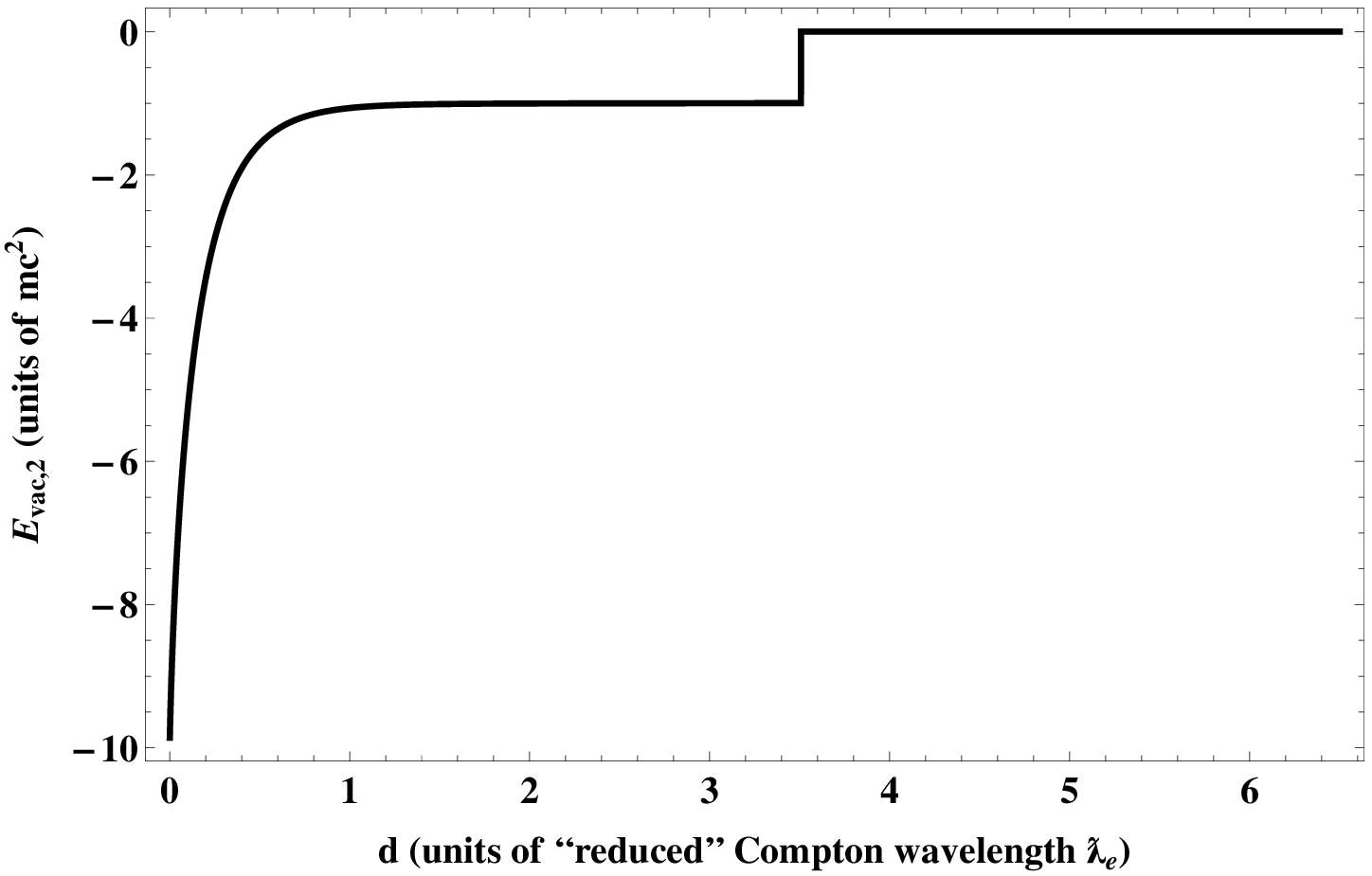}
}
\hfill
\subfigure[]{
		\includegraphics[width=\columnwidth]{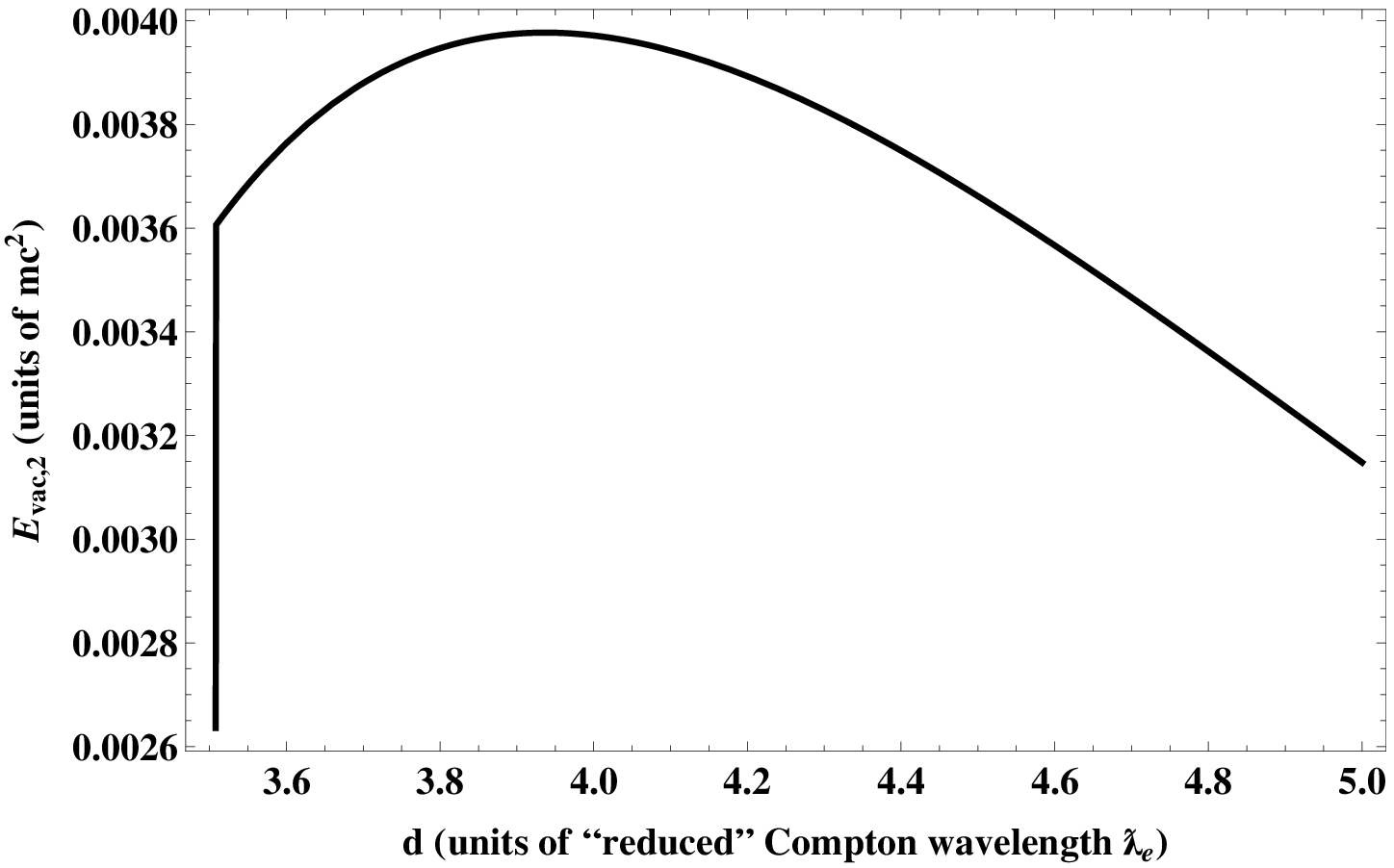}
}
\vfill
\subfigure[]{
		\includegraphics[width =\columnwidth]{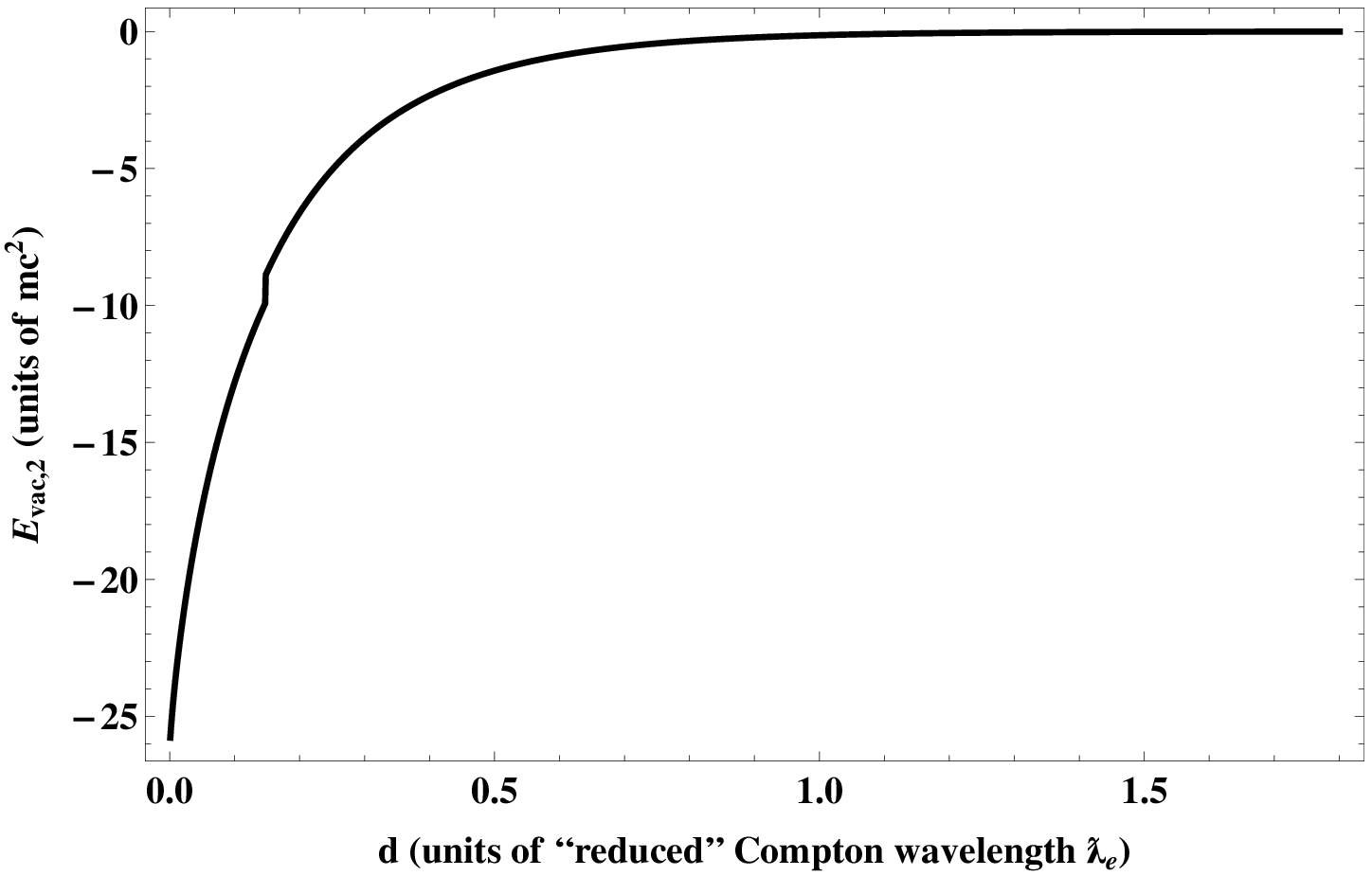}
}	
\hfill
\subfigure[]{
		\includegraphics[width=\columnwidth]{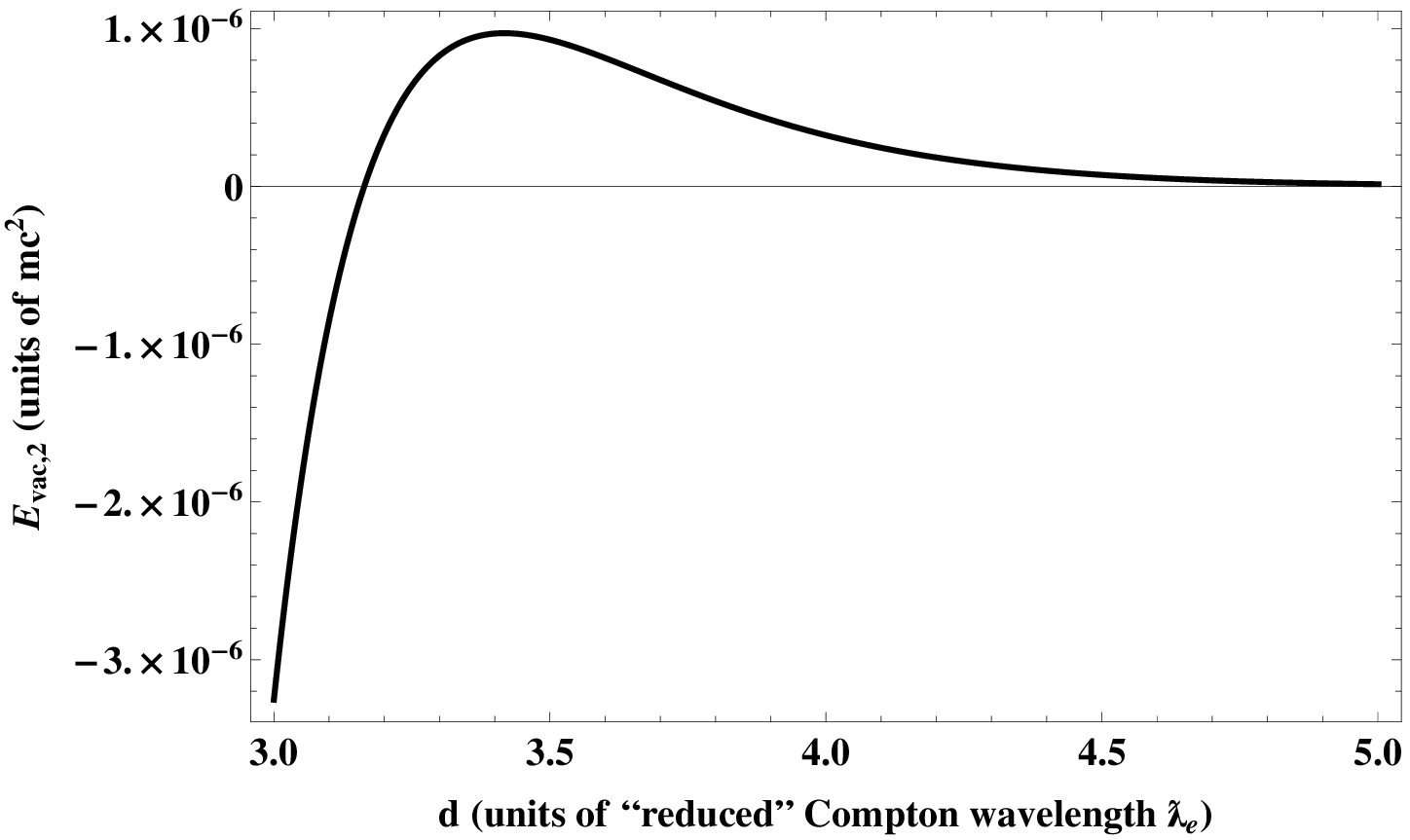}
}
\caption{Different types of the Casimir interaction energy between two $\d$-wells as functions of the distance $d$ between them for: (a) $C=1$, (b) $C=10$, (c,d) $C=3$, (e,f) $C=5$.}
	\label{EintD}	
\end{figure*}

If $\e_0<0$, the contribution from the discrete spectrum for $d \sqrt{1-\e^2_0} \gg 1$ equals to
\beq\label{delta-discr}\begin{gathered}
-S_{int}=-(\e_+ + \e_- -2\e_0)\simeq \\ \simeq e^{-4 d\sqrt{1-\e^2_0}}\,\e_0\, (1-\e^2_0)\,\left(1-4 d \sqrt{1-\e^2_0}\right)>0 \  ,
\end{gathered}\eeq
and due to the exponent $e^{-4 d\sqrt{1-\e^2_0}}$ turns out to be the leading term in $\E^{int}_{vac}(d)$, implying for $\e_0$ close to $\e_F$  the existence of long-range forces between such $\d$-wells quite similar to the case of finite wells.  In turn, this is the reason of the behavior of interaction energy between wells for $C=3$ and $C=5$ for large separation (see Figs. \ref{EintD}d,f below).
At the same time, if $\e_0>0$, then $S_{int}=0$, and the interaction energy $\E^{int}_{vac}(d)=I_{int}(d)+\Lambda_{int}(d) V^2_0$  decreases  with growing  $d$ much faster, namely as $O\(e^{-4 d}\)$.

If $\e_0=0$, i.e. for $C=\pi/2+\pi n$, the expression (\ref{eq:deltaIintBigd}) isn't valid, since an essential circumstance here is that  $\cos C$ entering the denominators in (\ref{delta-int1}) and (\ref{eq:deltaIintBigd})  should be non-zero. In this case the integral term transforms into
\beq
I_{int} \simeq -e^{-2 d}+ O\(e^{-4 d}\) \ ,
\eeq
while the contribution from the discrete spectrum contains now  the level $\e_-<0$ only and gives \beq
-S_{int} \simeq e^{-2 d}+ O\(e^{-6 d}\) \ .
\eeq
Therefore for $\e_0=0$ the interaction energy between two $\d$-wells decreases also as $O\(e^{-4 d}\)$.

In Figs.\ref{EintD} the dependence of the interaction energy between two  $\d$-wells on the distance  $d$ between them for a set of different values of the parameter  $C$ is shown. As it follows from Figs.\ref{EintD}c,d and e,f, depending on the concrete value of $C$  the nature of the Casimir force between wells   may change from attraction to repulsion with growing $d$. In the present case this effect takes place for $C=3$ and $C=5$. For other values of $C$, shown in Figs.\ref{EintD}, the interaction energy is strictly negative and grows with increasing $d$, so the wells  attract each other.
The jump-like behavior of energy at $d=3.5076$ for $C=3$ (Figs.\ref{EintD}c) and at $d=0.1479$ for $C=5$ (Fig.\ref{EintD}e) is caused by emergence of a new level at the lower threshold, provided the condition
\beq
d=-\ctg(C)/2>0 \ ,
\eeq
which follows from (\ref{95}) in the limit $\e_- \to -1$, is fulfilled. Another way to achieve this condition is to use the eq. (\ref{creven}) in the $\d$-limit.

With further removal of the wells from each other this level goes up, approaching from below  the unique level $\e_0$ in the single  $\d$-well (\ref{eq:root1}) (for $C=3$ and $C=5$ the latter is negative). Meanwhile the second level goes down, approaching the value $\e_0$ from above. For $C=1$ and $C=10$ there are no negative  $\e_0$, and so starting  from sufficiently large  $d$ the contribution from the discrete spectrum to $\E_{vac}^{int}(d)$ disappears.

\subsection*{6. Casimir forces in the source-anti-source system}

There exists only one exception, when the effect of long-range Casimir force in quasi-one-dimensional QED system with short-range Coulomb sources of the type considered above, cannot be able in principle. It is   the anti-symmetric configuration of  the type source-anti-source, where one of the wells is replaced by a barrier with the same width and height. For our purposes it would be  pertinent to consider an even more general situation, described by the external potential of the form
\beq
W_2(x) =-\[V_1 \, \tt\(x -d \) + V_2 \, \tt\(-x -d \)\]\,\tt\(d+a -|x|\) \ ,
\label{w2}
\end{equation}
although in what follows we'll be interested first of all in the antisymmetric case with  $V_1=-V_2=V_0>0$.

In the first step, for such a configuration of external short-range Coulomb sources,   the calculation of corresponding vacuum charge density will be  useful. For these purposes one needs to consider the trace of the Green function
\beq\label{trG}
\tr G(x,x;\e)={1\over J(\e)}\psi^T_L(x)\psi_R(x) \ ,
\eeq
with  $\psi_{R,L}(x)$ being the  solutions of DE (\ref{deq}) with $V(x)$ replaced by $W_2(x)$, which are regular at $\pm \infty$, respectively,
while $J(\e)$ is their Wronskian. As in eq. (\ref{F}), we use here the denotation
\beq
[f,g]_a=f_2(a)g_1(a)-g_2(a)f_1(a)
\eeq
for the Wronskian of functions $f(x)$ and $g(x)$, calculated at  the point $x=a$.
In terms of the latter definition the Wronskian $J(\e)$ in eq. (\ref{trG}) equals to
\beq
J(\e)=[\psi_L,\psi_R] \ .
\eeq
For the external potential $W_2(x)$ the pertinent solutions of DE are represented in the following form
\beq\label{solsRL}
\psi_L(x)=\left\{
\begin{array}{l}
\Phi(x) \ ,\quad x\leq -d-a \ ,\\
A_L u(V_2, x) + B_L v(V_2, x) \ , \ -d-a\leq x\leq -d \ ,\\
C_L \Phi(x) + D_L \Psi(x) \ , \quad |x|\leq d \ ,\\
E_L u(V_1, x) + F_L v(V_1, x) \ , \quad d\leq x\leq d+a \ ,\\
G_L \Phi(x) + H_L \Psi(x) \ ,\quad a+d \leq x \ ,
\end{array}
\right.
\eeq
\beq
\psi_R(x)=\left\{
\begin{array}{l}
G_R \Psi(x) + H_R \Phi(x) \ , \quad x\leq -d-a \ ,\\
E_R u(V_2, x) - F_R v(V_2, x) \ , \ -d-a\leq x\leq -d \ ,\\
C_R \Psi(x) + D_R \Phi(x) \ , \quad |x|\leq d \ ,\\
A_R u(V_1, x) - B_R v(V_1, x) \ , \quad d\leq x\leq d+a \ ,\\
 \Psi(x) \ , \quad a+d \leq x \ ,
\end{array}
\right.
\eeq
with the coefficients $A_{R,L}\, , B_{R,L}\, , C_{R,L}\, , D_{R,L}\, ,  E_{R,L}\, , F_{R,L}\, , $ $ G_{R,L}\, , H_{R,L}$ being obtained via the requirement of continuity of solutions  $\psi_{R,L}(x)$ at the points  $x=\pm d\, , \pm (d+a)$, while $\Phi(x)\, , \Psi(x)\, , u(V_i, x)\, , v(V_i, x)\, ,\ i=1, 2,$ are the linearly independent solutions of DE in the corresponding regions  of constant potential $W_2(x)$
\beq\begin{aligned}
&\Phi(x)=\begin{pmatrix}
\sqrt{1+\e}\ e^{x\sqrt{1-\e^2}}\\
\sqrt{1-\e}\ e^{x\sqrt{1-\e^2}}
\end{pmatrix} \ , \\
&\Psi(x)=\begin{pmatrix}
\sqrt{1+\e}\ e^{-x\sqrt{1-\e^2}}\\
-\sqrt{1-\e}\ e^{-x\sqrt{1-\e^2}}
\end{pmatrix} \ , \\
\end{aligned}
\label{tsols1}
\end{equation}
\beq\begin{aligned}
&u(V_i, x)=\\
&=\begin{pmatrix}
\cos\(x\sqrt{(\e+V_i)^2-1}\) \\
-\sqrt{\e+V_i-1}\, \sin\(x\sqrt{(\e+V_i)^2-1}\) / \(\e+V_i+1\)  \\
\end{pmatrix} \ , \\
&v(V_i, x)=\\
&=\begin{pmatrix}
\sqrt{\e+V_i +1}\, \sin\(x\sqrt{(\e+V_i)^2-1}\) / \(\e+V_i-1\)  \\
 \cos\(x\sqrt{(\e+V_i)^2-1}\)
\end{pmatrix} \ . \\
\end{aligned}
\label{tsols2}
\end{equation}
\normalsize
The cross-linking coefficients  with label $R$ take the form
\beq
\begin{gathered}
A_R={[\Psi,v(V_1)]_{a+d}\over [u(V_1),v(V_1)]_{a+d}}\ ,\\ B_R={[\Psi,u(V_1)]_{a+d}\over [u(V_1),v(V_1)]_{a+d}} \ ,\\
D_R={A_R[u(V_1),\Psi]_{d}-B_R[v(V_1),\Psi]_{d}\over [\Phi,\Psi]_{d}} \ , \\ C_R={A_R[\Phi,u(V_1)]_{d}-B_R[\Phi,v(V_1)]_{d}\over [\Phi,\Psi]_{d}} \ ,\\
F_R={C_R[\Phi,u(V_2)]_{d}+D_R[\Psi,u(V_2)]_{d}\over[v(V_2),u(V_2)]_{d}} \ , \\ E_R={C_R[v(V_2),\Phi]_{d}+D_R[v(V_2),\Psi]_{d}\over[v(V_2),u(V_2)]_{d}} \ ,
 \end{gathered}\label{CoefR}
\eeq
$$ H_R={E_R[u(V_2),\Phi]_{a+d}+F_R[v(V_2),\Phi]_{a+d}\over[\Psi,\Phi]_{a+d}} \ , $$
$$ G_R={E_R[\Psi,u(V_2)]_{a+d}+F_R[\Psi,v(V_2)]_{a+d}\over[\Psi,\Phi]_{a+d}} \ . $$
The corresponding coefficients with label $L$ are obtained from (\ref{CoefR}) by means of replacement $R \to L$ and $V_1 \leftrightarrow V_2$.

By means of (\ref{solsRL}-\ref{CoefR}) for the explicit form of  $J(\e)$ one finds
\beq\begin{gathered}
J(d,\e)=2\,{e^{-2 a \sqrt{1-\e^2}}\over \sqrt{1-\e^2}}\, \left[ f_1(V_1,\e) f_1(V_2,\e) - \right. \\ \left. - e^{-4 d \sqrt{1-\e^2}} f_2(V_1,\e) f_2(V_2,\e) \right] \ , \label{Wronskian}
\end{gathered}\eeq
where
\beq\begin{gathered}\label{f12}
f_1(V_i,\e)=\sqrt{1-\e^2}\,\cos\(a\sqrt{(V_i+\e)^2-1}\) - \\ \(\e^2-1+V_i \e\) \,\sin\(a\sqrt{(V_i+\e)^2-1}\)/\sqrt{(V_i+\e)^2-1} \ , \\
f_2(V_i,\e)=V_i\,\sin\(a\sqrt{(V_i+\e)^2-1}\) / \sqrt{(V_i+\e)^2-1} \ .
\end{gathered}\
\eeq

\begin{figure*}[ht!]
\subfigure[]{
		\includegraphics[width=\columnwidth]{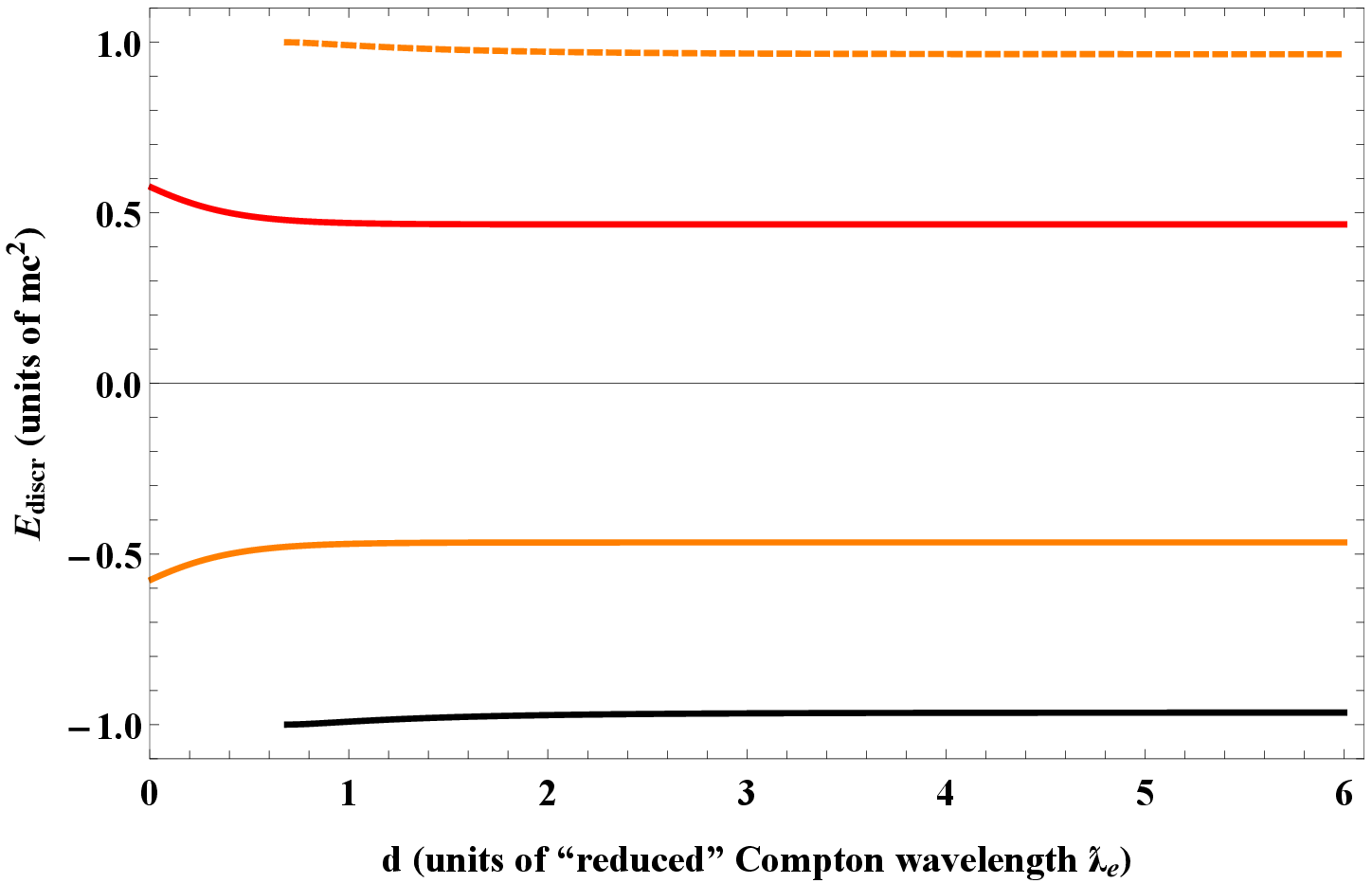}
}
\hfill
\subfigure[]{
		\includegraphics[width=\columnwidth]{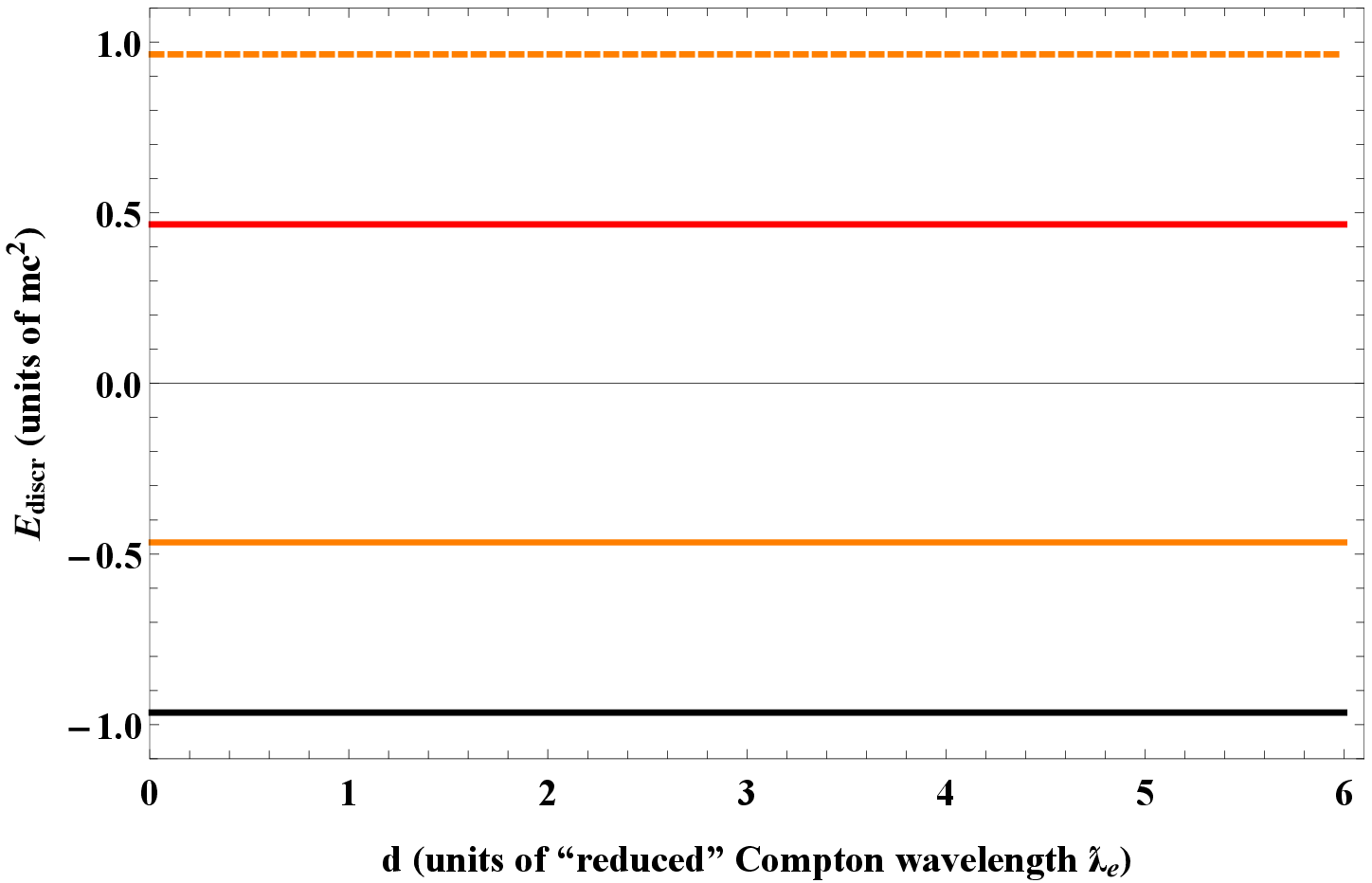}
}
\caption{(Color online). Energy levels in the case  $a=1$, $V_0=4.08$ for the configurations: (a) the antisymmetric one; (b) single well (black and red lines) and barrier (orange solid and dashed lines). }
	\label{Ediscr}	
\end{figure*}
Let us consider now more thoroughly the antisymmetric case of the configuration  barrier-well, when  $V_1=-V_2=V_0>0$. As it follows from the expressions  (\ref{Wronskian})-(\ref{f12}), in this case  $J(d,\e)$ turns out to be an even function of the energy
\beq\label{J}
J(d,\e)=J(d,-\e) \ .
\eeq
Therefore, the discrete spectrum of the problem should be sign-symmetric, i.e. the levels appear only in pairs with  $\pm \e$. Actually, the latter circumstance  is the general feature of the source-anti-source system, including both the discrete spectrum and  continua. Namely, all the energy eigenstates in such system are   related via (up to a phase factor)
\beq\label{oddness}
\p_{-\e}(x)=\a\, \p_\e (-x) \ .
\eeq
The typical behavior of levels for the antisymmetric case  is shown in Fig.\ref{Ediscr}a in dependence on the distance $d$ between sources for $a=1\, , V_0=4.08$. The set $(V_0\, , a)$ is taken with the same values as for the symmetric case containing two wells, considered in Sect.4. The symmetry of levels  relative to the zero energy line is apparent. Note also that  the highest and lowest levels appear only starting from certain $d>0$. With increasing $d$ all the levels tend to constant values, coinciding with those of the single well and barrier of the same width  $a$ and depth/hight $V_0$. Such behavior follows directly from the eq. $J(d,\e)=0$. For $d\gg 1$ the second term in the expression (\ref{Wronskian}) can be neglected, hence, the resulting equation for the levels  transforms into
\beq
 f_1(V_0,\e)f_1(-V_0,\e)=0 \ .
\eeq
In turn, the latter splits into two independent equations for the levels in the single well and barrier,  namely, $f_1(V_0,\e)=0$ for the well and $f_1(-V_0,\e)=0$ for the barrier, which are related by reflection $\e \to  -\e$. So the discrete spectra of the well and barrier differ only by the sign, as
expected. For more  clarity, in Fig.\ref{Ediscr}b the levels in the single well and barrier with the same parameters $a=1$, $V_0=4.08$, are shown.  There exist two levels with values $\e_1=-0.965$, $\e_2=0.466$ in the well, while for the barrier one finds two levels with opposite signs.

\begin{figure*}[ht!]
\subfigure[]{
		\includegraphics[width=\columnwidth]{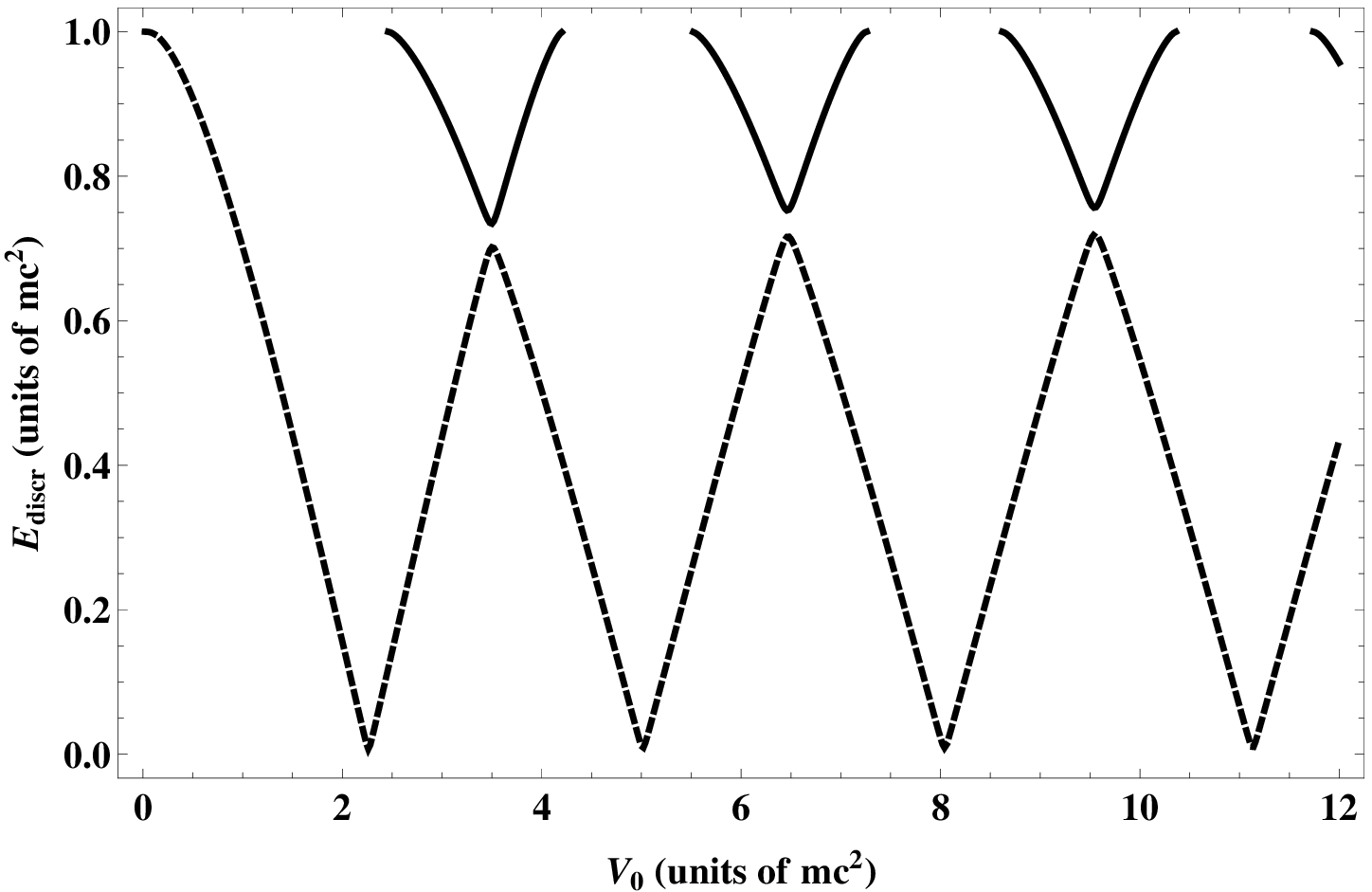}
}
\hfill
\subfigure[]{
		\includegraphics[width=\columnwidth]{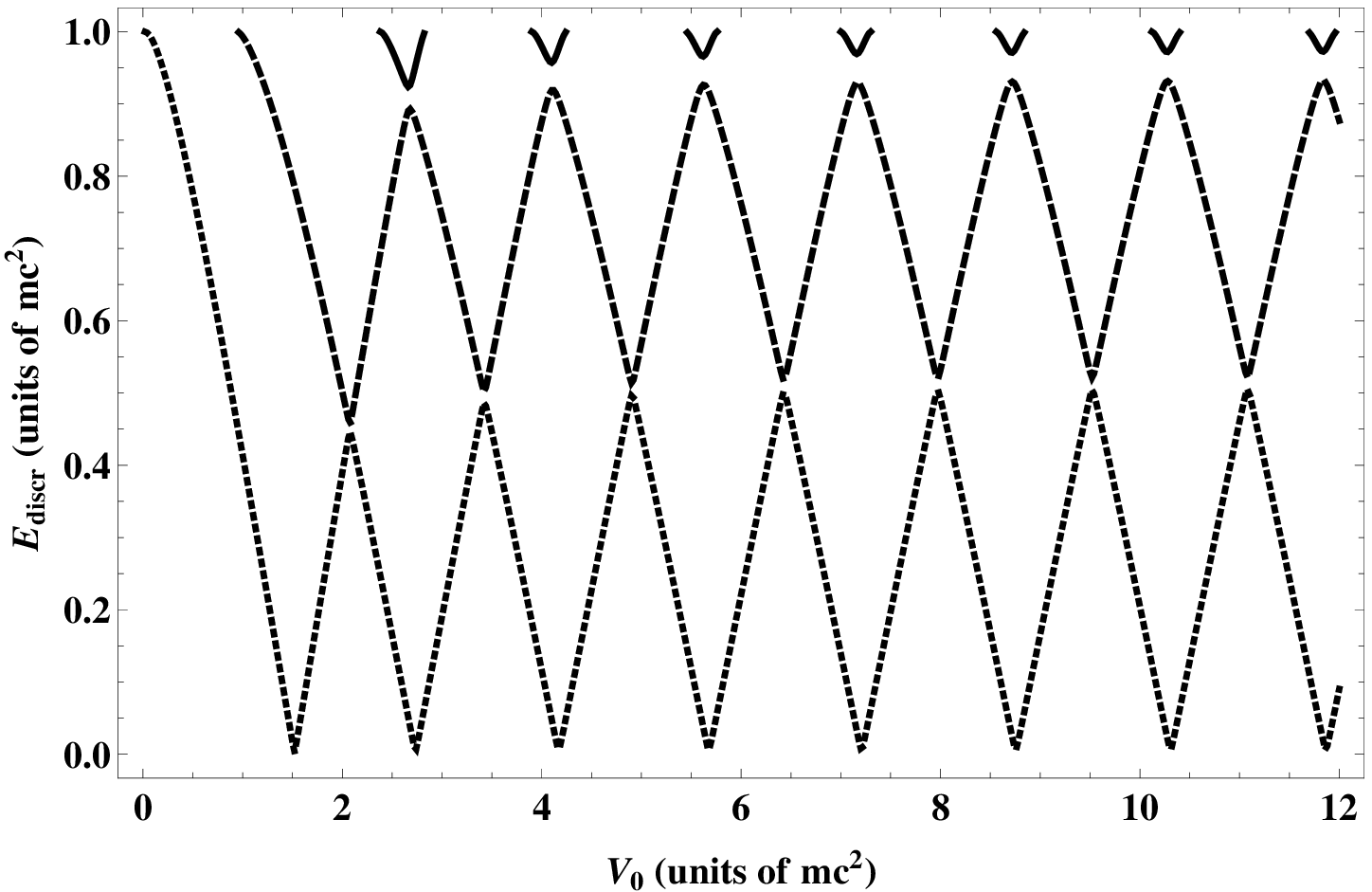}
}
\caption{The behavior of positive energy levels in the antisymmetric case of the type barrier-well in dependence on $V_0$: (a) for $d=2$, $a=1$; (b) for $d=2$, $a=2$. }
	\label{EdiscrV0}	
\end{figure*}

The sign symmetry of the energy spectrum in the source-anti-source systems leads to significant changes in the definition and properties of vacuum polarization density and energy. The most important point here is that due to sign symmetry of the levels  the whole spectrum splits into  two non-intersecting parts with positive and negative energies, respectively, since the levels cannot intersect, and hence, cannot cross the zero line (see Figs.\ref{EdiscrV0}).
Therefore, in this case the Fermi level, dividing the electronic and positronic (electron-hole) eigenstates in the initial expressions for the vacuum averages similar to (\ref{eterms}), should be chosen equal to zero, i.e. $\e_F=0$.

So the starting expression  for the  induced density should be written as
\beq \label{rhoVP}
\r_{vac}(x)=-\frac{|e|}{2}\(\sum\limits_{\e_{n}<0} \p_{n}(x)^{\dagger}\p_{n}(x)-\sum\limits_{\e_{n}>0} \p_{n}(x)^{\dagger}\p_{n}(x) \),
\eeq
where $\e_{n}$ and $\p_n(x)$ are the eigenvalues and the eigenfunctions of the corresponding DE for the antisymmetric case. Proceeding further, one finds that due to sign symmetry of the spectrum the WK-contour, shown in Fig.\ref{contour}, transforms now into the symmetric one with respect to reflection $\e \to -\e^{\ast}$, while  its separate parts $P(R)$ and $E(R)$ each lie in their half-planes $\mathrm{Re}\, \e <0$ for  $P(R)$ and $\mathrm{Re}\, \e >0$ for $E(R)$ and  don't intersect with the imaginary axis.

As it should be expected from general grounds, there follows from eq. (\ref{rhoVP}) combined with relation (\ref{oddness}) that the  vacuum density is an odd function
\beq
\rho_{vac}(x)=-\rho_{vac}(-x) \ ,
\eeq	
reproducing this way the similar property of the external potential (\ref{w2}) in the antisymmetric case.

Applying  further the same technique as in Refs.\cite{davydov2017}-\cite{voronina2017}, \cite{wk1956}-\cite{21} for the expression of the induced density in terms of $\tr G$, one finds
\beq\label{rhoVP1}
\rho_{vac}(x)={|e|\over 2 \pi}\int\limits_{-\infty}^{\infty} dy\, \tr G(x,x;i y) \ .
\eeq
Note that in the expression (\ref{rhoVP1}) there is no separate contribution from negative discrete levels, since the latter appears only in the case when the part $E(R)$ of the WK-contour  captures  a piece of negative real axis containing these discrete levels.

Since $\rho_{vac}(x)$ is odd from the very beginning, in contrast to symmetric case \cite{annphys} and all the more to the one-dimensional QED systems with long-range external Coulomb sources considered in Refs.\cite{davydov2017}-\cite{voronina2017}, the total induced charge vanishes now  without any additional renormalization
\beq
Q_{vac}=\int\limits_{-\infty}^{\infty} dx\, \rho_{vac}(x)=0 \ .
\eeq
Nevertheless, a finite renormalization is needed due to condition that in the perturbative region  $V_0 \to 0$ the renormalized vacuum density $\rho^R_{vac}(x)$ should reproduce the perturbative density $\rho^{(1)}_{vac}(x)$, calculated within the standard perturbation theory (PT) to the leading (one-loop) order \cite{davydov2017}-\cite{voronina2017}, \cite{annphys,tmf}. Actually,  this procedure is equivalent to a finite renormalization and normalization conditions as known from perturbative QED (see, e.g., Ref.\cite{itzykson2012}). The explicit expression for $\rho^{(1)}_{vac}(x)$ reads
\begin{widetext}
\beq\begin{gathered}
\rho^{(1)}_{vac}(x)=\\ -{|e| \over \pi^2}\int\limits_0^{\infty} {dq \over q}\, \left(1-2{ \mathrm{arcsinh} (q/2) \over q \sqrt{1+(q/2)^2}}\right)  \Big(V_1\[\sin (q(d-x)) - \sin (q(a+d-x))\] + V_2 \[\sin (q(d+x)) - \sin (q(a+d+x))\]\Big) \ .
\label{rho1}
\end{gathered}\eeq\end{widetext}
It should be quite clear without any additional comments that in the antisymmetric case the perturbative density is an odd function by construction, hence, in this case  the total induced charge $Q^{(1)}_{vac}$, calculated to the leading order of PT by means of $\rho^{(1)}_{vac}(x)$,  vanishes (actually this statement holds also for the non-symmetric case, for details see, e.g., Ref.\cite{davydov2018}).

Thus, by means of the standard renormalization procedure for the vacuum density considered in Refs. \cite{davydov2017}-\cite{voronina2017}, \cite{wk1956}-\cite{21}, one obtains
\beq\label{rhoR}
\rho^R_{vac}(x) = \rho^{(1)}_{vac}(x) + \rho^{(3+)}_{vac}(x) \ ,
\eeq
where
\beq\label{rho3+}
\rho^{(3+)}_{vac}(x)= {|e|\over 2 \pi}\int\limits_{-\infty}^{\infty} \! dy\, \[\tr G(x,x; i y)- \tr G^{(1)}(x,x; i y)\] \ .
\eeq
In the expression (\ref{rho3+})  the function  $\tr G^{(1)}(x,x; \e)$ is the first-order term in the expansion of the Green function in the Born series in powers of  $V_0$ (for the antisymmetric case).
\begin{figure*}[ht!]
\subfigure[]{
		\includegraphics[width=\columnwidth]{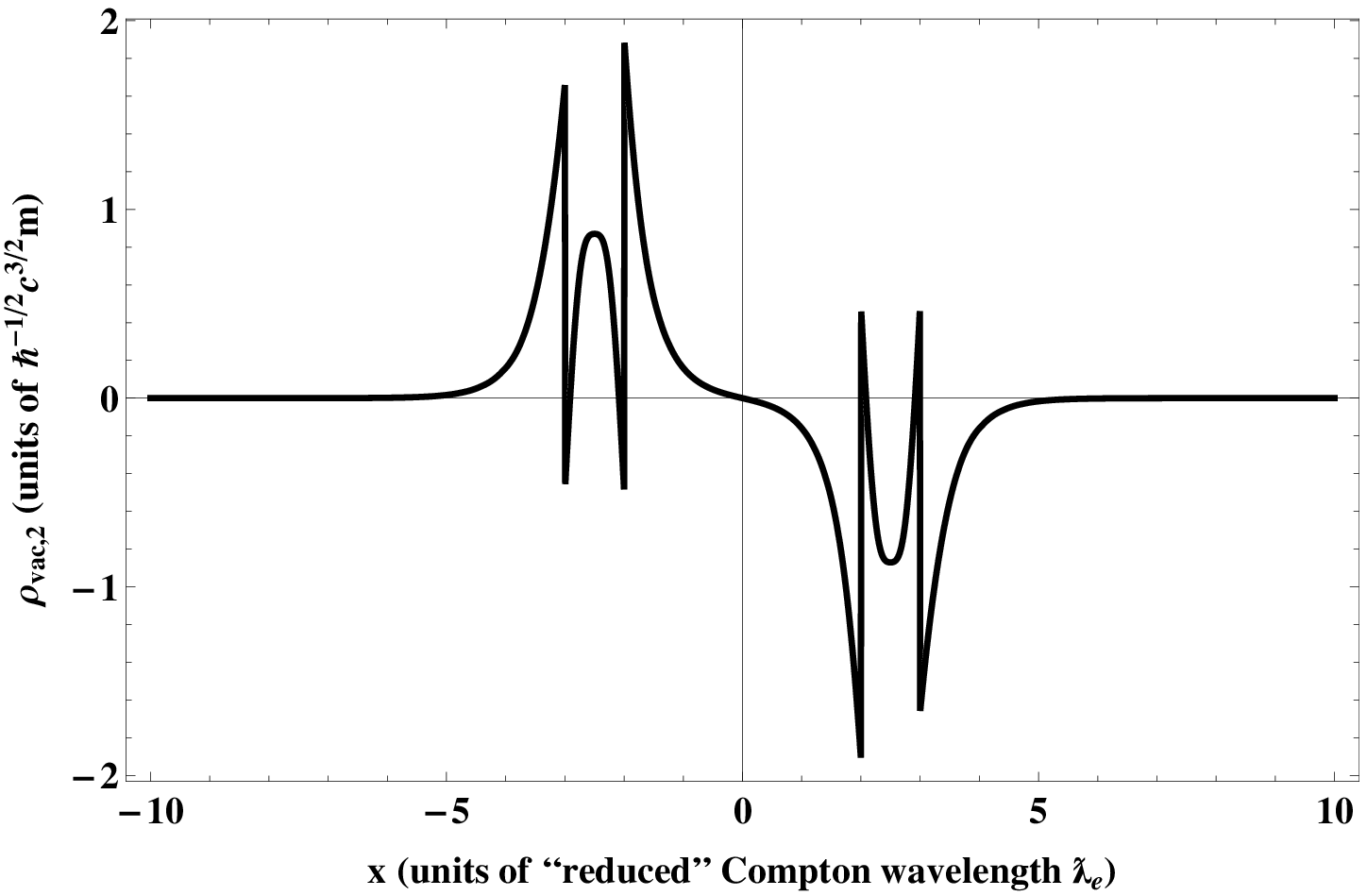}
}
\hfill
\subfigure[]{
		\includegraphics[width=\columnwidth]{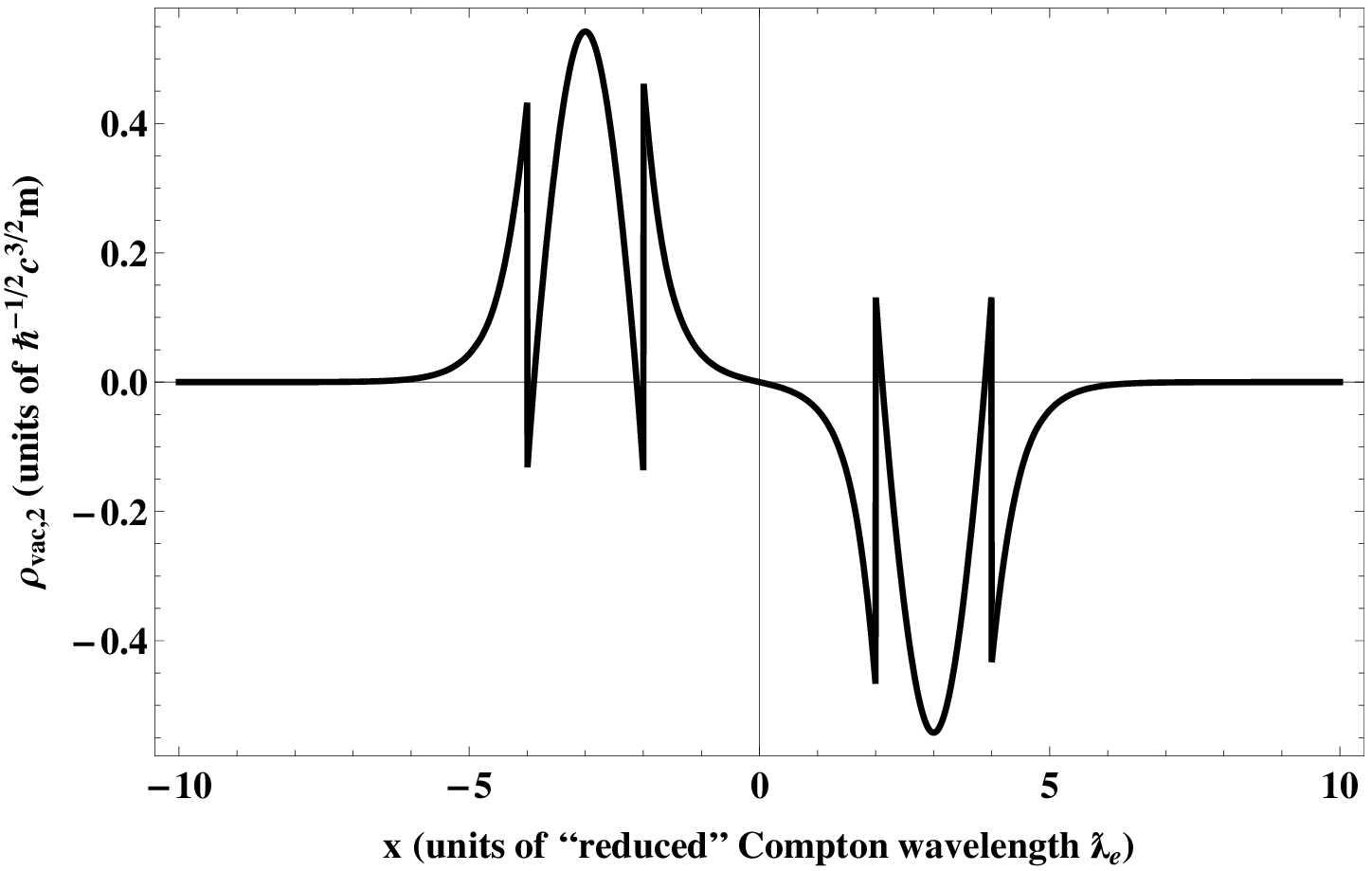}
}
\caption{The renormalized vacuum charge density in the antisymmetric case for the following sets of the system parameters:  (a) $d=2$, $a=1$, $V_0=8$; (b) $d=2$, $a=2$, $V_0=2$. }
	\label{rho}	
\end{figure*}
In Fig.\ref{rho} the renormalized vacuum charge density is shown for the following sets of the system parameters:  (a) $d=2$, $a=1$, $V_0=8$; (b) $d=2$, $a=2$, $V_0=2$. In general, the behavior of density is quite similar to those achieved for the case of two wells in Ref.\cite{annphys} with the main exception that now the density is odd.

After these preliminary considerations let us turn to calculation of the Casimir energy for the antisymmetric configuration. Repeating the procedure of passing from the initial definition of the vacuum energy by means of the Schwinger average (\ref{eterms}) to the integration over the imaginary axis in (\ref{IntWronskReg}), considered in detail for the symmetric case,  for the non-renormalized vacuum energy one obtains
\beq\label{IntWronskReg1}
\E_{vac}(d)=-{1\over \pi}\int\limits_0^{\infty} dy\, \mathrm{Re}\left[\ln J_{red}(d,i y)\right] \ ,
\eeq
with  the same definition of the ``reduced'' Wronskian $J_{red}(d,i y)$ as in (\ref{Jred}).
For the antisymmetric case one obtains
\beq\begin{gathered}
J_{red}(d, i y)={e^{-2 a \sqrt{1+y^2}}\over 1+y^2} \times \\ \times \left[ |f_1(V_0, i y))|^2 -e^{-4 d \sqrt{1+y^2}} |f_2(V_0, i y)|^2  \right] \ ,
\end{gathered}\eeq
with $f_i(V_0, iy)$ being defined in (\ref{f12}).  Note also that in contrast to (\ref{econt}), due to the same reasons as in (\ref{rhoVP1}), in the expression (\ref{IntWronskReg1})  there is no separate contribution from the negative discrete levels.

The renormalized vacuum energy is represented as
\beq
\E^R_{vac}(d)=\E_{vac}(d)+\lambda(d)\, V^2_0 \ ,
\eeq
where the renormalization coefficient $\lambda(d)=\lambda_1(d)-\lambda_2(d)$ contains two terms of the following form
\begin{widetext}\beq
\lambda_1(d)=\lim_{V_0\to 0}\, \E^{(1)}_{vac}(d)/V^2_0 \ , \quad
\lambda_2(d)=\lim_{V_0\to 0}\, \E_{vac}(d)/V^2_0 = {a\over \pi}-{1\over 8}+{1\over \pi}\,\int\limits_0^{\infty}  dy\, {1-2 e^{-4 d\sqrt{1+y^2}}\sh^2(a\sqrt{1+y^2}) \over 2 (1+y^2)^2}\,e^{-2 a\sqrt{1+y^2}} \ ,
\eeq
where the first-order perturbative vacuum energy $\E^{(1)}_{vac}(d)$ is given by the following expression, calculated within PT in the one-loop approximation for the antisymmetric case
\begin{multline}
\E^{(1)}_{vac}(d)={1\over 2}\int\limits_{-\infty}^{\infty} dx\, \rho^{(1)}_{vac}(x) A^{ext}_0(x)=-{1\over \pi^2}\,\int\limits_0^{\infty}\,{dq\over q^2}\, \left(1-2{ \mathrm{arcsinh} (q/2) \over q \sqrt{1+(q/2)^2}}\right) \times \\
\times \[-2+2 \cos(a q)-\cos(2 d q)-\cos(2 (a+d)q) + 2\cos((a+2 d)q)\] \ ,
\label{EvacV0}
\end{multline}\end{widetext}
where $A^{ext}_0(x)$ is related to the external potential $W_2(x)$ in DE via $W_2(x)=-|e|A^{ext}_0(x)$.

In the antisymmetric case there holds also the relation  $\lambda_1+\lambda_2=a/\pi$, which allows to represent the renormalization coefficient in a more convenient form, namely
\beq
\lambda(d)={a\over \pi}-2\lambda_2(d) \ .
\eeq

To explore the Casimir force in the source-anti-source system let us start with the behavior of non-renormalized vacuum energy $\E_{vac}(d)$ for large $d \gg 1$. In this case the expression (\ref{IntWronskReg1}) simplifies up to
\beq
\E_{vac}(d \gg 1)=-{1\over \pi}\,\int\limits_0^{\infty} dy\, \ln \left[ {e^{-2 a \sqrt{1+y^2}}\over 1+y^2} |f_1(V_0, i y))|^2\right] \ ,
\eeq
and coincides with the non-renormalized total energy of the system, containing infinitely separated barrier and well with the same width $a$ and depth/height $V_0$, but preserving the antisymmetry property (\ref{oddness}) of the whole configuration. Otherwise, considering the limiting configuration as a direct sum of the  single barrier and single well without antisymmetry property, we should deal with their contributions according to (\ref{econt}) for the well and to similar expression for the barrier, where the additional sum includes now positive discrete levels and enters with opposite sign. As a result, in this case the limiting vacuum energy will contain twice the sum over discrete levels, entering the expression (\ref{econt}). However, such a configuration cannot be considered as a physically correct limit for $\E_{vac}(d \gg 1)$, since the antisymmetry property is lost.

So the non-renormalized interaction  energy in the coupled barrier-well system equals to
\beq\label{Eint}\begin{gathered}
\E_{int}(d)=\E_{vac}(d)-\E_{vac}(d \to \inf)=\\ -{1\over \pi}\,\int\limits_0^{\infty} dy\, \ln \[J_{red}(d,i y)\, (1+y^2)\, {e^{2 a \sqrt{1+y^2}} \over |f_1(V_0,i y)|^2 }\] \ .
\end{gathered}\eeq
Expanding the integrand in the r.h.s. of (\ref{Eint}) for $d \gg 1 $ up to $O\(e^{-8 d \sqrt{1+y^2}}\)$, one obtains
\beq\label{EintApp}
\E_{int}(d) \simeq - {1\over \pi}\,\int\limits_0^{\infty} dy\, e^{-4 d \sqrt{1+y^2}} \ \Bigg| {f_2(V_0,i y) \over f_1(V_0,i y)}\Bigg|^2 \ .
\eeq
Further expansion of the expression (\ref{EintApp}) for large $d$ proceeds quite similar to the symmetric case and leads to the next answer
\beq
\E_{int}(d) \simeq -V^2_0\, {e^{-4 d} \over \sqrt{2 \pi d}}\,\left( {A^2 \over 2} +  {1\over 8 d} \left( {3 A^2 \over 8} + B \right)\right) + O\({1\over d^2}\) \ ,
\eeq
where $z_0$ and $A$ are defined as in (\ref{79}), while
\begin{multline}
B=A^3\Bigg( -V^2_0 \left( 1 - {\ctg(a z_0)\over z_0} + {a \over \sin^2(a z_0)} \right)^2 A \ - \\ - \ 2-{(1+z^4_0)\over z^3_0}\ctg(a z_0) \ +\\
+ \ {a\over z^2_0 \sin^2(a z_0)} \left( 1- 2 V^2_0(1- a z_0 \ctg(a z_0))\right) \Bigg) \ .
\end{multline}
As in the symmetric case, these expressions are  valid both for $V_0<1$ and $V_0>1$, while for $V_0=1$, when  $z_0=0$, they should be replaced by
\beq\begin{gathered}
A={a\over 1+ a} \ , \\
B=-a^2\, {45+135 a +165 a^2+90 a^3 + 28 a^4+ 8 a^5 \over 45(1+ a)^4} \ .
\end{gathered}\eeq
Moreover, the expansion of $\E_{int}(d)$ for large $d$, presented above, becomes invalid, when in the single well (or barrier) there exists the level with zero energy, since in this case the denominator in  $A$ vanishes, i.e. $ \sin(a z_0)+ z_0 \cos(a z_0)=0$. Therefore this case requires for a separate analysis, similar to considered for two wells in Sect.4.
\begin{figure*}[ht!]
\subfigure[]{
		\includegraphics[width=\columnwidth]{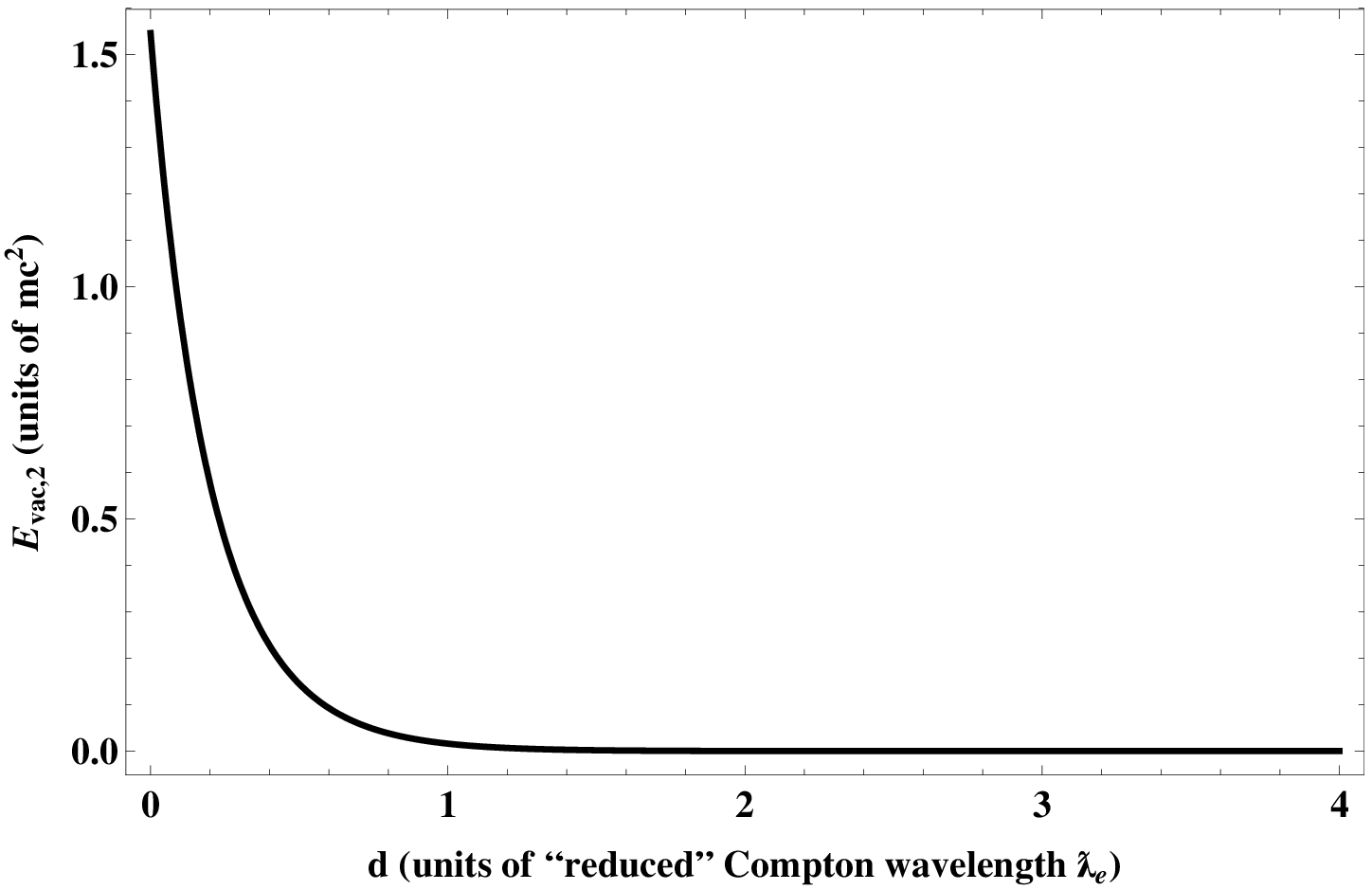}
}
\hfill
\subfigure[]{
		\includegraphics[width=\columnwidth]{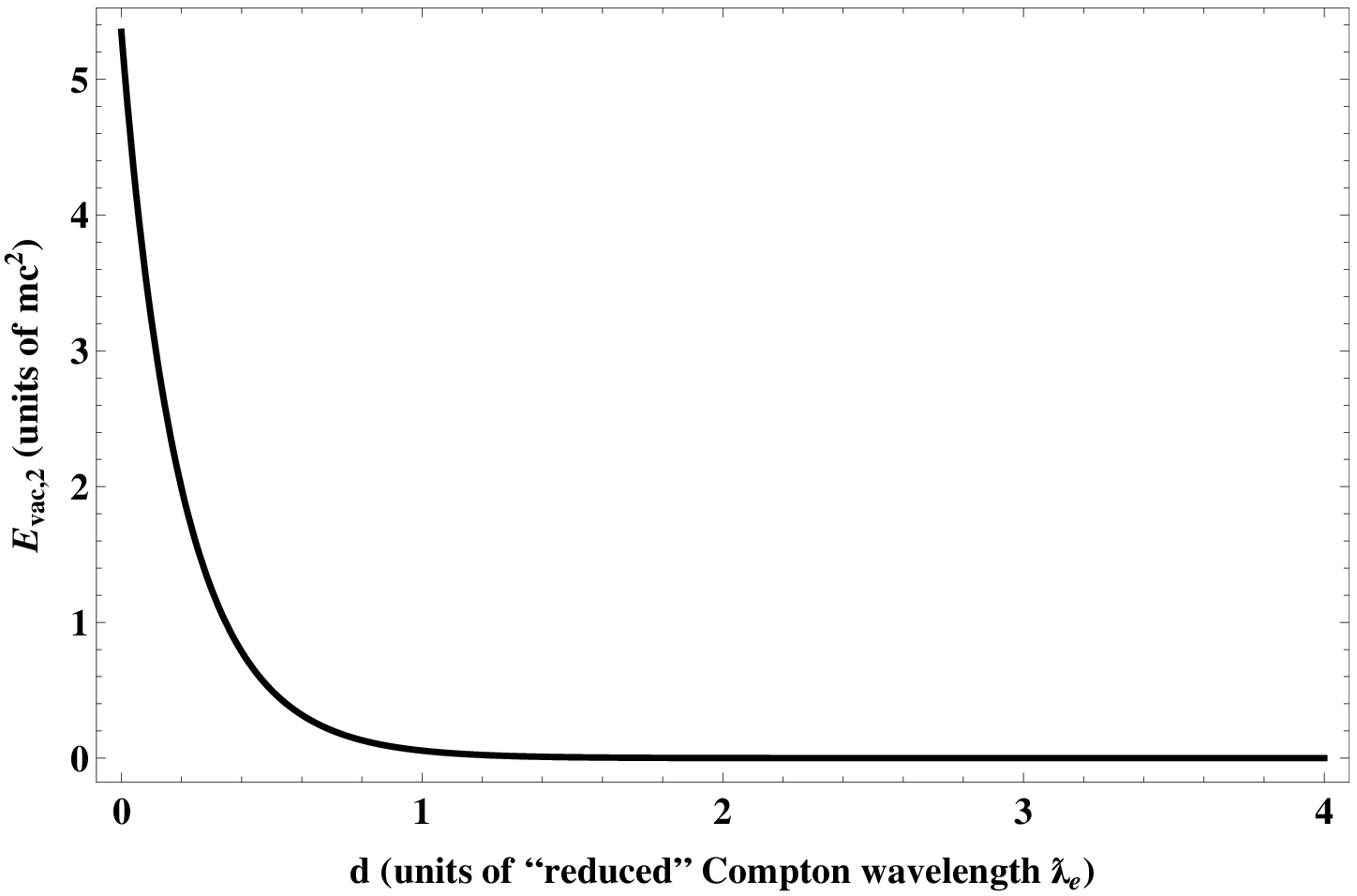}
}
\vfill
\subfigure[]{
		\includegraphics[width=\columnwidth]{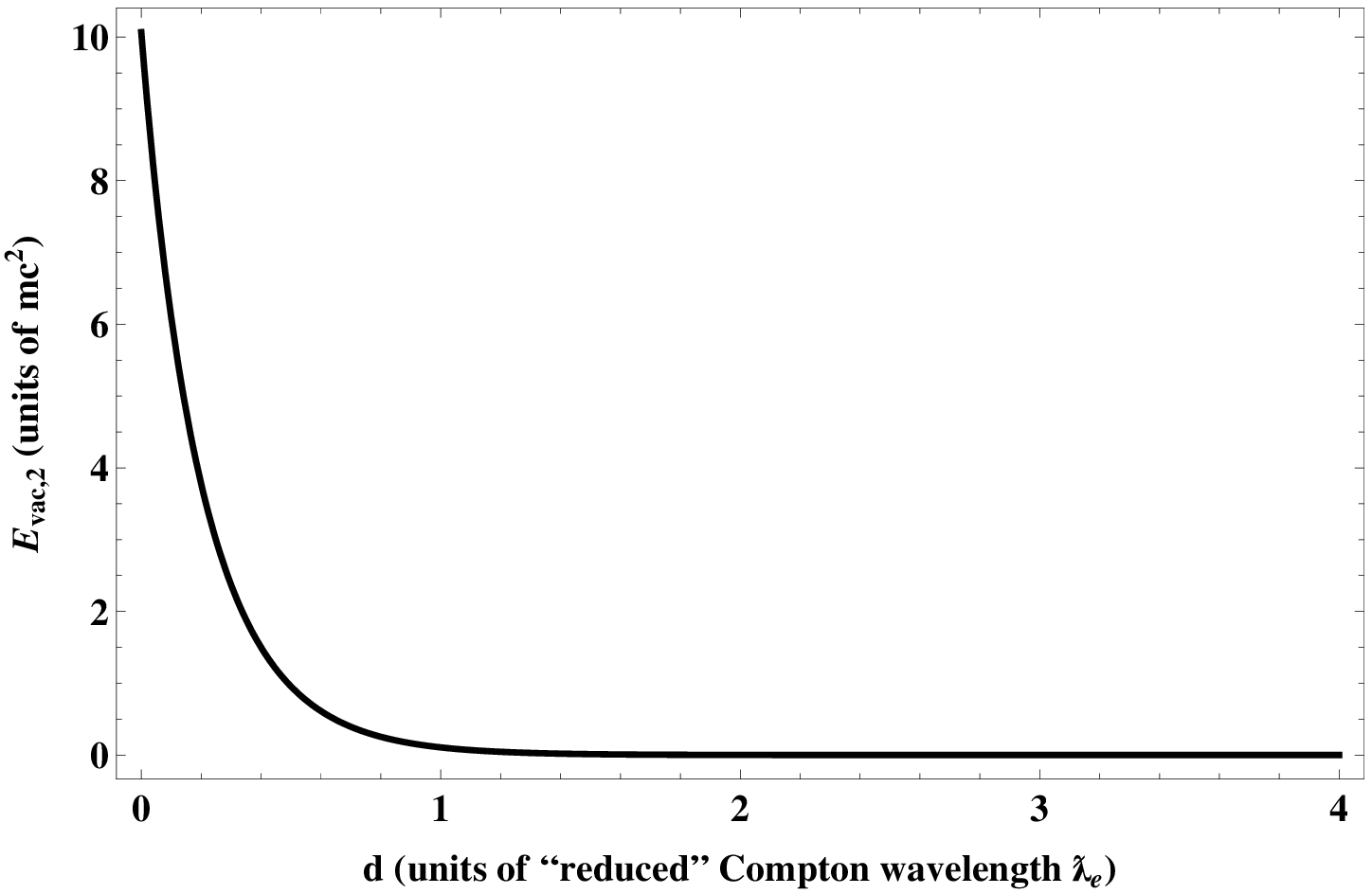}
}
\vfill
\subfigure[]{
		\includegraphics[width=\columnwidth]{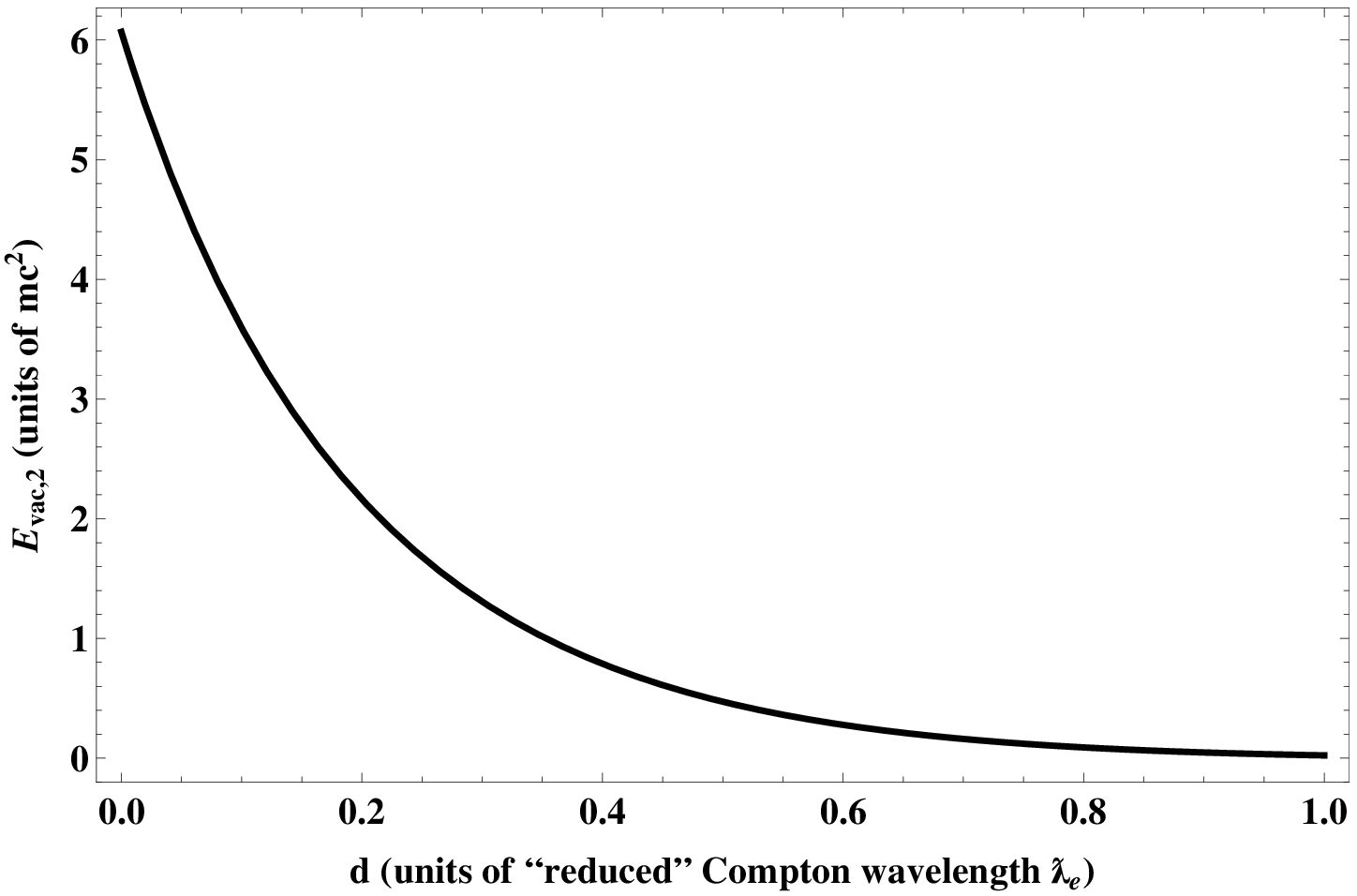}
}
\hfill
\subfigure[]{
		\includegraphics[width=\columnwidth]{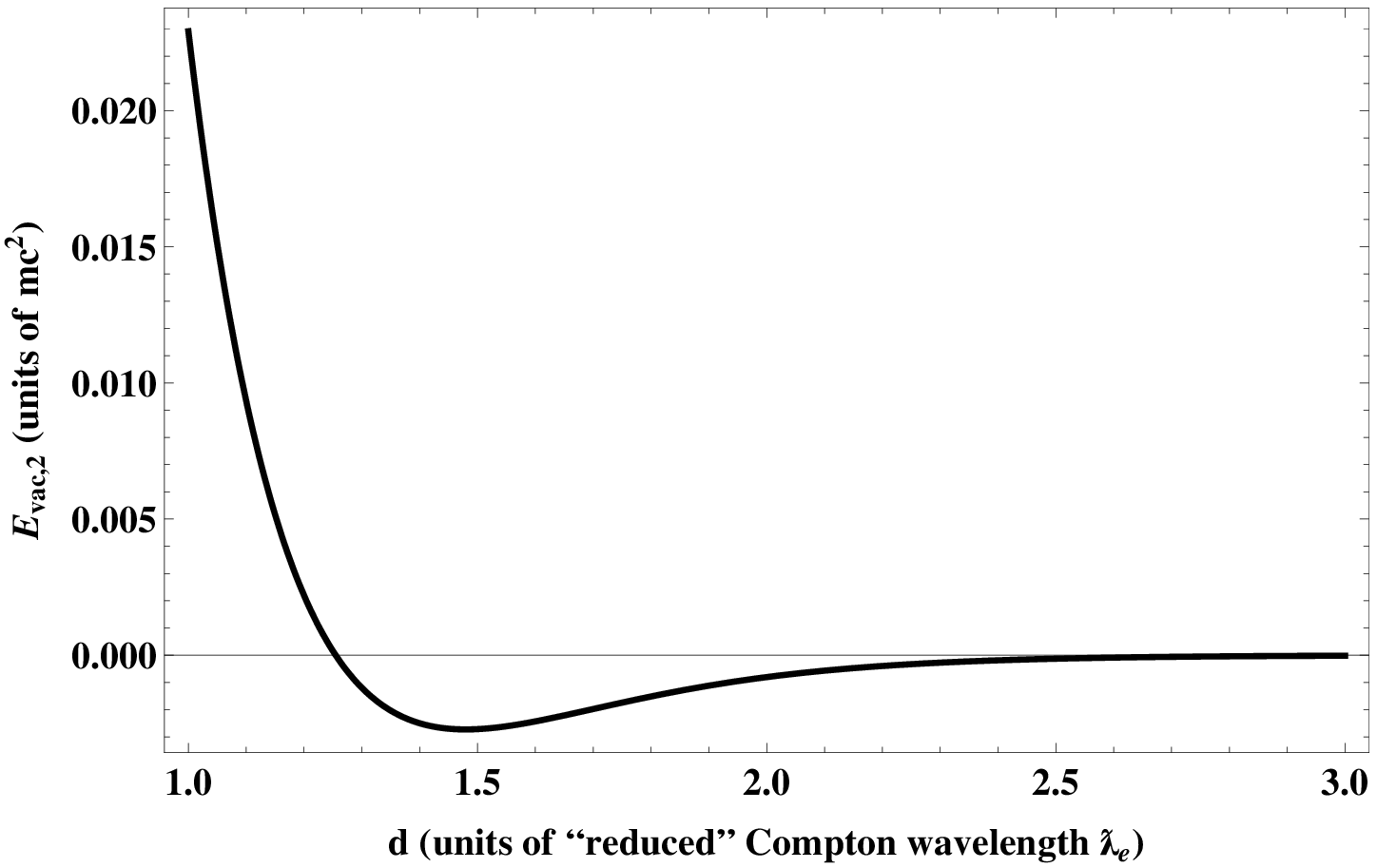}
}
\caption{The behavior of $\E_{int}^R(d)$ for $a=1$ and
(a) $V_0=4.08$; (b) $V_0=7.4$; (c) $V_0=10$; (d,e) $V_0=8$. }
	\label{Eint3}	
\end{figure*}

The renormalization coefficient for $\E_{int}^R(d)$ coincides with the corresponding one in the two-wells configuration up to the sign, namely
\beq
\L_{int}(d) ={1 \over 2\pi}\, \int\limits_0^{\infty} dy\, e^{-4 d \sqrt{1+y^2}}\, {\left(1- e^{-2 a \sqrt{1+y^2}}\right)^2 \over (1+y^2)^2} \geqslant 0 \ ,
\eeq
and so for large $d$ reveals the same asymptotics  as in (\ref{aslam}) with different sign.
\begin{figure*}[ht!]
\subfigure[]{
		\includegraphics[width=\columnwidth]{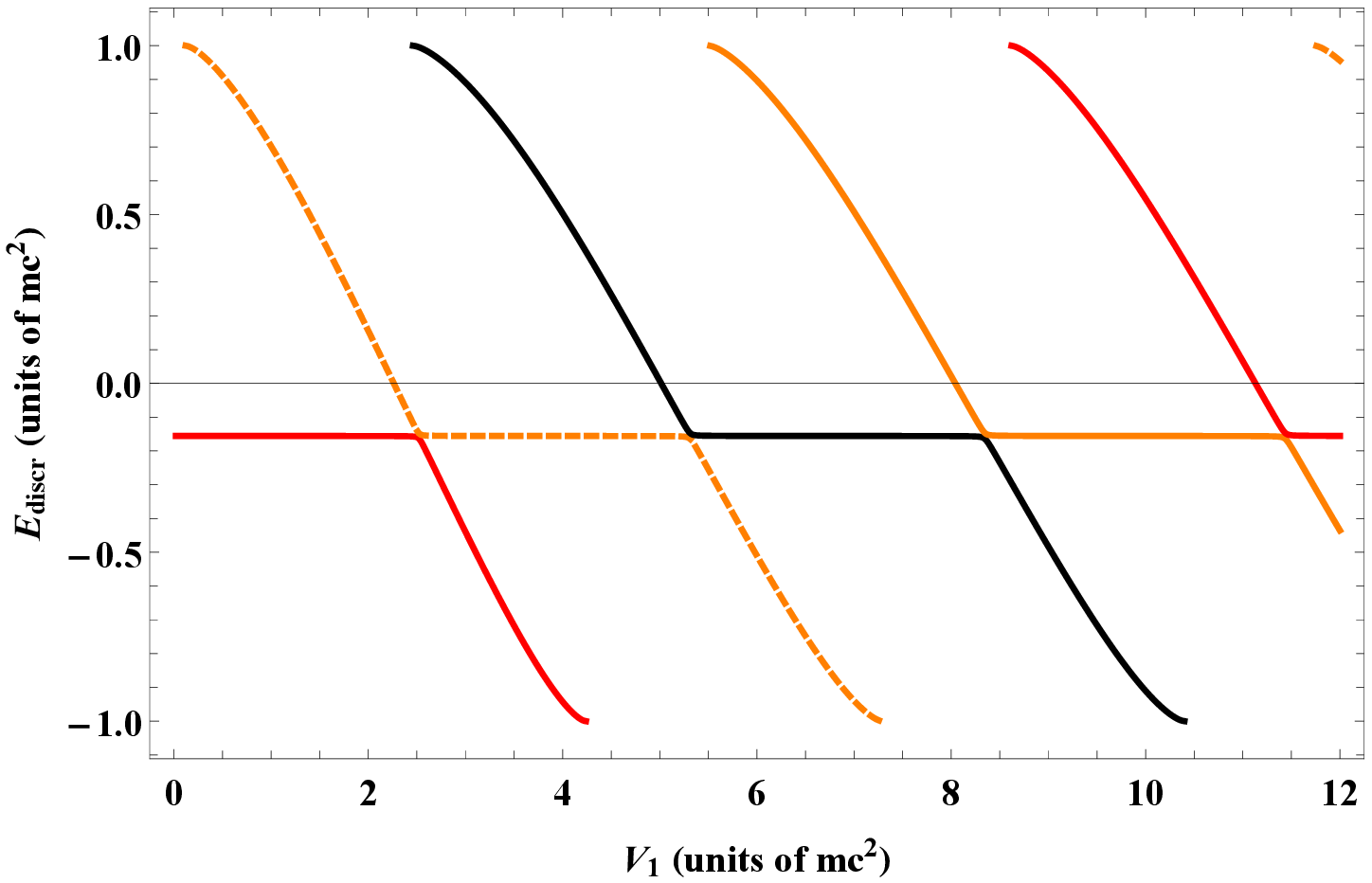}
}
\hfill
\subfigure[]{
		\includegraphics[width=\columnwidth]{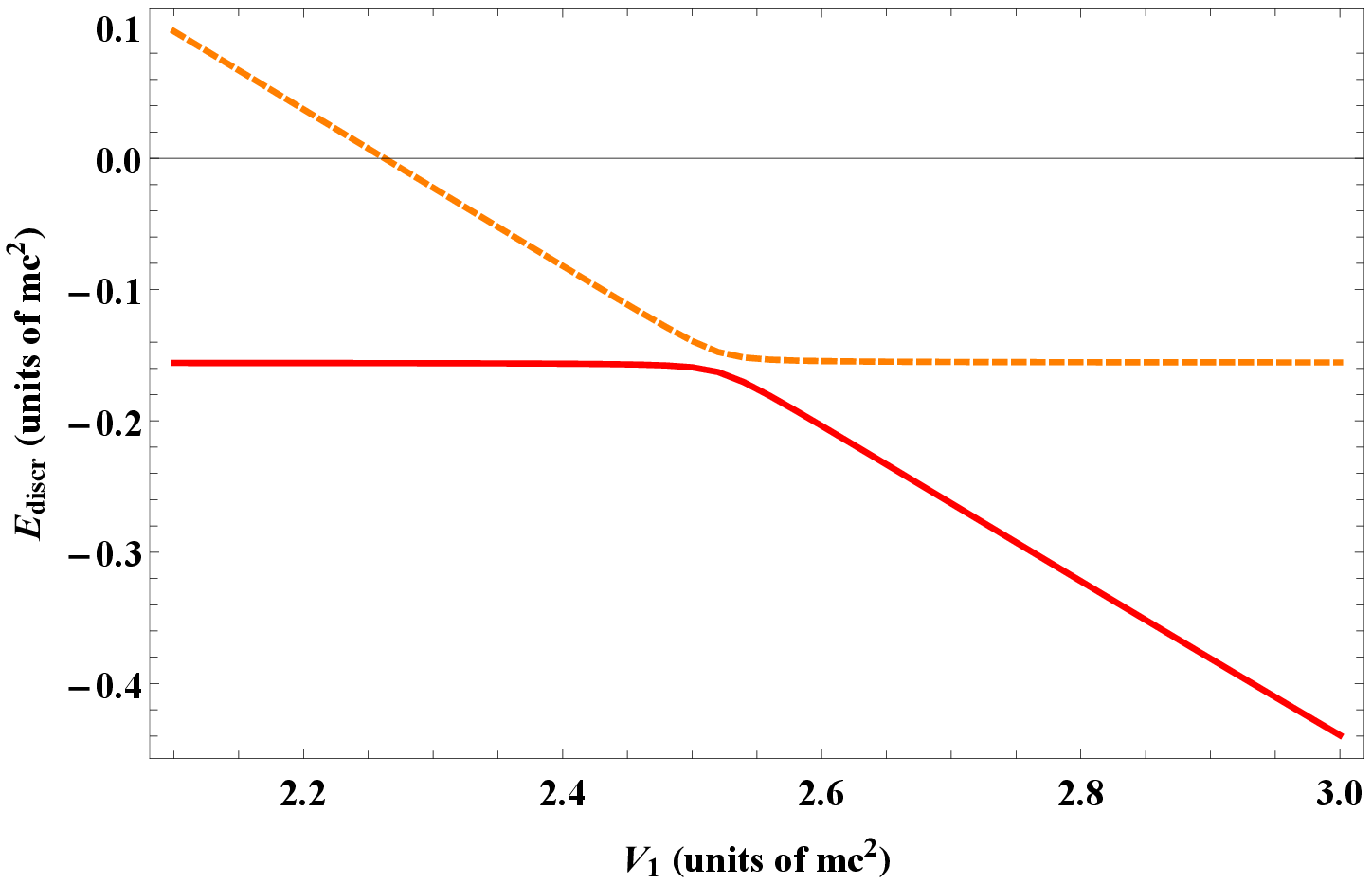}
}
\vfill
\subfigure[]{
		\includegraphics[width=\columnwidth]{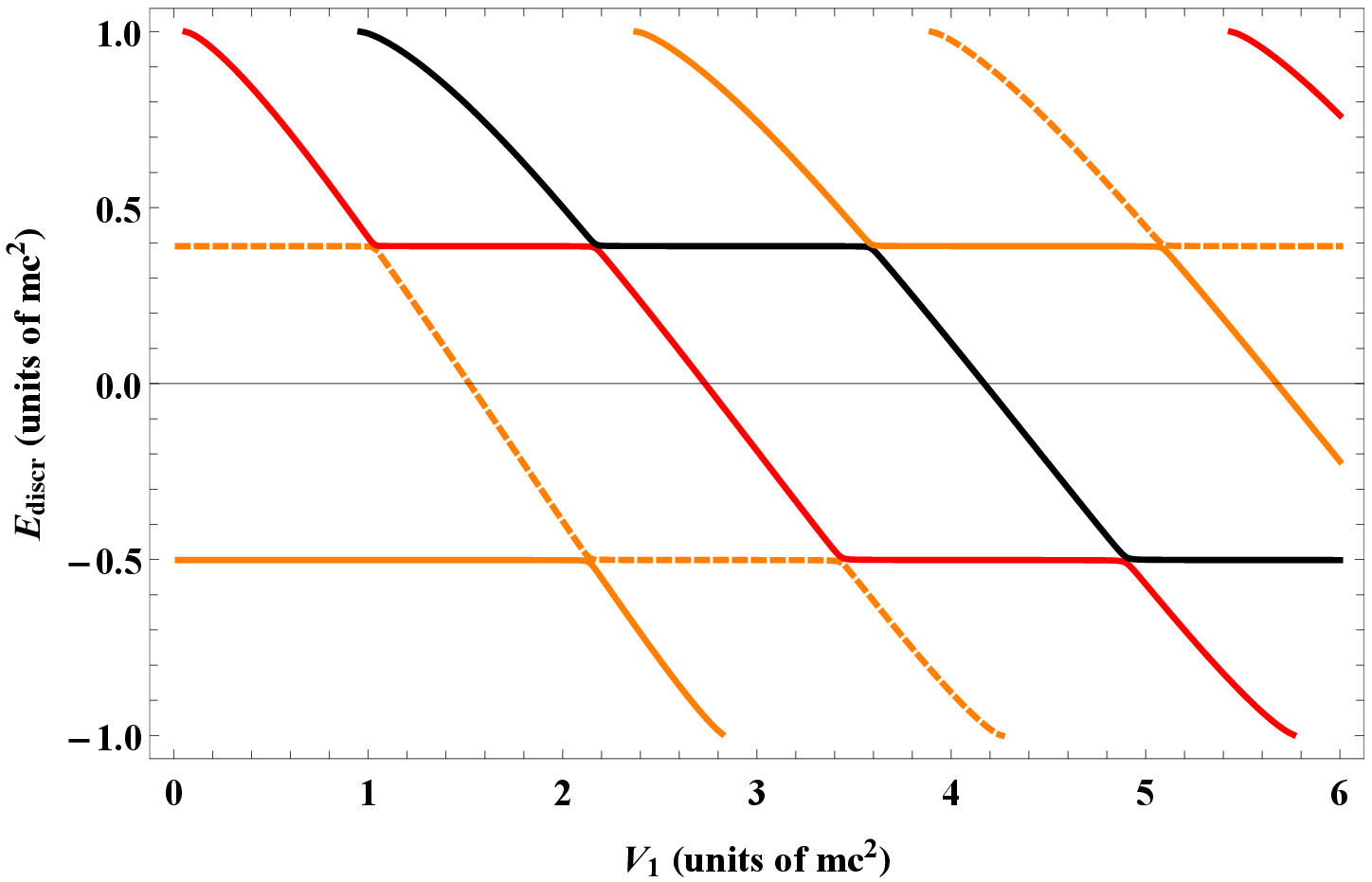}
}
\hfill
\subfigure[]{
		\includegraphics[width=\columnwidth]{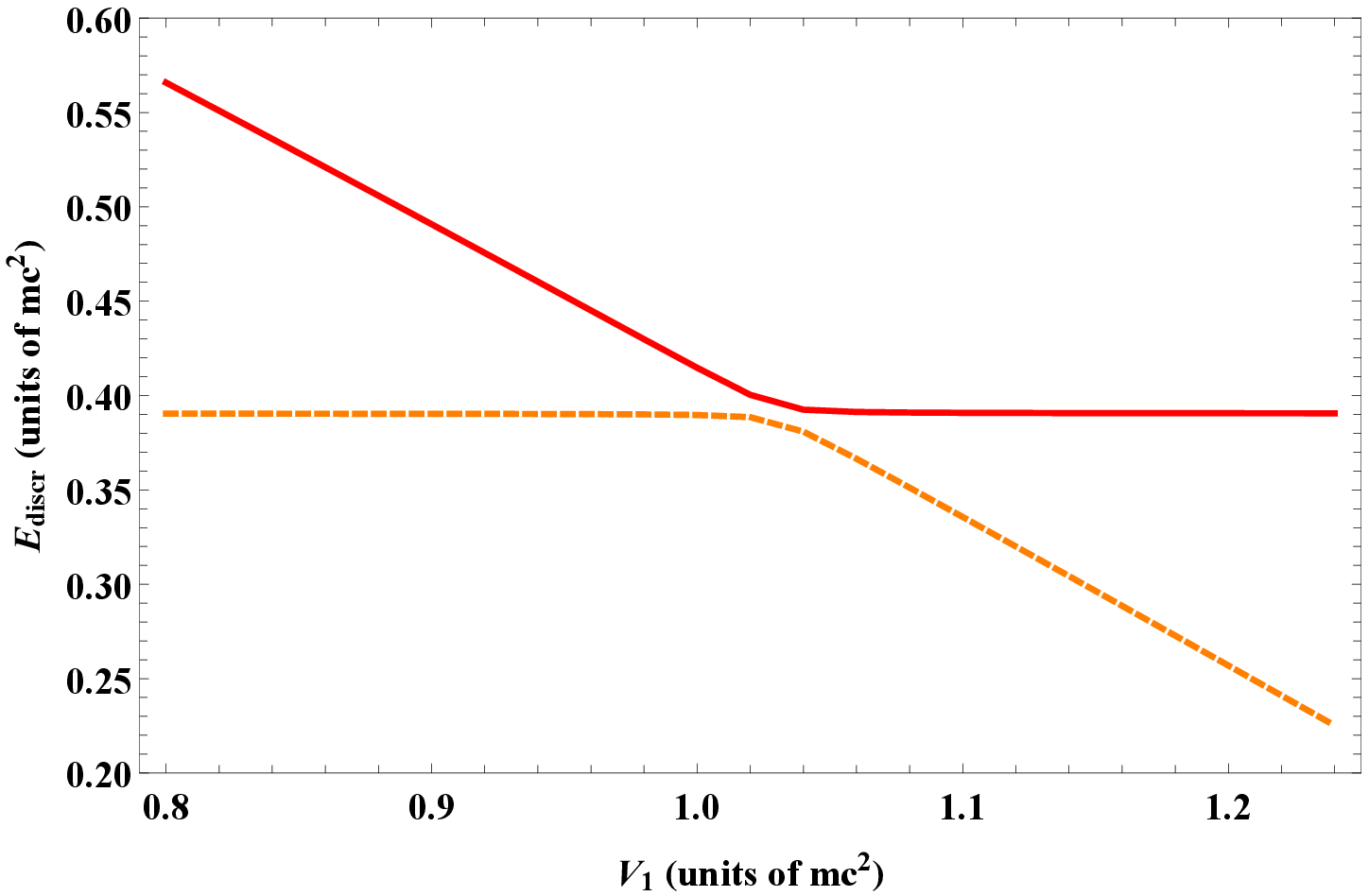}
}
\caption{(Color online). The behavior of the levels in the barrier-well system without antisymmetry of the potential in dependence on the well depth $V_1$ with fixed height of the barrier $V_2=-2$: (a,b) for $d=2$, $a=1$; (c,d) for $d=2$, $a=2$.  }
	\label{EdiscrV1}	
\end{figure*}
As a result, the leading term in the renormalized Casimir energy $\E_{int}^R(d)$ for the antisymmetric case  turns out to be the following
\beq\label{Casimir}\begin{gathered}
\E^R_{int}(d)=\E_{int}(d) + \L_{int}(d) V_0^2 \simeq \\ \simeq V^2_0\, {e^{-4 d} \over \sqrt{2 \pi d}}\, \left[ e^{-2 a} \sh^2 a -{A^2\over 2} \right] \ .
\end{gathered}\eeq
In (\ref{Casimir}) the multiplier in square brackets is sign-alternating in dependence on the single source parameters $(V_0\, , a)$. In particular, for the set  $a=1$ and $V_0=4.08, 7.4, 10$, considered in Sect.4, this multiplier is positive, hence, the sources reflect at large separations, whereas for $a=1$ and $V_0=8$ it is negative and so the sources attract. Note that in the last case the Casimir force changes from reflection to attraction by increasing  $d$. The behavior of $\E_{int}^R(d)$ starting from sufficiently small separations up to large $d$-asymptotics is shown in Figs.\ref{Eint3}.

Apart from these peculiar features, the general  answer for the Casimir force in the antisymmetric case is substantially different from the symmetric one, since now the asymptotics of the Casimir force for large separations between sources is subject of the standard $\exp (-2 m s)$ law. Moreover, it is the unique specifics of the source-anti-source system, since it is the only  case, when the symmetry between the positive and negative energy eigenstates according to (\ref{oddness}) takes place. The direct consequence of this symmetry is that the separate contributions from negative discrete levels are absent both in final expressions (\ref{rhoVP1}) and (\ref{IntWronskReg1}) for $\r_{vac}(x)$ and $\E_{vac}(d)$. Indeed this circumstance underlies the standard $\exp (-2 m s)$ fall-down of the Casimir force for large separations between sources, since in the symmetric case the breakdown of the latter is caused by the contribution from the negative discrete levels.  Namely, the main contribution to the asymptotics  of $\E_{vac}^{int}(d)$ will be given by the lowest  $\e_0<0$ according to eq. (\ref{e0<0}). As soon as the strict antisymmetry of the external potential (\ref{w2}) is broken, the spectrum immediately transforms into the standard non-symmetric form, where the levels are able to approach the threshold of the lower continuum, for instance, with growing depth of the well. This circumstance  reminds the well-known quantum-mechanical effect, when in the one-dimensional potential well with arbitrary small depth and size, but with equal height of both walls, there exists always at least one discrete level, which can be very shallow, but disappears as soon as the height of the walls becomes different. As an illustration of this property of the antisymmetric case in Figs.\ref{EdiscrV1} the behavior of the levels in the barrier-well system without antisymmetry of the potential $W_2(x)$ in dependence on the well depth $V_1$ with fixed height of the barrier $V_2$ is shown.

\subsection*{7. Conclusion}

To conclude, in this work by means of  the $\ln\text{[Wronskian]}$ contour integration techniques for calculating the Casimir effect we have shown the magnitude of the Casimir force variability for two short-range Coulomb sources, embedded in the background of one dimensional massive Dirac fermions. The main result is that  essentially non-perturbative vacuum QED-effects, including the effects of super-criticality, are able to add a set of new properties to Casimir forces between such sources, which turn out to be more diverse compared to the case of scattering potentials with scalar coupling to fermions, considered in Refs.~\cite{Jaffe2004, nanotubes}. In particular, we have shown  that   the interaction energy between two identical positively charged short-range Coulomb sources can exceed sufficiently large negative values and simultaneously reveal some  features similar to a long-range force, like the electronic Casimir force between two impurities on a one-dimensional semiconductor quantum wire  despite nonzero effective mass of the mediator ~\cite{tanaka2013}, which could significantly alter the properties of such  quasi-one-dimensional QED-systems.

The most intriguing circumstance here is that in the symmetric case  their mutual interaction is governed first of all by the structure of the discrete spectrum  of the single source, in dependence on which it can be tuned to give an attractive, a repulsive, or an (almost) compensated   Casimir force with various rates of the exponential fall-down, quite different from the standard $\exp (-2 m s)$ law. Let us mention once more that the essence of the long-range interaction between sources, which appears whenever the single well contains a level $\e_0$ close to the lower threshold,  is that under these conditions the exponential fall-down starts  at extremely large distances $d \gg \(1-\e_0^2\)^{-1/2}$ between sources, rather than by replacement of the exponential asymptotics by a power-like behavior, what could happen only for a massless mediator. No less interesting is the pattern of Casimir interaction observed  in the $\d$-limit with sources of negligible width, which  can also be explored in detail within the presented $\ln\text{[Wronskian]}$ contour integration approach. The latter circumstance could be quite important, since in some reasonable cases the best description for impurities is achieved indeed in the $\d$-limit.

A special attention should be paid to the antisymmetric source-anti-source system, which reveals quite different features. In particular, in this case there is no possibility for the long-range interaction between sources. The asymptotics of the Casimir force follows the standard $\exp (-2 m s)$ law. Moreover, the symmetric and antisymmetric cases are substantially different for small separations between sources. Namely, there follows from Figs.\ref{Eint1}-\ref{EintD},\ref{Eint3} which are calculated for the same sets of single source parameters up to replacement well-barrier, that in the symmetric case the Casimir interaction between sources is attractive, while in the antisymmetric one it turns into sufficiently strong repulsion. Remarkably enough, the classic electrostatic force for  such Coulomb sources should be of opposite sign. There is no evident explanation for this effect. However, the set of parameters used is quite wide to consider this effect as a general one.

These results may be relevant for indirect interactions between charged defects and adsorbed species in  quasi-one-dimensional QED systems mentioned above.

\subsection*{7. Acknowledgments}

The authors are very indebted to Prof. P.K.Silaev and Dr. O.V.Pavlovsky  from MSU Department of Physics for interest and helpful discussions. KS is especially grateful to Ms. A.Kondakova, MSU Department of Physics, Solid State Division for  information on the current situation in the research of quantum wires.  This work has been supported in part by the RF Ministry of Sc. $\&$ Ed.  Scientific Research Program, projects No. 01-2014-63889, A16-116021760047-5, and by RFBR grant No. 14-02-01261.

\end{document}